\documentclass[longauth]{aa}
\usepackage[utf8]{inputenc}
\usepackage{graphicx}
\usepackage{threeparttable}
\usepackage[varg]{txfonts}
\usepackage{amsmath}
\usepackage[breaklinks, colorlinks, citecolor=blue, linkcolor=blue]{hyperref}
\usepackage[english]{babel}
\usepackage[normalem]{ulem}
\usepackage{natbib}
\usepackage{bibspacing}
\usepackage{nicefrac}
\usepackage{fancyhdr}
\newcommand\corot{{CoRoT}}
\newcommand\tess{{TESS}}
\newcommand\fies{{FIES}}
\newcommand\sophie{{SOPHIE}}
\newcommand\cheops{{CHEOPS}}
\newcommand\kepler{{\it Kepler}}

\newcommand\harps{{HARPS}}
\newcommand\hires{{HIRES}}
\newcommand\gaia{{\it Gaia}}
\newcommand\lcogt{{LCOGT}}
\newcommand{\espresso}{{ESPRESSO}}
\newcommand\harpsn{{HARPS-N}}

\newcommand\target{{HD\,93963}}

\newcommand{\ms}{${\rm m\,s^{-1}}$}
\newcommand{\kms}{${\rm km\,s^{-1}}$}

\newcommand{\rrbTESSone}[1][ ]   {$0.01147 _{ - 0.00071 } ^ { + 0.00068 }$~#1} 
\newcommand{\rrbCHEOPSone}[1][ ]   {$0.01202 \pm 0.00040$~#1} 
\newcommand{\rpbTESSone}[1][$\mathrm{R}_{\oplus}$]   {$1.304 _{ - 0.082 } ^ { + 0.079 }$~#1}
\newcommand{\rpbCHEOPSone}[1][$\mathrm{R}_{\oplus}$]   {$1.366 \pm 0.047$~#1}

\newcommand{\depthbTESSone}[1][ppm]   {$131.5 \pm 16.2$~#1} 
 
\newcommand{\depthbCHEOPSone}[1][ppm]   {$144.37 _{ - 9.5 } ^ { + 9.6 }$~#1}

\newcommand{\rrcTESSone}[1][ ]   {$0.02850 \pm 0.00057$~#1} 
\newcommand{\rrcCHEOPSone}[1][ ]   {$0.02826 \pm 0.00056$~#1} 
\newcommand{\rpcTESSone}[1][$\mathrm{R}_{\oplus}$]   {$3.243 \pm 0.071$~#1} 
\newcommand{\rpcCHEOPSone}[1][$\mathrm{R}_{\oplus}$]   {$3.21 \pm 0.070$~#1}

\newcommand{\depthcTESSone}[1][ppm]   {$812.51 \pm 33.01 $~#1} 
 
\newcommand{\depthcCHEOPSone}[1][ppm]   {$798.5 \pm 32.3$~#1}

\newcommand{\smass}[1][$\mathrm{M}_{\odot}$]{ $ 1.109 \pm 0.043 $ #1} 
\newcommand{\sradius}[1][$\mathrm{R}_{\odot}$]{ $1.043 \pm 0.009$ #1}
\newcommand{\stemp}[1][$\mathrm{K}$]{ $ 5987 \pm 64$ #1}
\newcommand{\Tzerob}[1][days]   {$1901.2774 \pm 0.0016$~#1} 
\newcommand{\Pb}[1][days]   {$1.0391353 _{ -0.0000045} ^ { + 0.0000049 }$~#1} 
\newcommand{\bb}[1][ ]   {$0.284 _{ - 0.126 } ^ { + 0.081 }$~#1} 
 
\newcommand{\rrbTESS}[1][ ]   {$0.01190 \pm 0.00036$~#1} 
\newcommand{\rpbTESS}[1][$\mathrm{R}_{\oplus}$]   {$1.35 \pm 0.042$~#1} 
 
\newcommand{\ib}[1][deg]   {$86.21 _{ - 1.14 } ^ { + 1.69 }$~#1} 
\newcommand{\arb}[1][ ]   {$4.299 \pm 0.067$~#1} 
\newcommand{\ab}[1][AU]   {$0.02085 \pm 0.00037$~#1} 
\newcommand{\depthbTESS}[1][ppm]   {$141.5 _{ - 8.3 } ^ { + 8.5 }$~#1} 
 
\newcommand{\insolationb}[1][${\rm F_{\oplus}}$]   {$2898.0 _{ - 147.0 } ^ { + 157.0 }$~#1} 
 
\newcommand{\densspb}[1][${\rm g\,cm^{-3}}$]   {$1.379 \pm 0.065$~#1} 
\newcommand{\Teqb}[1][K]   {$2042 \pm 27$~#1} 
\newcommand{\ttotb}[1][hours]   {$1.812 \pm 0.040$~#1}

\newcommand{\Tzeroc}[1][days]   {$1902.87274 _{ - 0.00095 } ^ { + 0.00100 }$~#1} 
\newcommand{\Pc}[1][days]   {$3.6451398 _{ - 0.0000142 } ^ { +0.0000106}$~#1} 
\newcommand{\bc}[1][ ]   {$0.640 \pm 0.026$~#1} 
\newcommand{\rrcTESS}[1][ ]   {$0.02838 \pm 0.00045$~#1} 
\newcommand{\rpcTESS}[1][$\mathrm{R}_{\oplus}$]   {$3.228 \pm 0.059$~#1} 
 
\newcommand{\ic}[1][deg]   {$86.31 \pm 0.19$~#1} 
\newcommand{\arc}[1][ ]   {$9.92 \pm 0.15$~#1} 
\newcommand{\ac}[1][AU]   {$0.04813 \pm 0.00085$~#1} 
\newcommand{\depthcTESS}[1][ppm]   {$805.2 _{ - 25.5 } ^ { + 25.5 }$~#1} 
 
\newcommand{\insolationc}[1][${\rm F_{\oplus}}$]   {$543.8 _{ - 27.6 } ^ { + 29.4 }$~#1}

\newcommand{\Teqc}[1][K] {$1344 \pm 18$~#1} 
\newcommand{\ttotc}[1][hours]   {$2.268 _{ - 0.042 } ^ { + 0.045 }$~#1}

\newcommand{\qoneTESS}[1][]   {$0.30 \pm 0.09$~#1} 
\newcommand{\qtwoTESS}[1][]   {$0.24 \pm 0.10$~#1} 
\newcommand{\uoneTESS}[1][]   {$0.25 \pm 0.11 $~#1} 
\newcommand{\utwoTESS}[1][]   {$0.28 \pm 0.12 $~#1} 
\newcommand{\qoneCHEOPS}[1][]   {$0.52 \pm 0.09$~#1} 
\newcommand{\qtwoCHEOPS}[1][]   {$0.32 \pm 0.10$~#1} 
\newcommand{\uoneCHEOPS}[1][]   {$0.46 \pm 0.14$~#1} 
\newcommand{\utwoCHEOPS}[1][]   {$0.26 \pm 0.14$~#1} 
\newcommand{\jtrTESS}[1][]   {$0.0005461 \pm 0.0000032$~#1} 
\newcommand{\jtrCHEOPS}[1][]   {$0.0002106 \pm 0.0000024$~#1} 

\newcommand{\kb}[1][${\rm m\,s^{-1}}$]   {$4.6 \pm 1.9$~#1} 
\newcommand{\mpb}[1][$M_{\oplus}$]   {$7.8 \pm 3.2$~#1} 
\newcommand{\kc}[1][${\rm m\,s^{-1}}$]   {$7.4 \pm1.6$~#1} 
\newcommand{\mpc}[1][$\mathrm{M}_{\oplus}$]   {$19.2 \pm 4.1$~#1} 
 
\newcommand{\Pd}[1][days]   {$11.02_{-0.08}^{+0.05}$~#1} 
\newcommand{\kd}[1][${\rm m\,s^{-1}}$]   {$10.4 \pm 2.1$~#1} 
\newcommand{\SOPHIE}[1][${\rm m\,s^{-1}}$]   {$13.225_{-0.010}^{+0.009}$~#1} 
\newcommand{\jSOPHIE}[1][${\rm m\,s^{-1}}$]   {$4.78_{-1.04}^{+1.26}$~#1} 
\newcommand{\ltrend}[1][${\rm m\,s^{-1}\,d^{-1}}$]   {$0.134_{-0.022}^{+0.024}$~#1}

\makeatletter
\renewcommand*\aa@pageof{, page \thepage{} of \pageref*{LastPage}}
\def\instrefs#1{{\def\scsep{\def\scsep{,}}\@for\w:=#1\do{\scsep\ref{inst:\w}}}}
\makeatother

\begin{document} 
\title{The HD\,93963\,A transiting system: A 1.04\,d super-Earth \\ and a 3.65\,d sub-Neptune discovered by \tess\ and \cheops}
   \titlerunning{The \target\,A planetary system}
   \subtitle{}

   \author{L. M. Serrano\inst{1} $^{\href{https://orcid.org/0000-0001-9211-3691 }{\includegraphics[width= 2mm]{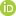}}}$,
D. Gandolfi\inst{1} $^{\href{https://orcid.org/0000-0001-8627-9628 }{\includegraphics[width= 2mm]{figures/orcid.jpg}}}$,
S. Hoyer\inst{2} $^{\href{https://orcid.org/0000-0003-3477-2466}{\includegraphics[width= 2mm]{figures/orcid.jpg}}}$,
A. Brandeker\inst{3} $^{\href{https://orcid.org/0000-0002-7201-7536}{\includegraphics[width= 2mm]{figures/orcid.jpg}}}$,
M. J. Hooton\inst{4} $^{\href{https://orcid.org/0000-0003-0030-332X}{\includegraphics[width= 2mm]{figures/orcid.jpg}}}$,
S. Sousa\inst{5} $^{\href{https://orcid.org/0000-0001-9047-2965}{\includegraphics[width= 2mm]{figures/orcid.jpg}}}$,
F. Murgas\inst{6,7} $^{\href{https://orcid.org/0000-0001-9087-1245}{\includegraphics[width= 2mm]{figures/orcid.jpg}}}$,
D. R. Ciardi\inst{8} $^{\href{https://orcid.org/0000-0002-5741-3047}{\includegraphics[width= 2mm]{figures/orcid.jpg}}}$,
S. B. Howell\inst{9 }, 
W. Benz\inst{4,10} $^{\href{https://orcid.org/0000-0001-7896-6479}{\includegraphics[width= 2mm]{figures/orcid.jpg}}}$,
N. Billot\inst{11} $^{\href{https://orcid.org/0000-0003-3429-3836}{\includegraphics[width= 2mm]{figures/orcid.jpg}}}$,
H.-G. Florén\inst{3,12 }, 
A. Bekkelien\inst{11 }, 
A. Bonfanti\inst{13} $^{\href{https://orcid.org/0000-0002-1916-5935}{\includegraphics[width= 2mm]{figures/orcid.jpg}}}$,
A. Krenn\inst{13} $^{\href{https://orcid.org/0000-0003-3615-4725}{\includegraphics[width= 2mm]{figures/orcid.jpg}}}$,
A. J. Mustill\inst{14} $^{\href{https://orcid.org/0000-0002-2086-3642}{\includegraphics[width= 2mm]{figures/orcid.jpg}}}$,
T. G. Wilson\inst{15} $^{\href{https://orcid.org/0000-0001-8749-1962}{\includegraphics[width= 2mm]{figures/orcid.jpg}}}$,
H. Osborn\inst{10,16} $^{\href{https://orcid.org/0000-0002-4047-4724}{\includegraphics[width= 2mm]{figures/orcid.jpg}}}$,
H. Parviainen\inst{6,7} $^{\href{https://orcid.org/0000-0001-5519-1391}{\includegraphics[width= 2mm]{figures/orcid.jpg}}}$,
N. Heidari\inst{17,18,2} $^{\href{https://orcid.org/0000-0002-2370-0187}{\includegraphics[width= 2mm]{figures/orcid.jpg}}}$,
E. Pallé\inst{6,7} $^{\href{https://orcid.org/0000-0003-0987-1593}{\includegraphics[width= 2mm]{figures/orcid.jpg}}}$,
M. Fridlund\inst{19,20} $^{\href{https://orcid.org/0000-0002-0855-8426}{\includegraphics[width= 2mm]{figures/orcid.jpg}}}$,
V. Adibekyan\inst{5} $^{\href{https://orcid.org/0000-0002-0601-6199}{\includegraphics[width= 2mm]{figures/orcid.jpg}}}$,
L. Fossati\inst{13} $^{\href{https://orcid.org/0000-0003-4426-9530}{\includegraphics[width= 2mm]{figures/orcid.jpg}}}$,
M. Deleuil\inst{2} $^{\href{https://orcid.org/0000-0001-6036-0225}{\includegraphics[width= 2mm]{figures/orcid.jpg}}}$,
E. Knudstrup\inst{21,22} $^{\href{https://orcid.org/0000-0001-7880-594X}{\includegraphics[width= 2mm]{figures/orcid.jpg}}}$,
K. A. Collins\inst{23} $^{\href{https://orcid.org/0000-0002-4317-142X}{\includegraphics[width= 2mm]{figures/orcid.jpg}}}$,
K. W. F. Lam\inst{24} $^{\href{https://orcid.org/0000-0002-9910-6088 }{\includegraphics[width= 2mm]{figures/orcid.jpg}}}$,
S. Grziwa\inst{25} $^{\href{https://orcid.org/0000-0003-3370-4058}{\includegraphics[width= 2mm]{figures/orcid.jpg}}}$,
S. Salmon\inst{11} $^{\href{https://orcid.org/0000-0002-1714-3513}{\includegraphics[width= 2mm]{figures/orcid.jpg}}}$,
S. H. Albrecht\inst{21} $^{\href{https://orcid.org/0000-0003-1762-8235}{\includegraphics[width= 2mm]{figures/orcid.jpg}}}$,
Y. Alibert\inst{4} $^{\href{https://orcid.org/0000-0002-4644-8818}{\includegraphics[width= 2mm]{figures/orcid.jpg}}}$,
R. Alonso\inst{6,26} $^{\href{https://orcid.org/0000-0001-8462-8126}{\includegraphics[width= 2mm]{figures/orcid.jpg}}}$,
G. Anglada-Escudé\inst{27,28} $^{\href{https://orcid.org/0000-0002-3645-5977}{\includegraphics[width= 2mm]{figures/orcid.jpg}}}$,
T. Bárczy\inst{29} $^{\href{https://orcid.org/0000-0002-7822-4413}{\includegraphics[width= 2mm]{figures/orcid.jpg}}}$,
D. Barrado y Navascues\inst{30} $^{\href{https://orcid.org/0000-0002-5971-9242}{\includegraphics[width= 2mm]{figures/orcid.jpg}}}$,
S. C. C. Barros\inst{5,31} $^{\href{https://orcid.org/0000-0003-2434-3625}{\includegraphics[width= 2mm]{figures/orcid.jpg}}}$,
W. Baumjohann\inst{13} $^{\href{https://orcid.org/0000-0001-6271-0110}{\includegraphics[width= 2mm]{figures/orcid.jpg}}}$,
M. Beck\inst{11} $^{\href{https://orcid.org/0000-0003-3926-0275 }{\includegraphics[width= 2mm]{figures/orcid.jpg}}}$,
T. Beck\inst{4 }, 
A. Bieryla\inst{32} $^{\href{https://orcid.org/0000-0001-6637-5401}{\includegraphics[width= 2mm]{figures/orcid.jpg}}}$,
X. Bonfils\inst{33} $^{\href{https://orcid.org/0000-0001-9003-8894}{\includegraphics[width= 2mm]{figures/orcid.jpg}}}$,
P. T. Boyd\inst{34} $^{\href{https://orcid.org/0000-0003-0442-4284}{\includegraphics[width= 2mm]{figures/orcid.jpg}}}$,
C. Broeg\inst{4,10} $^{\href{https://orcid.org/0000-0001-5132-2614}{\includegraphics[width= 2mm]{figures/orcid.jpg}}}$,
J. Cabrera\inst{24 }, 
S. Charnoz\inst{35} $^{\href{https://orcid.org/0000-0002-7442-491X}{\includegraphics[width= 2mm]{figures/orcid.jpg}}}$,
B. Chazelas\inst{11} $^{\href{https://orcid.org/0000-0002-9438-5057}{\includegraphics[width= 2mm]{figures/orcid.jpg}}}$,
J. L. Christiansen\inst{8} $^{\href{https://orcid.org/0000-0002-8035-4778}{\includegraphics[width= 2mm]{figures/orcid.jpg}}}$,
A. Collier Cameron\inst{15} $^{\href{https://orcid.org/0000-0002-8863-7828}{\includegraphics[width= 2mm]{figures/orcid.jpg}}}$,
P. Cortés-Zuleta\inst{2} $^{\href{https://orcid.org/0000-0002-6174-4666}{\includegraphics[width= 2mm]{figures/orcid.jpg}}}$,
Sz. Csizmadia\inst{24} $^{\href{https://orcid.org/0000-0001-6803-9698}{\includegraphics[width= 2mm]{figures/orcid.jpg}}}$,
M. B. Davies\inst{36} $^{\href{https://orcid.org/0000-0001-6080-1190 }{\includegraphics[width= 2mm]{figures/orcid.jpg}}}$,
A. Deline\inst{11 }, 
L. Delrez\inst{37,38,11} $^{\href{https://orcid.org/0000-0001-6108-4808}{\includegraphics[width= 2mm]{figures/orcid.jpg}}}$,
O. D. S. Demangeon\inst{5,31} $^{\href{https://orcid.org/0000-0001-7918-0355}{\includegraphics[width= 2mm]{figures/orcid.jpg}}}$,
B.-O. Demory\inst{10} $^{\href{https://orcid.org/0000-0002-9355-5165}{\includegraphics[width= 2mm]{figures/orcid.jpg}}}$,
A. Dunlavey\inst{39,8 }, 
D. Ehrenreich\inst{11} $^{\href{https://orcid.org/0000-0001-9704-5405}{\includegraphics[width= 2mm]{figures/orcid.jpg}}}$,
A. Erikson\inst{24 }, 
A. Fortier\inst{4,10} $^{\href{https://orcid.org/0000-0001-8450-3374}{\includegraphics[width= 2mm]{figures/orcid.jpg}}}$,
A. Fukui\inst{40,41} $^{\href{https://orcid.org/0000-0002-4909-5763}{\includegraphics[width= 2mm]{figures/orcid.jpg}}}$,
Z. Garai\inst{42,43,44,45} $^{\href{https://orcid.org/0000-0001-9483-2016}{\includegraphics[width= 2mm]{figures/orcid.jpg}}}$,
M. Gillon\inst{37} $^{\href{https://orcid.org/0000-0003-1462-7739}{\includegraphics[width= 2mm]{figures/orcid.jpg}}}$,
M. Güdel\inst{46 }, 
G. Hébrard\inst{47,48} $^{\href{https://orcid.org/0000-0001-5450-7067}{\includegraphics[width= 2mm]{figures/orcid.jpg}}}$,
K. Heng\inst{10,49} $^{\href{https://orcid.org/0000-0003-1907-5910 }{\includegraphics[width= 2mm]{figures/orcid.jpg}}}$,
C. X. Huang\inst{16} $^{\href{https://orcid.org/0000-0003-0918-7484}{\includegraphics[width= 2mm]{figures/orcid.jpg}}}$,
K. G. Isaak\inst{50} $^{\href{https://orcid.org/0000-0001-8585-1717}{\includegraphics[width= 2mm]{figures/orcid.jpg}}}$,
J. M. Jenkins\inst{51} $^{\href{https://orcid.org/0000-0002-4715-9460}{\includegraphics[width= 2mm]{figures/orcid.jpg}}}$,
L. L. Kiss\inst{52 }, 
J. Laskar\inst{53} $^{\href{https://orcid.org/0000-0003-2634-789X}{\includegraphics[width= 2mm]{figures/orcid.jpg}}}$,
D. W. Latham\inst{32} $^{\href{https://orcid.org/0000-0001-9911-7388}{\includegraphics[width= 2mm]{figures/orcid.jpg}}}$,
A. Lecavelier des Etangs\inst{54} $^{\href{https://orcid.org/0000-0002-5637-5253}{\includegraphics[width= 2mm]{figures/orcid.jpg}}}$,
M. Lendl\inst{11} $^{\href{https://orcid.org/0000-0001-9699-1459}{\includegraphics[width= 2mm]{figures/orcid.jpg}}}$,
A. M. Levine\inst{16} $^{\href{https://orcid.org/0000-0001-8172-0453}{\includegraphics[width= 2mm]{figures/orcid.jpg}}}$,
C. Lovis\inst{11} $^{\href{https://orcid.org/0000-0001-7120-5837}{\includegraphics[width= 2mm]{figures/orcid.jpg}}}$,
M. B. Lund\inst{8} $^{\href{https://orcid.org/0000-0003-2527-1598}{\includegraphics[width= 2mm]{figures/orcid.jpg}}}$,
D. Magrin\inst{55} $^{\href{https://orcid.org/0000-0003-0312-313X}{\includegraphics[width= 2mm]{figures/orcid.jpg}}}$,
P. F. L. Maxted\inst{56} $^{\href{https://orcid.org/0000-0003-3794-1317 }{\includegraphics[width= 2mm]{figures/orcid.jpg}}}$,
N. Narita\inst{40,57,41} $^{\href{https://orcid.org/0000-0001-8511-2981}{\includegraphics[width= 2mm]{figures/orcid.jpg}}}$,
V. Nascimbeni\inst{55} $^{\href{https://orcid.org/0000-0001-9770-1214}{\includegraphics[width= 2mm]{figures/orcid.jpg}}}$,
G. Olofsson\inst{3} $^{\href{https://orcid.org/0000-0003-3747-7120 }{\includegraphics[width= 2mm]{figures/orcid.jpg}}}$,
R. Ottensamer\inst{58 }, 
I. Pagano\inst{59} $^{\href{https://orcid.org/0000-0001-9573-4928 }{\includegraphics[width= 2mm]{figures/orcid.jpg}}}$,
A. C. S. V. Pessanha\inst{60 }, 
G. Peter\inst{61} $^{\href{https://orcid.org/0000-0001-6101-2513}{\includegraphics[width= 2mm]{figures/orcid.jpg}}}$,
G. Piotto\inst{55,62} $^{\href{https://orcid.org/0000-0002-9937-6387}{\includegraphics[width= 2mm]{figures/orcid.jpg}}}$,
D. Pollacco\inst{49 }, 
D. Queloz\inst{63,64} $^{\href{https://orcid.org/0000-0002-3012-0316}{\includegraphics[width= 2mm]{figures/orcid.jpg}}}$,
R. Ragazzoni\inst{55,62} $^{\href{https://orcid.org/0000-0002-7697-5555}{\includegraphics[width= 2mm]{figures/orcid.jpg}}}$,
N. Rando\inst{65 }, 
F. Ratti\inst{65 }, 
H. Rauer\inst{24,66,67} $^{\href{https://orcid.org/0000-0002-6510-1828 }{\includegraphics[width= 2mm]{figures/orcid.jpg}}}$,
I. Ribas\inst{27,28} $^{\href{https://orcid.org/0000-0002-6689-0312 }{\includegraphics[width= 2mm]{figures/orcid.jpg}}}$,
G. Ricker\inst{16} $^{\href{https://orcid.org/0000-0003-2058-6662}{\includegraphics[width= 2mm]{figures/orcid.jpg}}}$,
P. Rowden\inst{68} $^{\href{https://orcid.org/0000-0002-4829-7101}{\includegraphics[width= 2mm]{figures/orcid.jpg}}}$,
N. C. Santos\inst{5,31} $^{\href{https://orcid.org/0000-0003-4422-2919}{\includegraphics[width= 2mm]{figures/orcid.jpg}}}$,
G. Scandariato\inst{59} $^{\href{https://orcid.org/0000-0003-2029-0626}{\includegraphics[width= 2mm]{figures/orcid.jpg}}}$,
S. Seager\inst{16,69,70} $^{\href{https://orcid.org/0000-0002-6892-6948}{\includegraphics[width= 2mm]{figures/orcid.jpg}}}$,
D. Ségransan\inst{11} $^{\href{https://orcid.org/0000-0003-2355-8034 }{\includegraphics[width= 2mm]{figures/orcid.jpg}}}$,
A. E. Simon\inst{4} $^{\href{https://orcid.org/0000-0001-9773-2600}{\includegraphics[width= 2mm]{figures/orcid.jpg}}}$,
A. M. S. Smith\inst{24} $^{\href{https://orcid.org/0000-0002-2386-4341}{\includegraphics[width= 2mm]{figures/orcid.jpg}}}$,
M. Steller\inst{13} $^{\href{https://orcid.org/0000-0003-2459-6155}{\includegraphics[width= 2mm]{figures/orcid.jpg}}}$,
Gy. M. Szabó\inst{43,44 }, 
N. Thomas\inst{4 }, 
J. D. Twicken\inst{71} $^{\href{https://orcid.org/0000-0002-6778-7552}{\includegraphics[width= 2mm]{figures/orcid.jpg}}}$,
S. Udry\inst{11} $^{\href{https://orcid.org/0000-0001-7576-6236}{\includegraphics[width= 2mm]{figures/orcid.jpg}}}$,
B. Ulmer\inst{61 }, 
V. Van Grootel\inst{38} $^{\href{https://orcid.org/0000-0003-2144-4316}{\includegraphics[width= 2mm]{figures/orcid.jpg}}}$,
R. Vanderspek\inst{16} $^{\href{https://orcid.org/0000-0001-6763-6562}{\includegraphics[width= 2mm]{figures/orcid.jpg}}}$,
V. Viotto\inst{55} $^{\href{https://orcid.org/0000-0001-5700-9565}{\includegraphics[width= 2mm]{figures/orcid.jpg}}}$,
N. Walton\inst{72 }}
   \authorrunning{L. M. Serrano et al.}
   \institute{\label{inst:1}Dipartimento di Fisica, Università degli Studi di Torino, via Pietro Giuria 1, I-10125, Torino, Italy
   \email{luisamaria.serrano@unito.it}
\and
\label{inst:2} Aix Marseille Univ, CNRS, CNES, LAM, 38 rue Frédéric Joliot-Curie, 13388 Marseille, France \and
\label{inst:3} Department of Astronomy, Stockholm University, AlbaNova University Center, 10691 Stockholm, Sweden \and
\label{inst:4} Physikalisches Institut, University of Bern, Gesellsschaftstrasse 6, 3012 Bern, Switzerland \and
\label{inst:5} Instituto de Astrofisica e Ciencias do Espaco, Universidade do Porto, CAUP, Rua das Estrelas, 4150-762 Porto, Portugal \and
\label{inst:6} Instituto de Astrofisica de Canarias, 38200 La Laguna, Tenerife, Spain \and
\label{inst:7} Dept. Astrof\'isica, Universidad de La Laguna (ULL), E-38206 La Laguna, Tenerife, Spain \and
\label{inst:8} NASA Exoplanet Science Institute – Caltech/IPAC, 1200 E. California Blvd, Pasadena, CA 91125 USA \and
\label{inst:9} Space Science and Astrobiology Division  NASA Ames Research Center M/S 245-6, Moffett Field, CA 94035 \and
\label{inst:10} Center for Space and Habitability, Gesellsschaftstrasse 6, 3012 Bern, Switzerland \and
\label{inst:11} Observatoire Astronomique de l'Université de Genève, Chemin Pegasi 51, Versoix, Switzerland \and
\label{inst:12} Department of Astronomy, Stockholm University, SE-106 91 Stockholm, Sweden \and
\label{inst:13} Space Research Institute, Austrian Academy of Sciences, Schmiedlstrasse 6, A-8042 Graz, Austria \and
\label{inst:14} Lund Observatory, Dept. of Astronomy and Theoretical Physics, Lund University, Box 43, 22100 Lund, Sweden \and
\label{inst:15} Centre for Exoplanet Science, SUPA School of Physics and Astronomy, University of St Andrews, North Haugh, St Andrews KY16 9SS, UK \and
\label{inst:16} Department of Physics and Kavli Institute for Astrophysics and Space Research, Massachusetts Institute of Technology, Cambridge, MA 02139, USA \and
\label{inst:17} Department of Physics, Shahid Beheshti University, Tehran, Iran \and
\label{inst:18} 2. Laboratoire J.-L. Lagrange, Observatoire de la C\^ote d’Azur (OCA), Universite de Nice-Sophia Antipolis (UNS), CNRS, Campus Valrose, 06108 Nice Cedex 2, France \and
\label{inst:19} Leiden Observatory, University of Leiden, PO Box 9513, 2300 RA Leiden, The Netherlands \and
\label{inst:20} Department of Space, Earth and Environment, Chalmers University of Technology, Onsala Space Observatory, 43992 Onsala, Sweden \and
\label{inst:21} Stellar Astrophysics Centre, Department of Physics and Astronomy, Aarhus University, Ny Munkegade 120, DK-8000 Aarhus C, Denmark \and
\label{inst:22} Nordic Optical Telescope, Rambla José Ana Fernández Pérez 7, E-38711 Breña Baja, Spain \and
\label{inst:23} Center for Astrophysics \textbar \ Harvard \& Smithsonian, 60 Garden Street, Cambridge, MA 02138, USA \and
\label{inst:24} Institute of Planetary Research, German Aerospace Center (DLR), Rutherfordstrasse 2, 12489 Berlin, Germany \and
\label{inst:25} Rheinisches Institut f\"ur Umweltforschung an der Universit\"at zu K\"oln, Aachener Strasse 209, 50931 K\"oln, Germany \and
\label{inst:26} Departamento de Astrofisica, Universidad de La Laguna, 38206 La Laguna, Tenerife, Spain \and
\label{inst:27} Institut de Ciencies de l'Espai (ICE, CSIC), Campus UAB, Can Magrans s/n, 08193 Bellaterra, Spain \and
\label{inst:28} Institut d'Estudis Espacials de Catalunya (IEEC), 08034 Barcelona, Spain \and
\label{inst:29} Admatis, 5. Kandó Kálmán Street, 3534 Miskolc, Hungary \and
\label{inst:30} Depto. de Astrofisica, Centro de Astrobiologia (CSIC-INTA), ESAC campus, 28692 Villanueva de la Cañada (Madrid), Spain \and
\label{inst:31} Departamento de Fisica e Astronomia, Faculdade de Ciencias, Universidade do Porto, Rua do Campo Alegre, 4169-007 Porto, Portugal \and
\label{inst:32} Harvard-Smithsonian Center for Astrophysics, 60 Garden Street, Office: P-333, Cambridge, MA 02138, MS-16, USA \and
\label{inst:33} Université Grenoble Alpes, CNRS, IPAG, 38000 Grenoble, France \and
\label{inst:34} NASA, Goddard Space Flight Center 8800 Greenbelt Rd, Greenbelt, MD 20771, USA \newpage \and
\label{inst:35} Université de Paris, Institut de physique du globe de Paris, CNRS, F-75005 Paris, France \and
\label{inst:36} Centre for Mathematical Sciences Lund University Box 118 SE 221 00, Lund Sweden \and
\label{inst:37} Astrobiology Research Unit, Université de Liège, Allée du 6 Août 19C, B-4000 Liège, Belgium \and
\label{inst:38} Space sciences, Technologies and Astrophysics Research (STAR) Institute, Université de Liège, Allée du 6 Août 19C, 4000 Liège, Belgium \and
\label{inst:39} University of California at Santa Cruz \and
\label{inst:40} Komaba Institute for Science, The University of Tokyo, 3-8-1 Komaba, Meguro, Tokyo 153-8902 \and
\label{inst:41} Instituto de Astrof\'{i}sica de Canarias (IAC), 38205 La Laguna, Tenerife, Spain \and
\label{inst:42} Astronomical Institute, Slovak Academy of Science, Stellar Department, Tatranská Lomnica, 05960 Vysoké Tatry, Slovakia \and
\label{inst:43} ELTE Eötvös Loránd University, Gothard Astrophysical Observatory, 9700 Szombathely, Szent Imre h. u. 112, Hungary \and
\label{inst:44} MTA-ELTE Exoplanet Research Group, 9700 Szombathely, Szent Imre h. u. 112, Hungary \and
\label{inst:45} MTA-ELTE Lendület Milky Way Research Group, 9700 Szombathely, Szent Imre h. u. 112, Hungary \and
\label{inst:46} University of Vienna, Department of Astrophysics, Türkenschanzstrasse 17, 1180 Vienna, Austria \and
\label{inst:47} Institut d'astrophysique de Paris, UMR7095 CNRS, Universit\'e Pierre \& Marie Curie, 98bis boulevard Arago, 75014 Paris, France \and
\label{inst:48} Observatoire de Haute-Provence, CNRS, Universit\'e d'Aix-Marseille, 04870 Saint-Michel-l'Observatoire, France \and
\label{inst:49} Department of Physics, University of Warwick, Gibbet Hill Road, Coventry CV4 7AL, United Kingdom \and
\label{inst:50} Science and Operations Department - Science Division (SCI-SC), Directorate of Science, European Space Agency (ESA), European Space Research and Technology Centre (ESTEC),
Keplerlaan 1, 2201-AZ Noordwijk, The Netherlands \and
\label{inst:51} NASA Ames Research Center, Mail Stop 269-3, Bldg. T35A, Rm. 102, P.O. Box 1, Moffett Field, CA 94035-0001, USA \and
\label{inst:52} Konkoly Observatory, Research Centre for Astronomy and Earth Sciences, 1121 Budapest, Konkoly Thege Miklós út 15-17, Hungary \and
\label{inst:53} IMCCE, UMR8028 CNRS, Observatoire de Paris, PSL Univ., Sorbonne Univ., 77 av. Denfert-Rochereau, 75014 Paris, France \and
\label{inst:54} Institut d'astrophysique de Paris, UMR7095 CNRS, Université Pierre \& Marie Curie, 98bis blvd. Arago, 75014 Paris, France \and
\label{inst:55} INAF, Osservatorio Astronomico di Padova, Vicolo dell'Osservatorio 5, 35122 Padova, Italy \and
\label{inst:56} Astrophysics Group, Keele University, Staffordshire, ST5 5BG, United Kingdom \and
\label{inst:57} Astrobiology Center, 2-21-1 Osawa, Mitaka, Tokyo 181-8588, Japan \and
\label{inst:58} Department of Astrophysics, University of Vienna, Tuerkenschanzstrasse 17, 1180 Vienna, Austria \and
\label{inst:59} INAF, Osservatorio Astrofisico di Catania, Via S. Sofia 78, 95123 Catania, Italy \and
\label{inst:60} Citizen Scientist, Marica, Rio de Janeiro, Brazil \and
\label{inst:61} Institute of Optical Sensor Systems, German Aerospace Center (DLR), Rutherfordstrasse 2, 12489 Berlin, Germany \and
\label{inst:62} Dipartimento di Fisica e Astronomia "Galileo Galilei", Universita degli Studi di Padova, Vicolo dell'Osservatorio 3, 35122 Padova, Italy \and
\label{inst:63} ETH Zurich, Department of Physics, Wolfgang-Pauli-Strasse 27, 8093 Zurich, Switzerland \and
\label{inst:64} Cavendish Laboratory, JJ Thomson Avenue, Cambridge CB3 0HE, UK \and
\label{inst:65} ESTEC, European Space Agency, 2201AZ, Noordwijk, NL \and
\label{inst:66} Center for Astronomy and Astrophysics, Technical University Berlin, Hardenberstrasse 36, 10623 Berlin, Germany \and
\label{inst:67} Institut für Geologische Wissenschaften, Freie UniversitÃ¤t Berlin, 12249 Berlin, Germany \and
\label{inst:68} Royal Astronomical Society, Burlington House, Piccadilly, London W1J 0BQ, UK \and
\label{inst:69} Department of Earth, Atmospheric and Planetary Sciences, Massachusetts Institute of Technology, Cambridge, MA 02139, USA \and
\label{inst:70} Department of Aeronautics and Astronautics, MIT, 77 Massachusetts Avenue, Cambridge, MA 02139, USA \and
\label{inst:71} SETI Institute/NASA Ames Research Center \and
\label{inst:72} Institute of Astronomy, University of Cambridge, Madingley Road, Cambridge, CB3 0HA, United Kingdom}
  \abstract{We present the discovery of two small planets transiting \target\,A (TOI-1797), a G0\,V star (M$_\star$=1.109\,$\pm$\,0.043\,M$_\odot$, R$_\star$=1.043\,$\pm$\,0.009\,R$_\odot$) in a visual binary system. We combined \tess\ and \cheops\ space-borne photometry with MuSCAT\,2 ground-based photometry, `Alopeke and PHARO high-resolution imaging, TRES and \fies\ reconnaissance spectroscopy, and \sophie\ radial velocity measurements. 
  We validated and spectroscopically confirmed the outer transiting planet \target\,A\,c, a sub-Neptune with an orbital period of $\mathrm{P}_\mathrm{c}$\,$\approx$\,3.65\,d that was reported to be a \tess\ object of interest (TOI) shortly after the release of Sector 22 data.
  \target\,A\,c has a mass of $\mathrm{M}_\mathrm{c}$\,=\mpc\ and a radius of $\mathrm{R}_\mathrm{c} =$ \rpcTESS, implying a mean density of $\rho_\mathrm{c}$\,=\,3.1\,$\pm$\,0.7\,g\,cm$^{-3}$. The inner object, \target\,A\,b, is a validated 1.04\,d ultra-short period (USP) transiting super-Earth that we discovered in the \tess\ light curve and that was not listed as a TOI, owing to the low significance of its signal (\tess\ signal-to-noise ratio $\approx$\,6.7, \tess\ $+$ \cheops\ combined transit depth D$_\mathrm{b} = $\,\depthbTESS). We intensively monitored the star with \cheops\ by performing nine transit observations to confirm the presence of the inner planet and validate the system. \target\,A\,b is the first small ($\mathrm{R}_\mathrm{b} = $ \rpbTESS) USP planet discovered and validated by \tess\ and \cheops. Unlike planet c, \target\,A\,b is not significantly detected in our radial velocities (M$_{b} = $ \mpb). The two planets are on either side of the radius valley, implying that they could have undergone completely different evolution processes. We also discovered a linear trend in our Doppler measurements, suggesting the possible presence of a long-period outer planet. With a V-band magnitude of 9.2, \target\,A is among the brightest stars known to host a USP planet, making it one of the most favourable targets for precise mass measurement via Doppler spectroscopy and an important laboratory to test formation, evolution, and migration models of planetary systems hosting ultra-short period planets.}

   \keywords{Planets and Satellites: detection, fundamental parameters; instrumentation: photometers, spectrographs; methods: data analysis}

   \maketitle

\section{Introduction}

Following the discovery of the first planet orbiting a solar-like star \citep[51\,Peg\,b;][]{Mayor1}, the field of exoplanets has continuously evolved at a fast pace, moving from the mere exploratory objective of finding new planets, to the aim of measuring their radii and masses and, when possible, of detecting and characterizing their atmospheres. This transition has been possible thanks to high-precision ground-based facilities, such as \emph{Hi}gh \emph{R}esolution \emph{E}chelle \emph{S}pectrometer \citep[\hires,][]{Vogt1994}, \emph{H}igh \emph{A}ccuracy \emph{R}adial velocity \emph{P}lanet \emph{S}earcher \citep[\harps,][]{Mayor2003}, \emph{H}igh \emph{A}ccuracy \emph{R}adial velocity \emph{P}lanet \emph{S}earcher for the \emph{N}orthern emisphere \citep[\harpsn,][]{Cosentino2012}, \emph{E}chelle \emph{SP}ectrograph for \emph{R}ocky \emph{E}xoplanets and \emph{S}table \emph{S}pectroscopic \emph{O}bservations \citep[\espresso,][]{Pepe2020}, and space-based missions such as \emph{Co}nvection, \emph{Ro}tation et \emph{T}ransits planétaires \citep[\corot,][]{Baglin2006}, \textit{Kepler} \citep{Borucki2010}, \textit{Kepler}-2 \citep[K2,][]{Howell2014}, and, most recently, \emph{T}ransiting \emph{E}xoplanet \emph{S}urvey \emph{S}atellite \citep[\tess][]{Ricker15}. Data analysis techniques have also improved, providing an enhanced capability to disentangle the signature of an orbiting planet from the spectroscopic and photometric signals induced by stellar activity \citep[e.g.][]{Hatzes11, Haywood2014, Serrano, Hippke19b, Luque21}.

Statistical analyses on the population of exoplanets thus far discovered have shown that about 25-30\,\% of Sun-like stars in our Galaxy host super-Earths (R$_\mathrm{P}$\,=\,1\,--\,2\,R$_{\oplus}$, M$_\mathrm{P}$\,=\,1\,--\,10\,M$_{\oplus}$) and sub-Neptunes (R$_\mathrm{P}$\,=\,2\,--\,4~R$_{\oplus}$, M$_\mathrm{P}$\,=\,10\,--\,40~M$_{\oplus}$) in tightly packed systems with orbital periods shorter than $100$\,d \citep[see, e.g.][]{Silburt15, Mulders16}. One of the biggest surprises was the discovery of a population of exotic planets with orbital periods $\mathrm{P}_\mathrm{orb} \lesssim 1$\,d, the so-called ultra-short period (USP) planets. With the exception of a small subgroup of giant planets \citep[see, e.g.][]{Sahu06} and a couple of Neptune-sized objects \citep[e.g. TOI-849\,b and TOI-193\,b;][]{Armstrong20, Jenkins20}, most of the known USP planets are Earth-like objects or super-Earths ($\mathrm{R}_\mathrm{P} < 2~\mathrm{R}_{\oplus}$), with CoRoT-7\,b being the prototype of this class of objects \citep{Leger09}.

Launched in December 2019, the \emph{CH}aracterizing \emph{E}x\emph{OP}lanet \emph{S}atellite \citep[\cheops;][]{Benz21} is the first ESA small-class mission whose main goal is to perform high-precision photometry of bright stars ($\mathrm{V}$\,$<$\,12) known to host planets. \cheops\ is pursuing a variety of science goals, both in exoplanetary and stellar physics \citep[some references,][]{Lendl20, Leleu21, VanGrootel21, Delrez21, Borsato21, Morris21, Swayne2021, Szabo2021}. The mission is also discovering new transiting planets, such as HD\,108236\,f \citep{Bonfanti21} and TOI-178\,b, c, and f \citep{Leleu21}.

To help the community arrange follow-up observations and report the detection of \tess\ exoplanets, specific systems that pass through a validation process attain the status of \tess\ objects of interest (TOIs). As a first step, the process confirms that the signal is real and astrophysical. Then, it performs a Bayesian model comparison with other astrophysical signals to show that the planetary hypothesis is overwhelmingly preferred. Finally, only transit events that surpass the signal-to-noise ratio S/N\,$\sim$\,7.1 threshold are upgraded to the status of candidates and receive a TOI number \citep[see][for further details]{Guerrero21}. Some transit-like signals may not pass the validation process for various reasons. Simple examples may be single transits or two transits with a significant depth mismatch, as in the case of $\nu^2$\,Lupi\,b and c. Although the two planets already appeared transiting in \tess\ Sector 12, the star received a TOI number only one year after, when the data were visually inspected \citep{Kane20}. Another case of non-detection by the \tess\ pipeline involves small planets, which can generate shallow transit-like signals with S/N\,$<$\,7.1. Since small planets are common and the S/N threshold has been chosen to minimize the false alarm rate \citep{Jenkins16}, there may be transiting exoplanets not being identified and followed up, even if they are part of systems where TOIs have already been assigned.

The confirmation of less significant transit signals may be performed through other means, such as radial velocity (RV) follow-up observations of the star \citep[see, as an example, TOI-421;][]{Carleo20}. However, using Doppler spectroscopy to confirm a planet may be time-consuming. Transiting planet candidates with low significant transit signals in \tess\ light curves can be efficiently vetted against false alarm detections by performing space-based photometric follow-up observations of the predicted transits with a more precise instrument.

Within the \cheops\ Guaranteed Time Observation (GTO) programme, we perform \cheops\ photometric follow-up observations of mostly bright (V\,$<$\,11) TOIs with the immediate objective of identifying additional small planet candidates. This is done by performing high-precision photometry of transiting planet candidates with a non-significant transit detection in \tess\ light curves (S/N\,$<$\,7.1). As transiting planet candidates in multiple systems have a much lower probability of being false positives \citep{Latham2011}, we mainly focus on low S/N non-TOI candidates in systems known to host at least one TOI candidate. Identifying new non-TOIs orbiting a TOI target opens up the possibility of increasing both the number of known multi-planet systems, and the number of planets in an already known multi-planet system. Furthermore, by construction, this programme delivers viable targets for RV follow-up campaigns and improved transit parameters and ephemerides.

As part of \cheops\ GTO programs \cheops\ Objects of Interest (\texttt{CHOI}, PR110045) and \texttt{CHESS} (PR110031), we photometrically monitored the brightest component of the visual binary \target, also known as TOI-1797 (Table~\ref{tab:stellar_values}). \target\,A is a bright (V\,=\,9.2) G0\,V, nearby (82 pc) star found to host a 3.65\,d sub-Neptune candidate, which was reported as likely being a planet in \citet{Giacalone21} with the name TOI-1797.01. We independently searched the \tess\ light-curve for possible candidates, and identified an additional transit signal with a period of $1.04$\,d that was not included in the TOI candidates list due to its low S/N\,$\sim$\,6.7. Here, we report on the discovery of this additional ultra short-period transiting planet, as well as the validation and radius determination of both bodies. This result was made possible thanks to the photometric observations carried out with CHEOPS. With its high precision and high scheduling flexibility \cheops\ was the only instrument capable of clearly detecting the transit signal induced by the 1.04\,d planet, making \target{A} the first system with an ultra-short period planet discovered and validated by \cheops. 

\begin{table}[!t]
\centering
\begin{threeparttable}
\caption{Main identifiers, equatorial coordinates, optical and infrared magnitudes, and the fundamental parameters of \target\,A.}
\label{tab:stellar_values}
\begin{tabular}{lrr}
\hline
\hline
\noalign{\smallskip}
Parameter & Value & Source \\
\noalign{\smallskip}
\hline
\noalign{\smallskip}
\multicolumn{3}{l}{\it Main identifiers}  \\
\noalign{\smallskip}
\multicolumn{2}{l}{HD}{93963\,A} & ExoFOP\tnote{b} \\
\multicolumn{2}{l}{TOI}{1797} & TIC v8\tnote{a} \\
\multicolumn{2}{l}{TIC}{368435330} & TIC v8\tnote{a} \\
\multicolumn{2}{l}{TYC}{1977-02600-1} & ExoFOP \\
\multicolumn{2}{l}{2MASS}{J10510650+2538281}  & ExoFOP \\
\multicolumn{2}{l}{\gaia\ EDR3}{729899906357408768}  & \gaia\ EDR3\tnote{c} \\
\noalign{\smallskip}
\hline
\noalign{\smallskip}
\multicolumn{3}{l}{\it Equatorial coordinates, distance, and radial velocity}  \\
\noalign{\smallskip}
R.A. (J2000.0)	&    10$^\mathrm{h}$51$^\mathrm{m}$06.51$^\mathrm{s}$	& \gaia\ EDR3 \\
Dec. (J2000.0)	& 25$\degr$38$\arcmin$28.19$\arcsec$	                & \gaia\ EDR3 \\
Distance (pc) & $82.34 \pm 0.39$ & TIC v8 \\
Radial Velocity (km\;s$^{-1}$) & $13.28 \pm 0.02$ & This work \\ 
\noalign{\smallskip}
\hline
\noalign{\smallskip}
\multicolumn{3}{l}{\it Optical and near-infrared photometry} \\
\noalign{\smallskip}
\tess\              & $8.6259\pm0.006$     & TIC v8\tnote{c}         \\
\noalign{\smallskip}
$B$              & $9.771\pm0.042$   & TIC v8 \\
$V$              & $9.182 \pm 0.003$   & TIC v8 \\
\noalign{\smallskip}
$J$ 			&  $8.108\pm0.034$      & TIC v8 \\
$H$				&  $7.807\pm0.034$      & TIC v8 \\
$K$			&  $7.776\pm0.024$      & TIC v8 \\
\noalign{\smallskip}
$W1$			&  $7.68\pm0.03$      & All{\it WISE}\tnote{d} \\
$W2$			&  $7.73\pm0.02$      & All{\it WISE} \\
$W3$             & $7.709\pm0.019$      & All{\it WISE} \\
$W4$             & $7.549\pm0.158$      & All{\it WISE} \\
					
\noalign{\smallskip}
\hline
\noalign{\smallskip}
\multicolumn{3}{l}{\it Fundamental parameters}   \\
T$_\mathrm{eff}$\ (K) & $5987 \pm 64$ & This work \\
$\log \mathrm{g}_{\star}$ (cgs) & $4.49 \pm 0.11$ & This work \\
$\log\mathrm{R}^{\prime}_{\mathrm{HK}}$ & $-4.63 \pm 0.05$ & This work\\
$[\mathrm{Fe}/\mathrm{H}]$ (dex) & $0.10 \pm 0.04$ & This work \\
$[\mathrm{Mg}/\mathrm{H}]$ (dex) & $0.08 \pm 0.06$ & This work \\
$[\mathrm{Si}/\mathrm{H}]$ (dex) & $0.08 \pm 0.04$ & This work \\
$\mathrm{v}_{\mathrm{mic}}$\ (km\;s$^{-1}$)     & $1.15 \pm 0.03$ & This work \\
$\mathrm{v}_{\mathrm{mac}}$\ (km\;s$^{-1}$)     & $4.1 \pm 0.4$ & This work \\
$\mathrm{v} \sin \mathrm{i}_{\star}$\ (km\;s$^{-1}$)  & $5.9 \pm 0.8$       & This work \\
$\mathrm{M}_{\star}$ ($\mathrm{M_{\odot}}$) & $1.109 \pm 0.043$  
& This work \\
$\mathrm{R}_{\star}$ ($\mathrm{R_\odot}$) & $1.043 \pm 0.009$  & This work \\
$\mathrm{P}_{\star}$ (days) & $12.8 \pm 1.8$  & This work \\
$\rho_{\star}$ ($\rho_{\odot}$) & $1.387 \pm 0.064$ & This work \\
$\mathrm{L}_{\star}$ ($\mathrm{L}_{\odot}$) & $1.254 \pm 0.058$ & This work \\
Age (Gyr) & $1.4^{+0.8}_{-0.4}$ & This work \\
$\mathrm{A_V}$ (mag) & $0.10 \pm 0.05$ & This work \\
\noalign{\smallskip}
\hline
\end{tabular}
\begin{tablenotes}
\item[a] \citet{Stassun2018}.
\item[b] \href{https://exofop.ipac.caltech.edu/tess}{https://exofop.ipac.caltech.edu/tess.}
\item[c] \citet{GaiaDR2, gaiadr3}.
\item[d] \citet{cutri2013}.
\end{tablenotes}
\end{threeparttable}
\end{table}

The present paper is organized as follows. In Section~\ref{sec:tess} we present the \tess\ data and our independent search for transit signals. Section~\ref{cheops} briefly describes the \cheops\ mission and presents the strategy of our observations and the data reduction. Section~\ref{imaging} presents the high-resolution imaging observations of \target\,A\ and its stellar companion \target\,B. Section~\ref{spectroscopic_data} presents the spectra used for the stellar characterization (Section~\ref{star}). Section~\ref{Joint} is dedicated to the analysis of the \tess\ and \cheops\ light curves and the determination of the parameters of \target\,A\,b and c. The validation of both planets is discussed in Section~\ref{validation}. In Section~\ref{RV_analysis}, we report the analysis of \sophie\ RV follow-up data and the measurement of \target\,A\,c mass. Section~\ref{system} describes the planetary system orbiting \target\,A, with empirical estimates of the masses of the two discovered planets and it discusses some of the possible migration theories that may be relevant to our system. In this section, we also present the importance and the impact of our discovery on future science.

\section{\tess\ observations}
\label{sec:tess}

\subsection{Data collection}

\begin{figure*}
    \centering
    \includegraphics[width = 20cm]{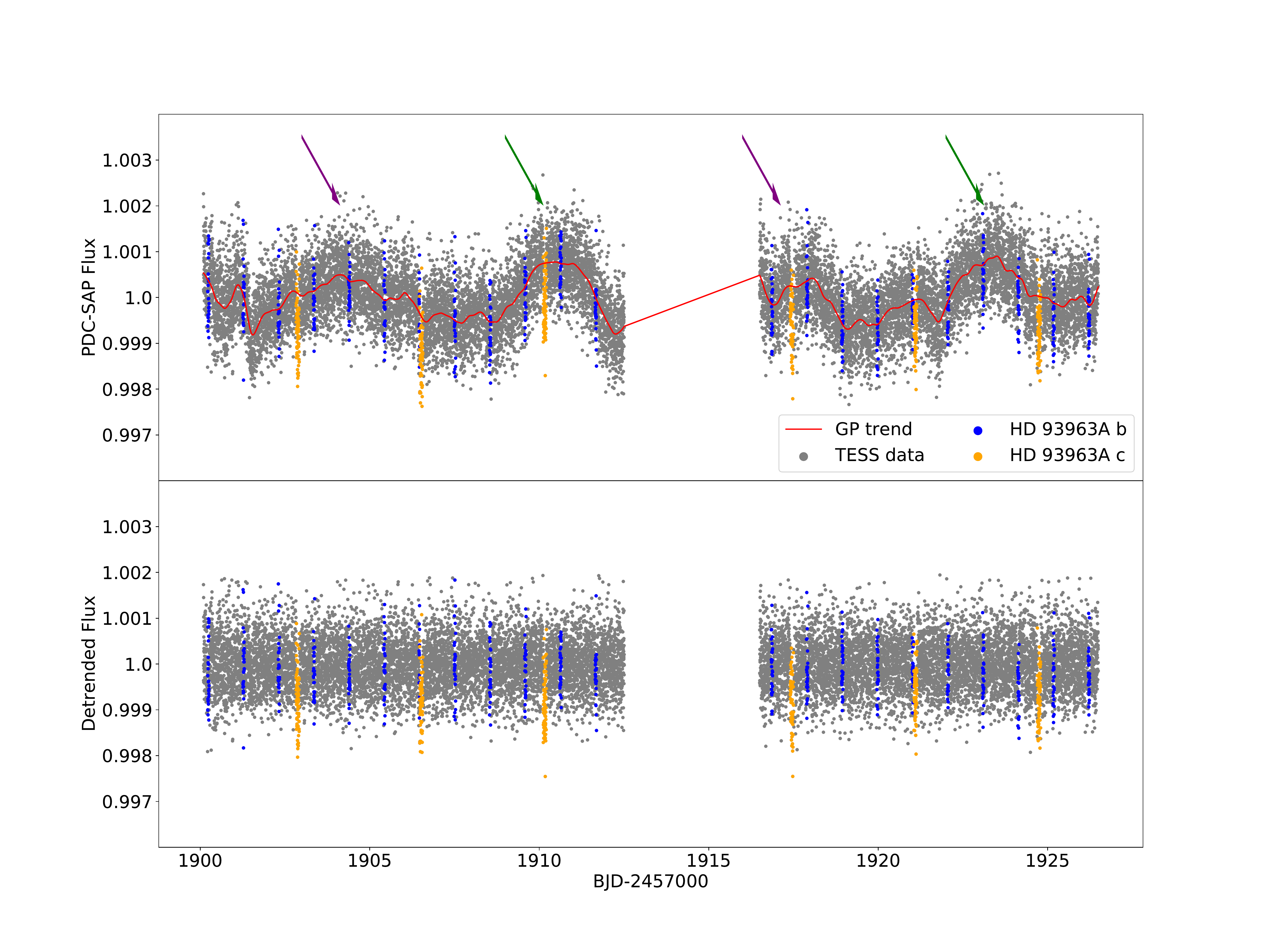}
    \caption{\tess\ light curve of \target\,A. In-transit data points of the 3.65\,d and the 1.04\,d candidate are marked in yellow and blue, respectively. \textit{Upper panel}: PDC-SAP flux (grey) and detrending model (red), as obtained with \texttt{citlalicue} (Sect.~\ref{search_for_transits}). The arrows point to the repeated maxima we identified in the light curve: the purple arrows points to the maximum at $\sim$1905 (BJD-2457000) and at its repetition at $\sim$1918 (BJD-2457000). The green arrows point to the second maxima, that appears at $\sim$1911 (BJD-2457000) and at $\sim$1924 (BJD-2457000). \textit{Lower panel}: detrended flux.\label{Tess_data}}
\end{figure*}

\tess\ observed \target\,A in Sector 22, during year 2 of its nominal mission, from 18 February to 18 March 2020, with data from the target stacked and telemetered at 2-minute cadence. The space telescope monitored the star using camera \#1 and CCD \#3 \citep[each CCD has a field of view of FoV = 12\degr$\,\times$\,12\degr;][]{Ricker15}. In addition to a 1\,day gap due to the data downlink during perigee passage, the \tess\ observations of \target\,A were affected by significant moon-light contamination, resulting in the removal of 2 additional days of data at the beginning of each half sector. Overall, a total of 5\,d of data was removed from the Sector. The \tess\ target pixel files were processed and calibrated by the Science Processing Operation Center \citep[SPOC;][]{SPOC} at the NASA Ames Research Center. The light curve was extracted using simple aperture photometry \citep[SAP; ][]{Twicken10, Morris20} and processed using the Presearch Data Conditioning (PDC) algorithm, which uses a Bayesian maximum a posteriori approach to remove the majority of instrumental artefacts and systematic trends \citep{Smith2012, Stumpe2012, Stumpe2014}. We retrieved the PDC-SAP \tess\ light curve of \target\,A from the Mikulski Archive for Space Telescope (MAST\footnote{\url{https://mast.stsci.edu}.}; see the upper panel of Fig.~\ref{Tess_data}).

The SPOC team searched the PDC-SAP light curve for transit-like signals, using a pipeline that iteratively performs multiple transiting planet searches and stops when it fails to find subsequent transit-like signatures above the detection threshold of S/N\,$\sim$\,7.1. On 28 March 2020, SPOC published the results in the Data Validation Report \citep[DVR;][]{Twicken18, Li19}. The \tess\ vetting team at the Massachusetts Institute of Technology (MIT) inspected the DVR to review the Threshold Crossing Events (TCEs), and later announced the detection of a transiting planetary candidate \citep[TOI 1797.01][]{Guerrero21} with a period of $\mathrm{P}_{\mathrm{orb}} \approx 3.65$\,d, a depth of about $800$\,ppm, and a duration of T$_{14} \approx 2.0$\,h. The candidate passed all the automatic validation tests from the TCE, such as odd-even transit depth variation and ghost diagnostic tests.

\subsection{Independent transit search}
\label{search_for_transits}

In order to confirm the presence of the 3.65\,d TOI announced by the \tess\ team and look for additional planet candidates, we independently searched the PDC-SAP \tess\ light-curve for transit signals. In order to do so, we applied different detrending algorithms and methodologies as described below.

\textit{Method 1}: We filtered out the variability of the light-curve using the Savitzky-Golay method \citep{Savitzky1964, Press02} and applied the Détection Spécialisé de Transits \citep[\texttt{DST};][]{Cabrera2012}, which searches for transits using a parabolic function as a transit model. In addition to the detection algorithm, we also used the information retrieved from the \tess\ data release notes and DVR, to assess and remove signals arising from possible astrophysical false positives or systematics.

\textit{Method 2}: We modelled the stellar activity modulation using the Gaussian Process (GP) implemented within the public code \texttt{citlalicue} \citep{Barragan21b}. Briefly, the routine masks out the planetary transits and bins the data using 3\,h bins. It performs an iterative maximum likelihood optimization coupled with a 5-sigma clipping algorithm, in order to fit a GP with quasi-periodic kernel and remove possible outliers. To maximize the performance of \texttt{citlalicue}, we masked out the transits of the 3.65\,d planet, as they can affect the modelling of the stellar activity signal. We flattened the light curve, dividing the full PDC-SAP time series (including the data points previously masked out) by the retrieved correlation model and we searched the \tess\ detrended light curve for transits, by repeatedly applying the Transit Least Square algorithm \citep[TLS,][]{Hippke19b}. \texttt{TLS} performs a least square minimization on the entire non binned light-curve, modelling the transit with a quadratic limb darkening law, as described in \citet{Mandel02}.

\textit{Method 3}: We used an out-of-box polynomial fit (which by its nature leaves transits untouched) to remove variability from the light curve. Then, as in the case of \textit{Method 2}, we iteratively performed a \texttt{TLS} periodic search and masked the detected transit events, until no significant signal remained.

\textit{Method 4}: We applied the detection pipeline \texttt{EXOTRANS} to the \tess\ light curve. \texttt{EXOTRANS} removes discontinuities  and stellar variability with the wavelet-based filter \texttt{VARLET} \citep{Grziwa16}, and routinely searches for transits applying an advanced version of the Box Least-Squares algorithm \citep[\texttt{BLS;}][]{Kovacs16}. 
\texttt{EXOTRANS} can detect multiple transit signals and-or transits masked by other strong periodic events (such as systematics, background binaries). 

Each of the four different methods confirmed the 3.65\,d transit candidate announced by the \tess\ team and unveiled the presence of a possible additional transit signal with a period of $1.04$\,d, a depth of $\sim$140\,ppm, and a duration of $\sim$2\,h. This transit has a S/N\,$\sim$\,6.7, which is below the detection threshold adopted by the \tess\ team (S/N\,$\sim$\,7.1). We thus focused our efforts on understanding the nature of this signal performing further photometric observations and complementary high-resolution imaging and Doppler spectroscopy.

\section{\cheops\ observations}
\label{cheops}

\begin{table*}
		\centering
		\caption{List of \cheops\ observations for \target\,A with start and end date, number of data points, and terms used to decorrelate the respective light curves using \texttt{pycheops}. The parameters $\sin{\phi}$ and $\cos{2\phi}$ are roll angle corrections, dx and dy are the $\mathrm{x}$ and $\mathrm{y}$-axis centroid position, while dt and dt$^2$ are a first and second order time trend, respectively \citep[please refer to][for a detailed explanation of the decorrelation terms]{Maxted21}.}
		\label{visits}
			 \scalebox{0.85}{\begin{tabular}{cccccc}
				\hline\hline
				\noalign{\smallskip}
				Planet & File key & Start date (UTC) & End date (UTC) & Frames & \texttt{pycheops} correlation terms \\
				\noalign{\smallskip}
				\hline
				\noalign{\smallskip}
				b & \texttt{CH\_PR110045\_TG001301\_V0200} & 2021-02-07 09:14 & 2021-02-08 16:31 & 1346 & $\sin{\phi}$, dt, dx, dt$^2$ \\
				b & \texttt{CH\_PR110045\_TG001601\_V0200} & 2021-02-24 11:11 & 2021-02-24 20:57 & 439 & background, dt \\
				b & \texttt{CH\_PR100031\_TG037201\_V0200} & 2021-03-02 17:06 & 2021-03-03 01:12 & 381 & background, dt\\
				b & \texttt{CH\_PR110045\_TG001801\_V0200} & 2021-03-05 15:41 & 2021-06-03 04:15 & 641 & background, dt, dt$^2$ \\
				b & \texttt{CH\_PR110045\_TG001901\_V0200} & 2021-03-07 19:28 & 2021-03-08 08:31 & 598 & background, dt, dt$^2$ \\
				b & \texttt{CH\_PR110045\_TG002001\_V0200} & 2021-03-13 00:53 & 2020-06-24 13:57 & 596 & dt, dt$^2$\\
				b & \texttt{CH\_PR110045\_TG002301\_V0200} & 2021-03-16 23:43 & 2021-03-18 02:06 & 1219 & background, dt, dt$^2$, dy\\
				b & \texttt{CH\_PR100031\_TG039701\_V0200} & 2021-04-19 12:21 & 2021-04-19 19:35 & 358 & background, $\mathrm{dx}$, dt\\
                c & \texttt{CH\_PR100031\_TG039601\_V0200} & 2021-04-20 20:38 & 2021-04-21 09:53 & 618 & background, $\sin{\phi}$, $\cos{2\phi}$, dt \\
                \noalign{\smallskip}
				\hline
			\end{tabular}}
\end{table*}

\begin{figure*}
    \centering
    \includegraphics[width=0.8\linewidth]{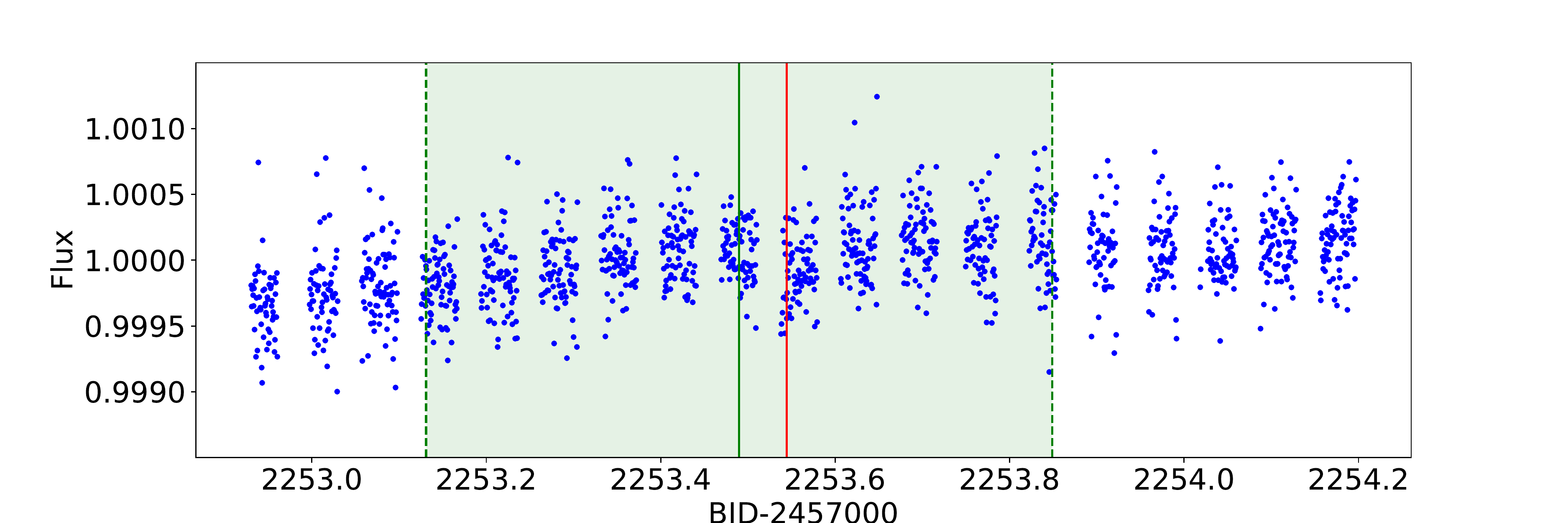}
    \caption{Raw data points of the first \cheops\ visit (blue circles). The transit of the 1.04\,d planet is visible at the centre of the visit. The mid-transit time, as predicted by the \tess\ ephemeris, is highlighted with a vertical green line, while the area coloured in light green is the corresponding 3\,$\sigma$ interval of confidence. The mid-transit time predicted with the ephemeris in the third column of Table~\ref{table_results} is marked with a vertical red line. The time difference between the red and the green lines is $\sim$1\,h.}
    \label{first_visit_raw}
\end{figure*}

In order to confirm the transit signal at 1.04\,d, validate the planetary system, and precisely determine the radius of the two planets, we performed \cheops\ photometric follow-up observations of \target\,A between February and April 2021. We started by verifying whether we could detect the additional planet candidate at 1.04\,d. A modelling of \tess\ transits\footnote{We modelled the detrended \tess\ light curve following the same methodology described in Sect.~\ref{Joint}.} provided the following ephemeris: $\mathrm{P} = 1.03897 \pm 0.00035$\,d and $\mathrm{T}_0 (\mathrm{BJD}\footnote{throughout the paper BJD stands for BJD\_TDB})= 2458901.279 \pm 0.004$\,d. Due to the relatively short baseline of the \tess\ light curve ($\sim$28 d) and the low significance of our detection (S/N\,$\approx$\,6.7), the propagated uncertainty on the predicted time of transit in February 2021 (nearly one year after the \tess\ observations) amounted to $\sim$3\,h. Given the short orbital period of the candidate (1.04\,d), we decided to follow a conservative approach and continuously observe \target\,A for nearly 1.3\,d\footnote{Corresponding to one visit of 19 \cheops\ orbits around the Earth.}, centring our observations around the transit that was expected to occur on 07 February 2021 (UTC). Taking into account the transit duration ($\sim$2\,h), this would have allowed us to cover at least one full transit, regardless of the uncertainties on the transit time, while having a long out-of-transit baseline to properly detrend the data. 

Figure~\ref{first_visit_raw} shows the raw \cheops\ light curve of the first visit of \target\,A. The gaps in the time series are due to the Earth's occultation and the passage of the spacecraft above the South Atlantic Anomaly (SAA)\footnote{Data collected during an SAA passage are usually not transmitted to Earth, owing to the large number of cosmic ray hits.}. A visual inspection of the light curve revealed the presence of a transit-like feature with a duration and depth compatible with those of the transits observed by \tess\ and occurring $\sim$1\,h after the predicted time of mid-transit. We jointly modelled the \tess\ and \cheops\ light curve following the same methods described in Sect.~\ref{Joint} and refined the period (P\,=\,$1.0391581 \pm 0.0000387$\,d). 

With one single \cheops\ visit we 'upgraded' the 1.04\,d signal to planetary candidate and improved the precision on its period estimate by one order of magnitude. We decided to keep observing the candidate, scheduling seven shorter \cheops\ visits of the 1.04\,d transit candidate and stopped the follow-up when we reached a precision on the planetary radius of $\sim$3\,\%. For the sake of completeness, we also observed one transit of the 3.65\,d TOI candidate and verified the achromaticity of the event. The observations were performed with a global efficiency (duty cycle) higher than 50\,\%. Table~\ref{visits} reports the starting and ending time of each \cheops\ visit, along with the number of data points.

\subsection{\cheops\ data reduction}
\label{DRP}

\begin{table*}
  \footnotesize
  \caption{G-band magnitude, parallax, proper motion, and renormalized unit weight error (RUWE) of \target\,A and its companion, as retrieved from the \gaia\ EDR3.\label{companion}}  
  \centering
  \begin{tabular}{lccccccc}
  \hline
  \hline
  \noalign{\smallskip}
  Star & \gaia\ ID & Distance & G magnitude & Parallax & PM(RA) & PM(Dec) & RUWE \\
  & & (arcsec) & & (mas) & (mas/yr) & (mas/yr) \\
  \hline
  \noalign{\smallskip}
  \multicolumn{7}{l}{\target\,A and its companion} \\
  TIC\,368435330 & 729899906357408768 & $\cdots$ & $9.040 \pm 0.003$ & $12.151 \pm 0.017$ & $-92.833 \pm 0.014$ & $-22.911 \pm 0.014$ & 1.04 \\
TIC\,368435331 & 729899906357408640 &  $5.9\,\arcsec$ & $16.906 \pm 0.004$ & $11.940 \pm 0.128$ & $-90.705 \pm 0.110$ & $-22.993 \pm 0.093$ & 0.95 \\
\hline
\noalign{\smallskip}
  \multicolumn{7}{l}{Nearby brightest stars} \\
TIC~368435329 & 729899902062379776 & $16.8\,\arcsec$ & $17.506 \pm 0.003$ & $0.045 \pm 0.136$ & $-5.122 \pm 0.108$ & $-6.519 \pm 0.113$ & 1.177 \\
TIC~900011362 & 729712126092040192 & $72.7\,\arcsec$ & $19.154 \pm 0.004$ & $0.709 \pm 0.310$ & $-5.287 \pm 0.246$ & $-7.272 \pm 0.231$ & $0.983$\\
TIC~900011363 & 729712160451779200 & $76.7\,\arcsec$ & $18.719 \pm 0.004$ & $0.308 \pm 0.222$ & $-15.072 \pm 0.183$ & $-9.396 \pm 0.185$ & $0.977$ \\
TIC~900012677 & 729900211299738240 & $110.1\,\arcsec$ & $19.322 \pm 0.004$ & $-0.474 \pm 0.326$ & $-0.858 \pm 0.266$ & $-3.560 \pm 0.236$ & $0.907$ \\
TIC~368435333 & 729876606159465088 & $124.9\,\arcsec$ & $17.764 \pm 0.003$ &$0.343 \pm 0.138$ & $-0.913 \pm 0.103$ & $-7.669 \pm 0.101$ & 1.020 \\
    \noalign{\smallskip}
    \hline
   \end{tabular}
\end{table*}

\cheops\ data for each visit is downlinked to the ground as packages of circular images with a diameter of 200\,px centred around the target star. Those images, referred to as ``subarrays'', are small windows of the \cheops\ CCD, whose sky-projected area has a diameter of about 200\,$\arcsec$, which corresponds to $\sim$4\,\% of the full \cheops\ field of view (FoV). The subarrays were automatically processed with the latest version of the Data Reduction Pipeline \citep[\texttt{DRP};][]{Hoyer20}, \texttt{DRP\,v13}, which includes bias and dark subtraction, non-linearity and flat field correction. In addition, the \texttt{DRP} corrects for sky-background, cosmic-ray hits, and smearing trails of nearby stars. The \texttt{DRP} performs aperture photometry on the processed \cheops\ images using different circular masks centred around the target. The mask follows the movements of the star due to the spacecraft pointing jitter and telescope rolling around the optical axis, which ensures a thermally stable environment of the payload radiators. The circular photometric masks can have three different standard sizes, namely, the \texttt{RINF}, \texttt{DEFAULT}, and \texttt{RSUP} apertures with radii of 22.5, 25.0, 30.0 pixels, respectively. In addition, the \texttt{DRP} also extracts the photometric flux using an \texttt{OPTIMAL} aperture which radius is estimated independently for single visits. The \texttt{OPTIMAL} aperture is intended to minimize the instrumental noise and contamination level from nearby stars. The \texttt{DRP} extraction provides the user with a series of vectors that allow one to model instrumental and environmental effects. Users can retrieve from the data the orbital roll-angle $\phi$ of the telescope, the $\mathrm{x}$ and $\mathrm{y}$ position in the CCD of the centre of the target's point spread function (PSF), the estimated sky background level, the smear curve, and the contamination by PSFs of background stars \citep[please refer to][for additional details on how the contamination is treated by the DRP]{Hoyer20}.

\label{PIPE}
We also extracted PSF photometry on the \cheops\ images, using the python module\footnote{Available at \url{https://github.com/alphapsa/PIPE}.} \texttt{PIPE} (Brandeker et al., in prep.). Point-spread function photometry has the advantage of being less sensitive to contaminants than aperture photometry, which makes it particularly well suited for crowded fields or when there is a suspected contaminant. In addition, with \texttt{PIPE} we can also remove the smear effect in advance. We used the observations of \target\,A to derive an empirical PSF, used by \texttt{PIPE} to extract the photometry. As such, \texttt{PIPE} provides as output a FITS file that includes the extracted light curve and the same decorrelation vectors provided by the \texttt{DRP}, with the exception of smearing and contamination. The difference in RMS between \texttt{DRP} and \texttt{PIPE} light curves is only $\sim$30 ppm, but as the PSF photometry can better account for the noise due to the closest contaminants, we decided to use \texttt{PIPE} extracted light curves for our analysis. 

\begin{figure*}
    \centering
    \includegraphics[width = 18cm]{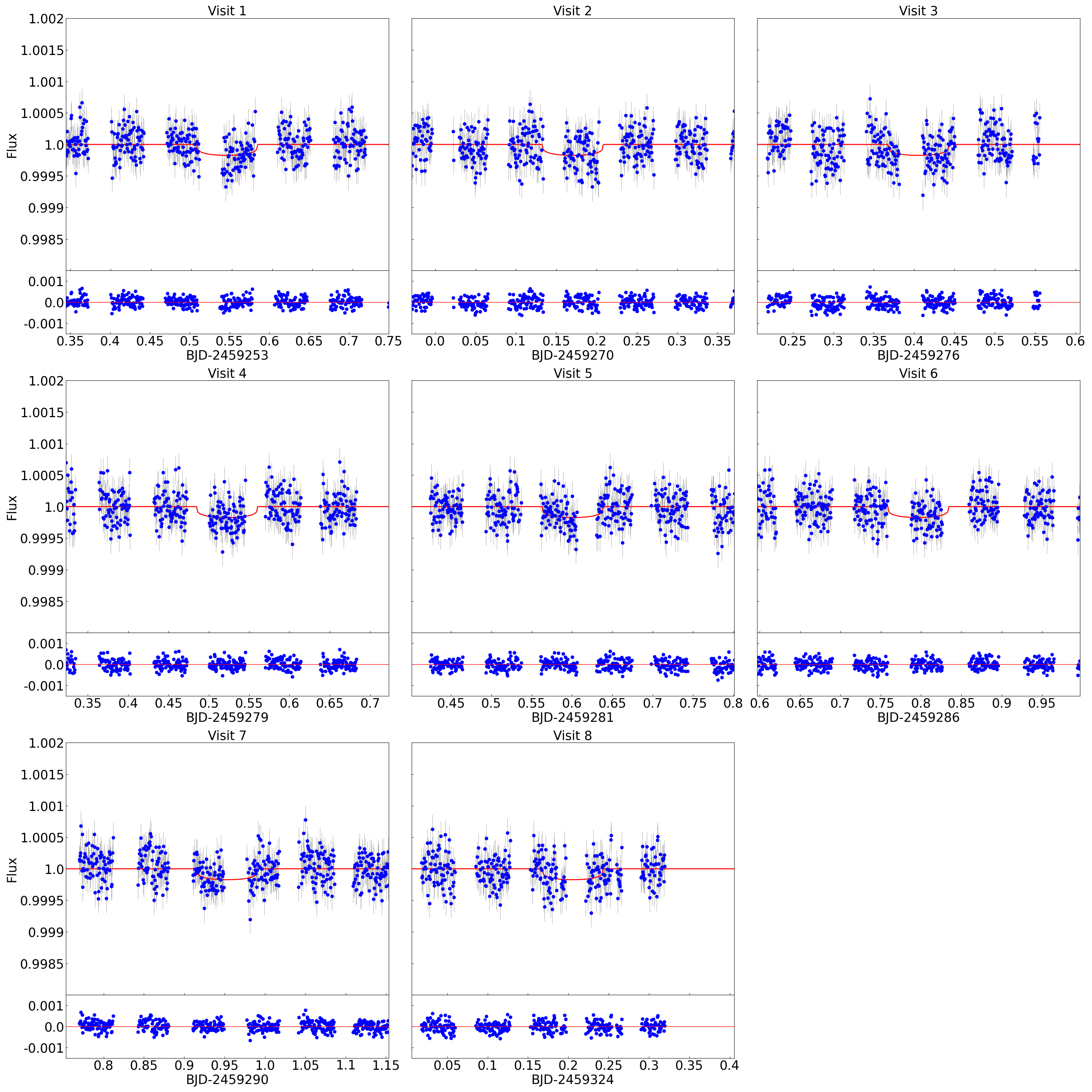}
    \caption{The eight detrended \cheops\ light curves of planet b. In each plot, the upper panels show the detrended flux (blue circles) with the best-fitting transit model (red line). The lower panels display the residuals, with the zero-level marked as a red line.}
    \label{CHEOPS_LIGHTCURVES_B}
\end{figure*}

\begin{figure}
    \centering
    \includegraphics[width = 7cm]{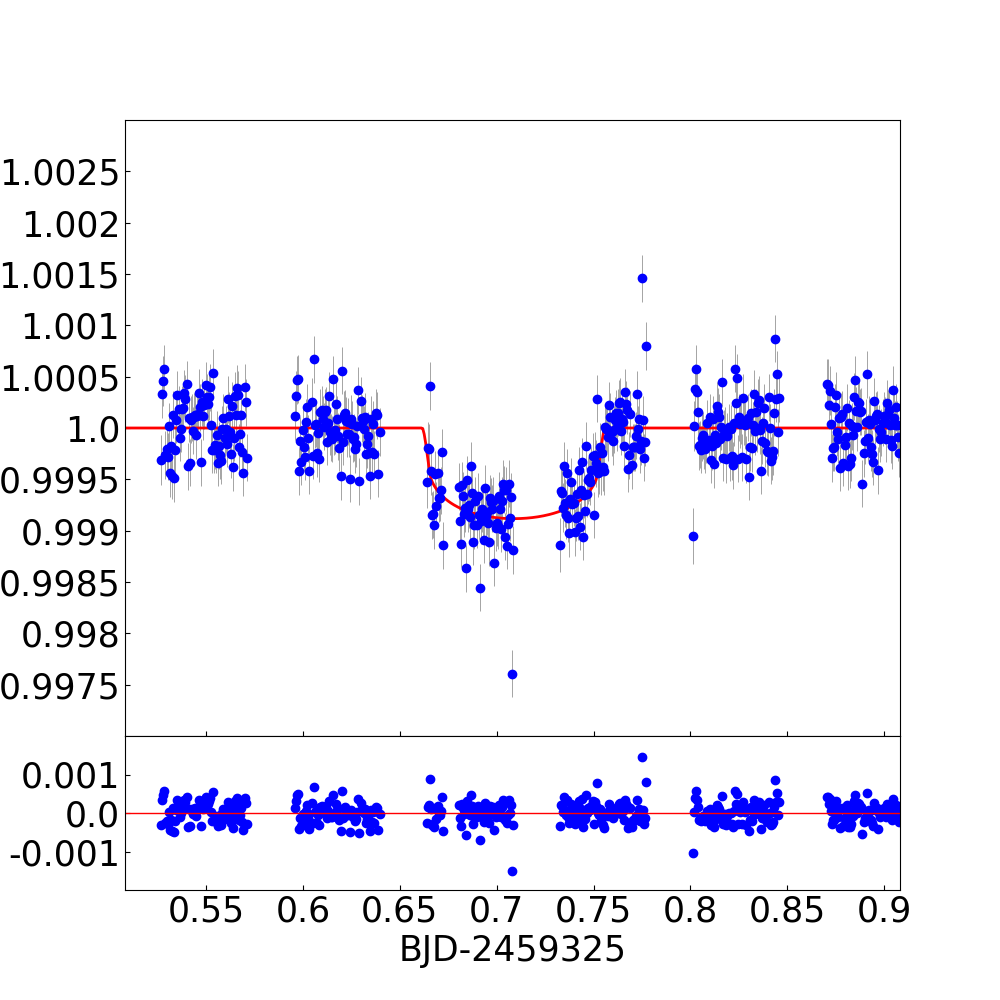}
    \caption{The detrended \cheops\ light curve of planet c and best fitting transit model (upper panel). Residuals are displayed in the lower panel.}
    \label{CHEOPS_LIGHTCURVES_C}
\end{figure}

\subsection{\cheops\ light curve detrending}
\label{CHEOPS_detrend}

We detrended the nine \cheops\ visits with the code \texttt{pycheops} \citep{Maxted21}, which allows one to de-correlate the data using the vectors that describe the short-term instrumental photometric trends: roll angle, centroid movements, smear curve, and contamination by background stars. \texttt{pycheops} can also detrend the \cheops\ light curves using simple linear or quadratic trends. 

We first applied a 3\,$\sigma$ clipping algorithm as implemented in \texttt{pycheops} to remove outliers from the \cheops\ data and analysed each visit individually to check if it is affected by any of the known \cheops\ instrumental systematics. We excluded from the process the smear and contamination effects, because they are already accounted for by the \texttt{PIPE} extraction. We chose the detrending vectors via a Bayes factor pre-selection method \citep{Trotta07}, which routinely fits each visit with a least mean square method, accounting for the transit model and one of the possible decorrelation terms. In Table~\ref{visits}, we report those terms for which we obtained a Bayes factor lower than 1. We performed preliminary modelling of the eight \cheops\ visits of the 1.04\,d transit signal using Markov Chain Monte Carlo simulations implemented in \texttt{pycheops}, including the effects of both the transits and the decorrelation terms listed in Table~\ref{visits}. For the 3.65\,d candidate, we applied the same methodology adopted for the visits of the 1.04\,d transit signal. We finally flattened the light curves, dividing the data by the inferred model of the instrumental noise. Figures~\ref{CHEOPS_LIGHTCURVES_B} and \ref{CHEOPS_LIGHTCURVES_C} show the 8 detrended transit light curves of the 1.04\,d candidate and the detrended single visit of the 3.65\,d candidate, respectively, as observed with \cheops. 

\section{Ground-based photometry and imaging}
\label{imaging}

Stellar companions, projected closely on the sky, whether bound or unbound (such as a faint nearby eclipsing binary (NEB)) can create false positive signals and provide ``third-light” that will dilute the transit signal. Unaccounted for ``third-light” in a transit fit will cause the derived planet radius to be underestimated \citep{ciardi2015} and can even lead to non-detection of small planets \citep{Lester2021}. Given that nearly one-half of solar-like stars are binary \citep{Matson18}, high-resolution imaging may help avoid incorrect interpretations of any detected exoplanet.

\subsection{LCOGT 1m observations}

\begin{figure}
    \centering
    \includegraphics[width=0.95\linewidth]{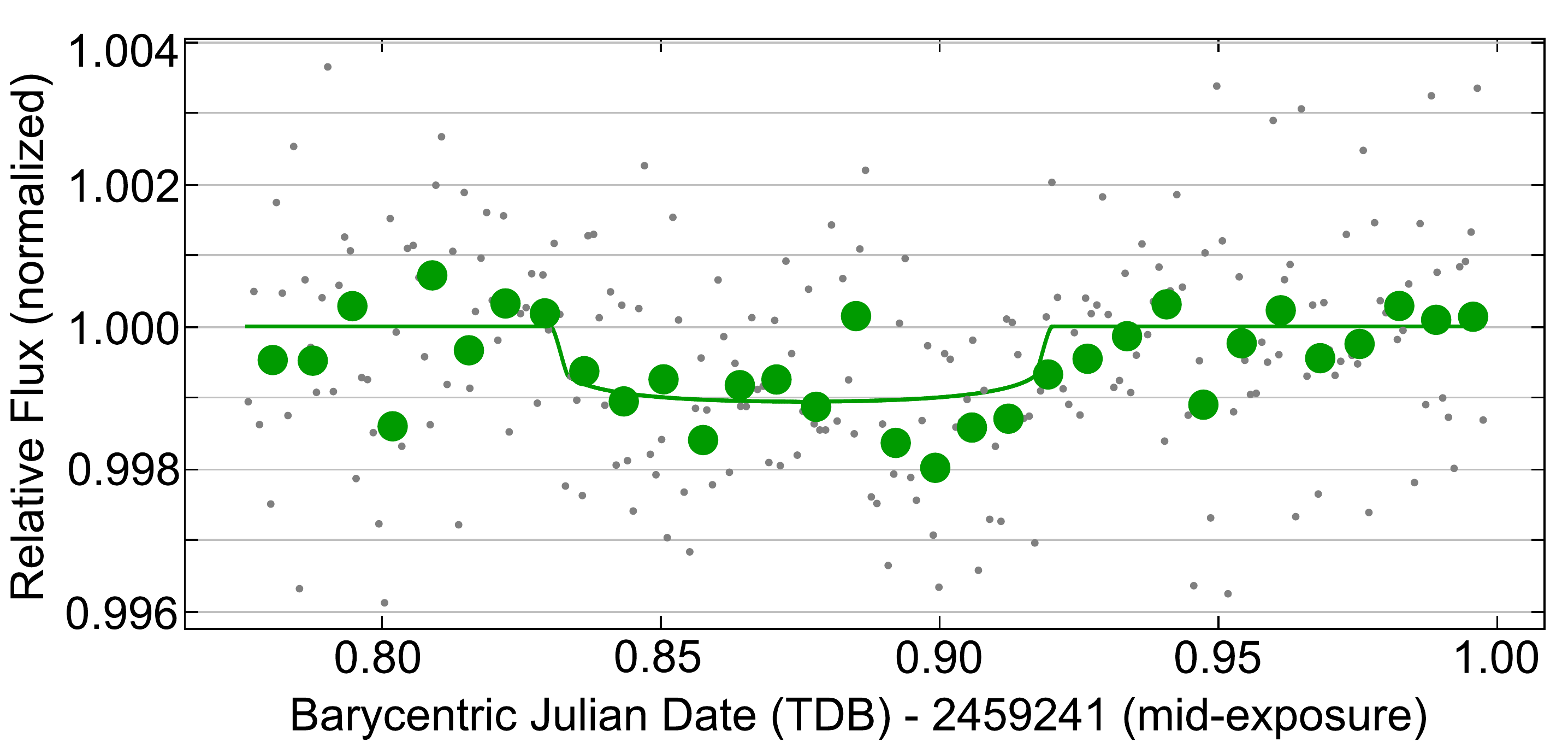}
    \caption{Ground-based follow-up light curve of TOI-1797 b on UTC 2021 January 27 from the LCOGT 1\,m network in PanSTARRS Y-band. The grey filled circles show the photometry from the individual exposures. The green filled circles show the results in 10 minute bins. The fitted transit centre time is 2459241.875 BJD, and is consistent with the ephemeris extracted from the TESS and CHEOPS data in the work.}
    \label{LCOGT_tr}
\end{figure}

We observed a full transit of \target\,c in Pan-STARRS Y-band on UT 2021 January 27 from the Las Cumbres Observatory Global Telescope \citep[LCOGT;][]{Brown:2013} 1.0\,m network node at McDonald Observatory in near Fort Davis, Texas. The telescope is equipped with the $4096\times4096$ SINISTRO camera having an image scale of $0.389\arcsec$ per pixel, resulting in a $26\arcmin\times26\arcmin$ field of view. The images were calibrated by the standard LCOGT {\tt BANZAI} pipeline \citep{McCully:2018}, and photometric data were extracted using {\tt AstroImageJ} \citep{Collins:2017}. Circular photometric apertures with radius of $9.7\arcsec$ were used to extract the differential photometry. The target star aperture also includes most of the flux from the companion TIC\,368435331, which is the nearest TESS Input Catalog and \gaia\ EDR3 neighbour and which we refer to as \target\,B. We detected a low signal-to-noise, roughly 900\,ppt event, having transit timing consistent with the ephemeris derived from the TESS and CHEOPS data presented in this work (see Figure~\ref{LCOGT_tr}).

\subsection{MuSCAT 2 observations}
\label{MUSCAT_2}

\begin{figure}
    \centering
    \includegraphics[width=1\linewidth]{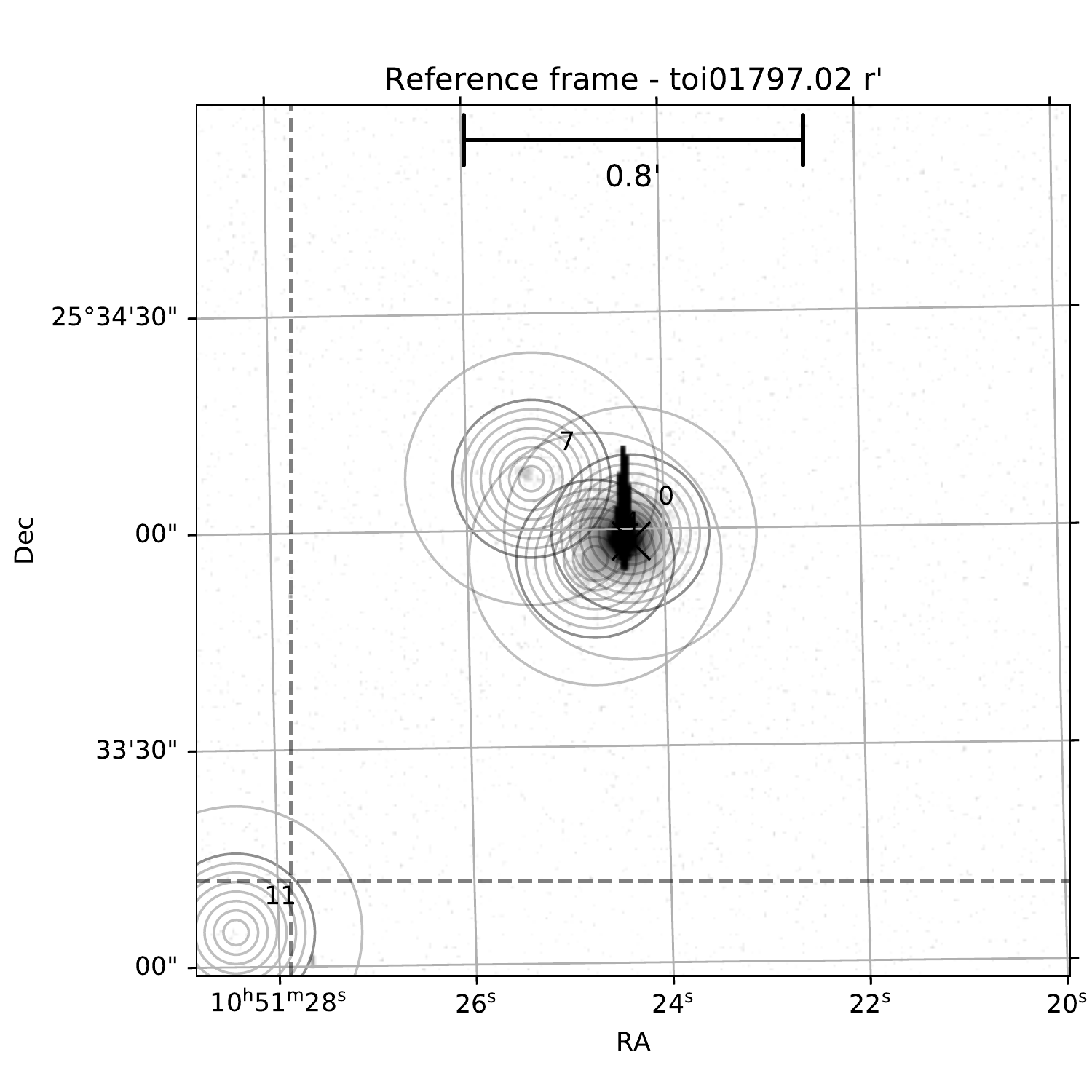}
    \caption{MuSCAT\,2 zoomed \textit{r}-band field of view. \target\,A (star ID 0) was saturated on purpose to search for transit event around nearby stars. MuSCAT\,2 observations ruled out star ID 7 (TIC 368435329) as the one with the 1.04\,d transit signal.}
    \label{FoV_MUSCAT}
\end{figure}

On 21-22 May 2021, \target\,A was observed  with the MuSCAT\,2 multi-colour imager \citep{Narita19} mounted at the 1.52\,m Telescopio Carlos Sanchez (TCS) of Teide Observatory (Tenerife, Spain). MuSCAT\,2 can perform simultaneous photometry in 4 broad passbands, namely, the \textit{g} (400-550\,nm), \textit{r} (550-700\,nm), \textit{i} (700-820\,nm), and $\mathrm{\textit{z}}_\mathrm{\textit{s}}$ (820-920\,nm) bands. The data calibration and photometric extraction was done using the dedicated MuSCAT\,2 pipeline \citep{Parviainen2019}. The observation aimed at verifying that the 1.04\,d signal is not caused by a nearby eclipsing binary (NEB). To this purpose, we used the \textit{z$_s$} and \textit{i} filters to observe \target\,A with 2 and 3\,s exposure times, respectively, and the other 2 filters to take 30\,s exposures to increase the S/N of nearby stars. Of the two observing runs, only the observations of 21 of May 2021 were useful to search for NEBs. While the 1.04\,d transit could not be confirmed to be on target due to the shallow signal, the MuSCAT\,2 observations rule out a scenario where the nearby star TIC 368435329 (star 7 in Fig. \ref{FoV_MUSCAT}) would be an eclipsing binary causing the 1.04\,d transit signal. The star is not resolved in TESS photometry; on the contrary, it is resolved in MuSCAT\,2 photometry, and an eclipse deep enough to cause the transit signal candidate would have been detectable from the MuSCAT\,2 photometry. The \gaia\ catatalogue contains two additional nearby stars that are not resolved in the MuSCAT\,2 photometry. One of these two is too faint (G magnitude above 19.5) to produce the 1.04\,d signal, but the brighter one, the companion \target\,B (see Table~\ref{companion}), can cause an event with a maximum depth of $\approx$1800~ppm (corresponding to complete disappearance of the star). In total, the MuSCAT\,2 photometry does not allow to rule out a scenario where the transit signal arises from an eclipse of the unresolved bright nearby star, but this would require the star to present eclipses with a depth of 20\,\%-40\,\%.

\begin{figure}
    \centering
    \includegraphics[width=0.9\linewidth]{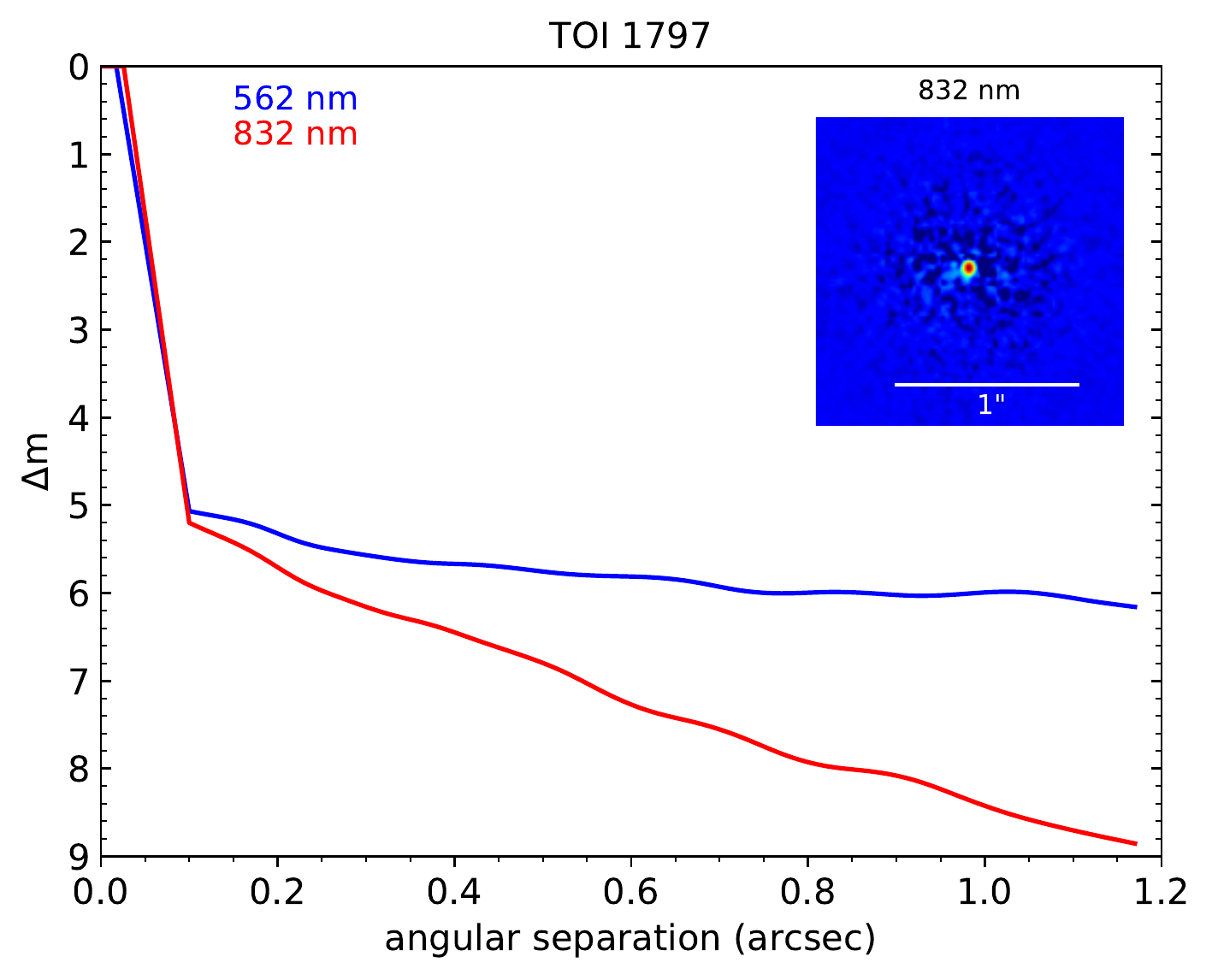}
    \caption{Gemini North/Alopeke 5-sigma contrast curves and the 832 nm reconstructed speckle image}
    \label{GeminiObs}
\end{figure}

\subsection{‘Alopeke speckle imaging}

\target\,A was observed on 09 June 2020 UT using the ‘Alopeke speckle instrument on the Gemini North 8-m telescope\footnote {\url{https://www.gemini.edu/sciops/instruments/alopeke-zorro/}.}. ‘Alopeke provides simultaneous speckle imaging in two bands (562\,nm and 832\,nm) with output data products including a reconstructed image with robust contrast limits on companion detections \citep{Howell2016}. Five sets of $1000\times 0.06$\,s exposures were collected and subjected to Fourier analysis in our standard reduction pipeline \citep{Howell2011}. Figure \ref{GeminiObs} shows our 5\,$\sigma$ contrast curves and the 832\,nm reconstructed speckle image. The analysis of Gemini data showed that \target\,A has no companion brighter than 5-9 magnitudes below that of the target star from the diffraction limit (20 mas) out to 1.2”. At the distance of \target\,A ($\mathrm{d} = 82$\,pc) these angular limits correspond to spatial limits of 1.8 to 98\,AU. Please refer to Section~\ref{contamination} for further analyses on the contamination induced by more distant stars.

\begin{figure}
    \centering
    \includegraphics[width=0.92\linewidth]{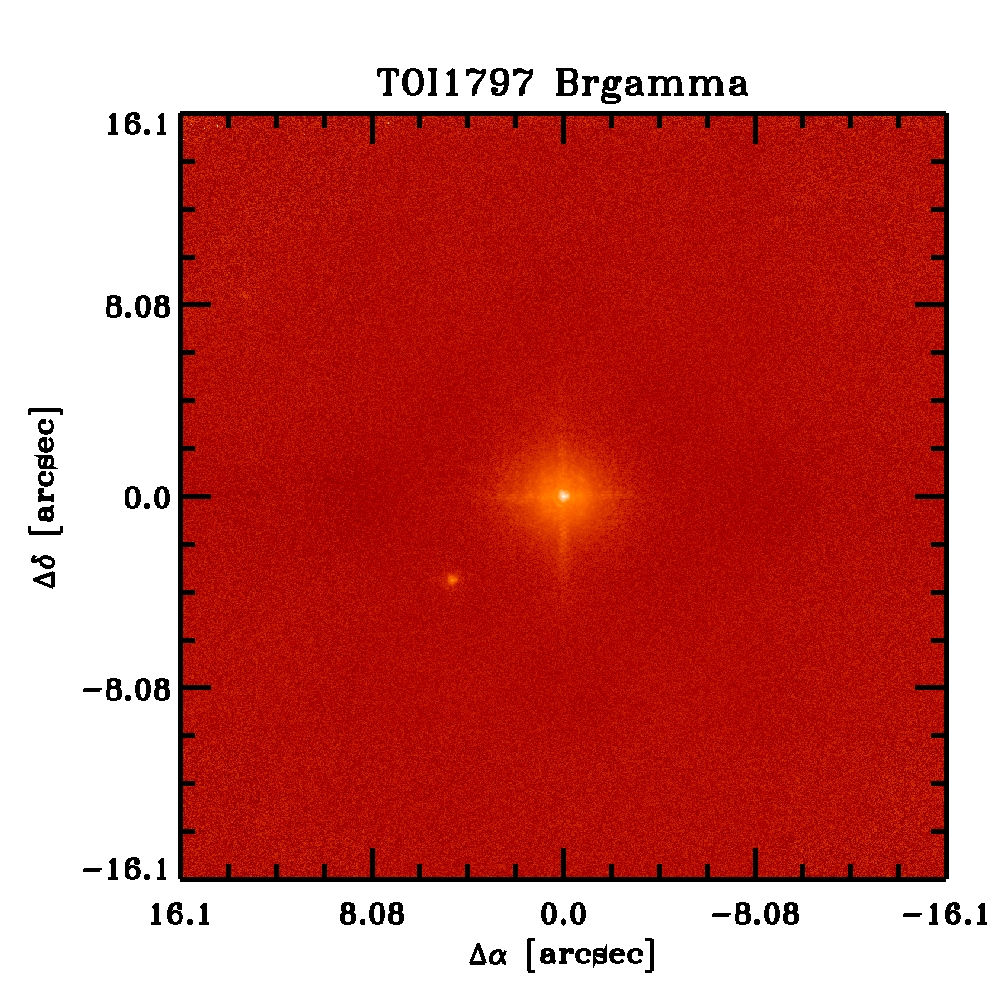}
    \caption{Final combined mosaic of NIR adaptive optics image constructed from the 15-point dither pattern. The nearby companion TIC~368435331 (Gaia~DR3~729899906357408640) is clearly detected $\sim 6\arcsec$ to the south-west of \target\,A}
    \label{fig:ao_fullfov}
\end{figure}

\begin{figure}
    \centering
    \includegraphics[width=0.9\linewidth]{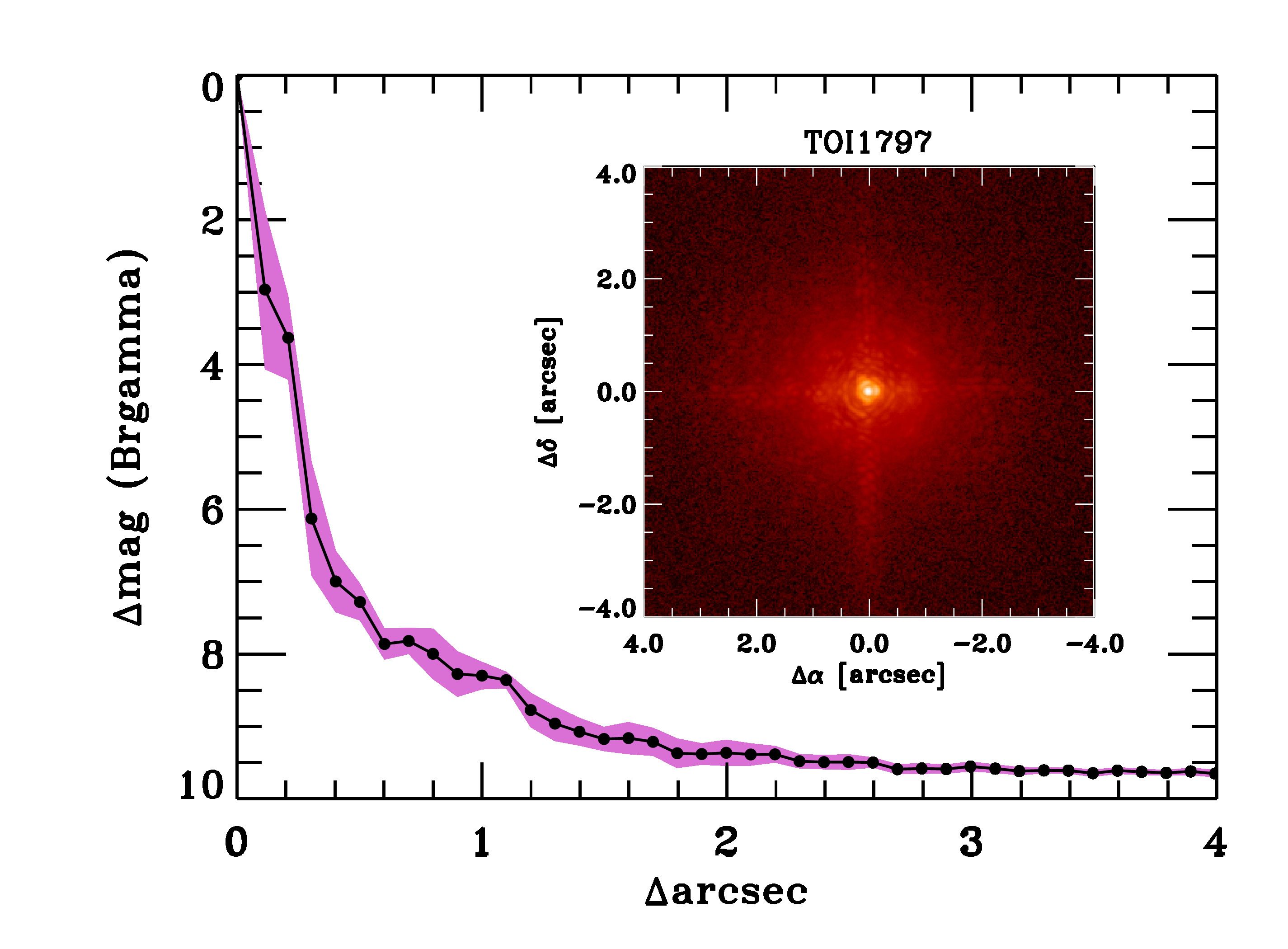}
    \caption{Companion sensitivity for the Palomar adaptive optics imaging.  The black points represent the 5\,$\sigma$ limits and are separated in steps of 1 FWHM ($\sim 0.097$\arcsec); the purple represents the azimuthal dispersion (1\,$\sigma$) of the contrast determinations (see text). The inset image is of the primary target showing no addtional companions to within 3\arcsec\ of the target.}
    \label{fig:ao_contrast}
\end{figure}

\subsection{PHARO adaptive optics imaging}
\label{imaging_palomar}

We observed \target\,A with infrared high-resolution adaptive optics (AO) imaging at Palomar Observatory. On 23 February 2021, we used the PHARO instrument \citep{hayward2001} behind the natural guide star AO system P3K \citep{dekany2013} in a standard 5-point dither pattern with steps of 5\arcsec. The dither positions were offset from each other by 0.5\arcsec, and three observations were made at each position for a total of 15 frames. The camera was in the narrow-angle mode with a full field of view of $\sim$25$\arcsec$ and a pixel scale of approximately 0.025$\arcsec$ per pixel. Observations were made in the narrow-band $Br$-$\gamma$ filter $(\lambda_o = 2.1686; \Delta\lambda = 0.0326\,\mu$m) with an integration time of 4.2 seconds per frame (63 seconds total on-source). 

We processed and analysed the AO data with a custom set of IDL tools. The science frames were flat-fielded and sky-subtracted. We generated the flat fields from a median average of dark subtracted flats taken on-sky and we normalized them such that the median value of the flats is unity. We then generated the sky frames from the median average of the 15 dithered science frames and we sky-subtracted each flat-fielded science image. Finally, we combined the reduced science frames into a single combined image using an intra-pixel interpolation that conserves flux, shifts the individual dithered frames by the appropriate fractional pixels, and median-coadds the frames (Fig.~\ref{fig:ao_fullfov}).

We determined the final resolution of the combined dither ($0.0998\arcsec$) from the full-width at half-maximum of the point spread function (Fig.~\ref{fig:ao_contrast}). We retrieved the sensitivities of the final combined AO image by injecting simulated sources azimuthally around the primary target every $20^\circ $ at separations of integer multiples of the central source's full width at half maximum \citep[FWHM;][]{furlan2017}. The brightness of each injected source was scaled until standard aperture photometry detected it with 5\,$\sigma $ significance. The resulting brightness of the injected sources relative to the target sets the contrast limits at that injection location. We finally calculated the 5\,$\sigma$ limit at each separation by averaging all of the determined limits at that separation and we set the uncertainty on the limit as the root mean square (RMS) dispersion of the azimuthal slices at a given radial distance (Fig.~\ref{fig:ao_contrast}). We confirmed the presence of the single companion to \target\,A, located $5.9$\,\arcsec\ to the south-west (position angle PA\,=\,$127\,\pm\,1^\circ$). The stars were first catalogued as a binary by \citet{Mugrauer21}, due to the consistency between their attributed distances and proper motions (see Table~\ref{companion} for Gaia astrometry). Identified as TIC~368435331 (Gaia~DR3~729899906357408640), \target\,B is $\Delta Ks = 5.21 \pm 0.02$ and $\Delta T = 7.21\pm0.01$ magnitudes fainter than \target\,A and it is separated from \target\,A by at least 484\,AU. Based upon the measured Gaia distance and the observed magnitudes of TIC~368435331, \target\,B is consistent with being an M5\,V star \citep{Mamajek08}. There are no additional companions to the primary detected within the limits of the instruments, that is, within $\Delta \mathrm{K} \leq 3-4$,mag inside of 0.2\arcsec\ and within $\Delta\ K \leq 6$\,mag outside of 0.3\arcsec, which correspond approximately to spectral type limits of M4\,V and M6\,V, respectively, in these angular ranges. Any undetected companions, at the inner angles, would need to have a lower mass than these limits. This result is consistent with both stars having a Gaia DR3 renormalized unit weight error (RUWE\footnote{See \href{https://gea.esac.esa.int/archive/documentation/GDR2/Gaia\_archive/chap\_datamodel/sec\_dm\_main\_tables/ssec\_dm\_ruwe.html}{https://gea.esac.esa.int/archive/documentation/GDR2/Gaia\_\\ archive/chap\_datamodel/sec\_dm\_main\_tables/ssec\_dm\_ruwe.html} for more details.}) around 1 (see Table~\ref{companion}), which indicates a good-quality, single-star astrometric solution for each component in the system. In the case in which the RUWE values had been higher than 1.4 we could have worried about the quality of the astrometric analysis.

\section{Ground-based spectroscopy}
\label{spectroscopic_data}

\subsection{TRES}
\label{TRES}

We obtained a single reconnaissance high-resolution spectrum of \target\,A on 25 November 2020 using the Tillinghast Reflector Echelle Spectrograph \citep[TRES;][]{Furesz08} mounted on the 1.5\,m Tillinghast Reflector telescope at the Fred Lawrence Whipple Observatory (FLWO) atop Mount Hopkins, Arizona. TRES is a fibre-fed, optical spectrograph covering the wavelength range 390-910\,nm and it has a resolving power of R\,=\,44\,000. The exposure time was 195 seconds, which resulted in an S/N of 35. We extracted the spectrum as described in \citet{Buchhave10} and we derived the stellar parameters using the Spectral Parameter Classification tool \citep[\texttt{SPC};][]{Buchhave12, Buchhave14}. SPC cross-correlates an observed spectrum with a grid of synthetic templates to derive stellar parameters. The grid is based on Kurucz atmospheric models \citep{Kurucz93}. We obtained an effective temperature of T$_\mathrm{eff}$\,=\,5883\,$\pm$\,50\,K, a surface gravity of log\,g$_\star$\,=\,4.38\,$\pm$\,0.10 (cgs), an iron content of [Fe/H]\,=\,0.02\,$\pm$\,0.08, and a projected rotational velocity of $\mathrm{v} \sin \mathrm{i}_{\star}$\,=\,5.3\,$\pm$\,0.5\,\kms. 

\subsection{FIES observations}
\label{FIES}

We used the FIbre-fed \'Echelle Spectrograph \citep[FIES;][]{Frandsen1999,Telting14} mounted on the 2.56\,m Nordic Optical Telescope (NOT) of Roque de los Muchachos Observatory (La Palma, Spain) to acquire three high-resolution (R\,$\approx$\,67\,000) spectra of \target\,A over a time base of nearly 15 days. The observations were carried out under good sky and seeing conditions ($\sim$1\,$\arcsec$), as part of the observing programme 62-506 (PIs: E. Knudstrup and L.\,M. Serrano). We set the exposure time to 1800 s, which led to an S/N$\,\sim\,$70-100 per pixel at 550\,nm. Following the same observing strategy adopted in \citet{Gandolfi2013,Gandolfi2015}, we removed cosmic ray hits by splitting each epoch observation in 3 consecutive sub-exposures of 600\,s, and traced the drift of the instrument by acquiring long-exposed ($\sim$140\,s) ThAr spectra immediately before and after each target observation. We reduced the data using standard IRAF and IDL routines and extracted the RVs via multi-order cross-correlation using the stellar spectrum with the highest S/N as a template. The three FIES RV measurements show no significant variation at a level of 10\,m\,s$^{-1}$ ruling out a short-period binary or giant planet scenario.

\subsection{SOPHIE observations}
\label{SOPHIE}

Twenty-nine high-resolution spectra of \target\,A were obtained with the \sophie\ spectrograph \citep{Perruchot2008, Bouchy2013} installed on the 1.93\,m reflector telescope of Haute-Provence Observatory (France), between 17 November 2020 and 26 May 2021. The observations were collected as part of two programs devoted to radial velocity follow-up observations of transiting planet candidates (programme IDs: 20B.PNP.HEBR and 21A.PNP.HEBR; PI: G. H\'ebrard). On each of six nights, we secured 2 observations about two hours apart, in order to attempt a mass measurement of the 1.04~d planetary candidate. The observations were performed using the \sophie\ high-resolution mode (R\,$\approx$\,75\,000) with simultaneous sky monitoring and an exposure time of 730\,s, which led to a typical S/N\,=\,53 per pixel at 550\,nm (Table~\ref{tab:sophie_rvs}). Additionally, for the characterization of the stellar host, we obtained two high S/N spectra using the \sophie\ high-efficiency mode which has a resolving power of R\,$\approx$\,40\,000 (programme ID: 20B.PNP.HOYE; PI: S. Hoyer). The exposure was 1800\,s, leading to S/N\,$\approx$\,128 and 156 per pixel at 550\,nm (Table~\ref{tab:sophie_rvs}).

The data were reduced using the \sophie\ data reduction pipeline \citep{Bouchy2009}. Radial velocities were measured by cross-correlating the extracted spectrum with a G2 binary mask and fitting a Gaussian function to the resulting cross-correlation function (CCF). We corrected the RVs for instrumental drifts by interpolating to the mid-time of exposure the drift estimated between two Fabry-Perot calibrations performed less than 2 hours before and after each exposure. The \sophie\ spectrograph also presents long-term variations of the zero-point \citep{Courcol2015}. To correct for this effect, we monitored RV standard stars every night, making a master time series and subtracting them from our RVs \citep[more details may be found in][]{Courcol2015, Hobson2018}. Finally, the data were also corrected for Moon light contamination and charge transfer inefficiency. Two measurements were not taken into account in the analysis presented in Sect.~\ref{RV_analysis} due to their low S/N ($<$\,30). One measurement was obtained under bad weather conditions, whereas the other has a short exposure time, owing to a technical failure during the integration. The \sophie\ RV measurements of \target\,A are listed in Table~\ref{tab:sophie_rvs}, along with the full width at half maximum (FWHM) and bisector inverse slope (BIS) of the CCF, and the Ca\,{\sc II}\,H\,\&\,K chromospheric activity indicator log\,R$^\prime_\mathrm{HK}$, as extracted using the \sophie\ data reduction pipeline \citep{Bouchy2009}.

\section{Stellar properties}
\label{star}

\subsection{Spectroscopic parameters}
 
We performed the spectral analysis of \target\,A using the combined \sophie\ spectra, as obtained from the co-addition of the two high-efficiency mode spectra (last two lines in Table\ref{RV_table}). The co-added spectrum has a $\mathrm{S}/\mathrm{N}$\,$\approx$\,200 per pixel at 550\,nm. We derived the stellar atmospheric parameters, namely, the effective temperature $\mathrm{T}_{\mathrm{eff}}$, surface gravity $\log \mathrm{g_\star}$, microturbulence velocity v$_\mathrm{mic}$, and iron abundance [Fe/H], using the \texttt{ARES}+\texttt{MOOG} codes, following the methodology described in \citet{Santos13} and \cite{Sousa14}. Briefly, we first computed the equivalent widths of the iron lines with the code \texttt{ARES}\footnote{The latest version of ARES code (ARES v2) can be downloaded at \url{http://www.astro.up.pt/$\sim$sousasag/ares}.} \citep{Sousa07, Sousa15}. We then applied a minimization process to find ionization and excitation equilibrium, and to converge to the best set of spectroscopic parameters. This process makes use of a grid of Kurucz's model atmospheres \citep{Kurucz93} and the radiative transfer code \texttt{MOOG} \citep{Sneden73}. We obtained $\mathrm{T}_{\mathrm{eff}}$\,=\,5987\,$\pm$\,64\,K, $\log \mathrm{g_\star}$\,=\,4.49\, $\pm$\,0.11\,(cgs), [Fe/H]\,=\,0.10\,$\pm$\,0.04 (in agreement with the TRES results presented in Sect.~\ref{TRES}), and $\mathrm{v}_\mathrm{mic}$\,=\,1.15\,$\pm$\,0.03\,\ms. Using the same tools, we also determined the abundance of Mg ([Mg/H]\,=\,0.08\,$\pm$\,0.06) and Si ([Si/H]\,=\,0.08\,$\pm$\,0.04), closely following \citet{Adibekyan12, Adibekyan15}.

As a sanity check, we independently analysed the co-added FIES spectrum using the \texttt{Specmatch-emp} package \citep{Yee2017}, which compares the observed spectrum with a library of over 400 template spectra of stars of all types with well determined physical parameters. After formatting our co-added FIES spectrum to a compatible format \citep{Hirano2018}, a $\chi^2$ minimization procedure provides an estimate of the following parameters: $\mathrm{T}_{\mathrm{eff}} = 5875 \pm 110~\mathrm{K}$, $[\mathrm{Fe}/\mathrm{H}] = 0.10 \pm 0.09$, and $\mathrm{R}_{\star}\,=\,1.10 \pm 0.18\,\mathrm{R}_{\oplus}$. These results are compatible within 1\,$\sigma$ with the parameters retrieved with \texttt{ARES}/\texttt{MOOG}. We finally utilized the package \texttt{Spectroscopy Made Easy} \citep[\texttt{SME}][]{Valenti96, Piskunov2017} to determine the projected rotational velocity of the star v$_{\star} \sin \mathrm{i}_{\star}$. Based on $\mathrm{T}_{\mathrm{eff}}$, $\log \mathrm{g}_{\star}$ and $[\mathrm{Fe}/\mathrm{H}]$ as derived with \texttt{Specmatch-emp}, $\mathrm{v}_{\mathrm{mic}}$ as determined with \texttt{ARES}/\texttt{MOOG}, and  $\mathrm{v}_{\mathrm{mac}} = 4.1 \pm 0.4$\,\kms\ from the \citet{Doyle}'s empirical equation, we found v$_{\star} \sin \mathrm{i}_{\star} = 5.9 \pm 0.8$\,\kms, in very good agreement with the TRES results (see Section~\ref{TRES}).

Using the \citet{Pecaut2013}'s calibration for main sequence stars\footnote{Available at \url{https://www.pas.rochester.edu/~emamajek/EEM_dwarf_UBVIJHK_colors_Teff.txt}.}, we found that the effective temperature implies that \target\,A is a G0 dwarf star. The complete set of spectroscopic parameters adopted in the current work are listed in Table~\ref{tab:stellar_values}.

\subsection{Radius}
\label{radiusmassage}

To determine the stellar radius of \target\,A we used a modified version of the infrared flux method (\texttt{IRFM}; \citealt{Blackwell1977}) in a Markov chain Monte Carlo (MCMC) approach \citep{Schanche2020}. Within the \texttt{IRFM} MCMC, we first constructed synthetic spectral energy distributions (SEDs) from \textsc{atlas} stellar atmospheric models \citep{Castelli2003}, using the photospheric parameters and their uncertainties derived with \texttt{ARES}+\texttt{MOOG}. We derived the stellar angular diameter by fitting the observed fluxes with synthetic photometry computed by convolving the \textsc{atlas} SEDs with the broadband response functions of each bandpass. The angular radius was then converted to stellar radius using the offset-corrected {\it Gaia} EDR3 parallax \citep{GaiaCollaboration2021,Lindegren2021}. In this study, we retrieved the broadband fluxes and their uncertainties for \target\,A from the most recent data releases for the following bandpasses: {\it Gaia} G, G$_{\rm BP}$, and G$_{\rm RP}$, {\it 2MASS} J, H, and Ks, and {\it WISE} W1 and W2 \citep{GaiaCollaboration2021, Wright2010,Skrutskie2006}. We found that \target\,A has a radius of $\mathrm{R}_{\star}$\,=\,1.043\,$\pm$\,0.009\,R$_{\odot}$.

\subsection{Rotation period}
\label{rotation}

The \tess\ light curve of \target\,A shows quasi-periodic photometric variability with a peak-to-peak amplitude of $\sim$0.2\,\% (Figure~\ref{Tess_data}). Given the G0\,V spectral type of the star, this is likely induced by the presence of active regions (spots and plages) carried around by stellar rotation. 

\begin{figure}
    \centering
    \includegraphics[width=\linewidth]{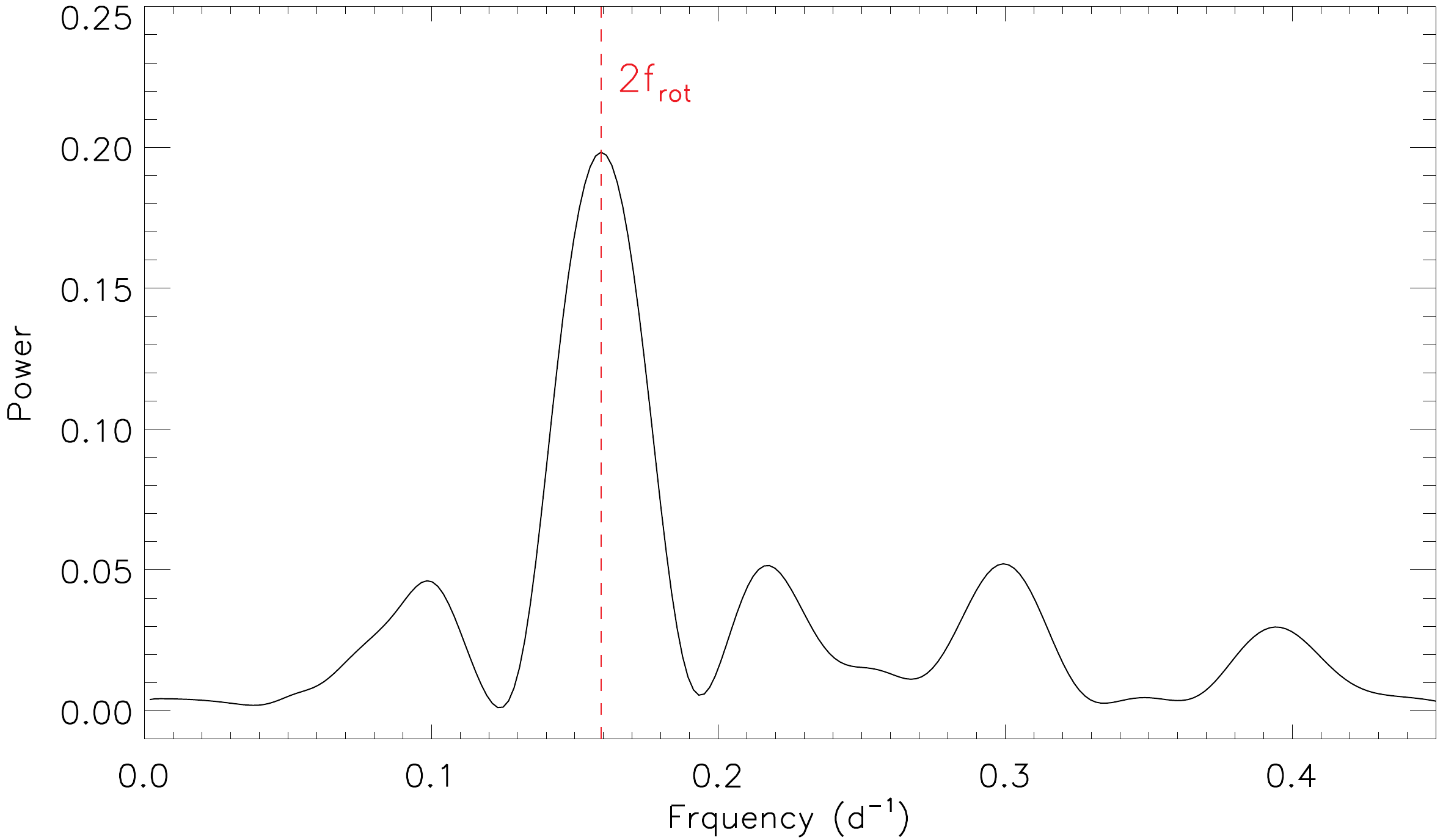}
     \includegraphics[width=\linewidth]{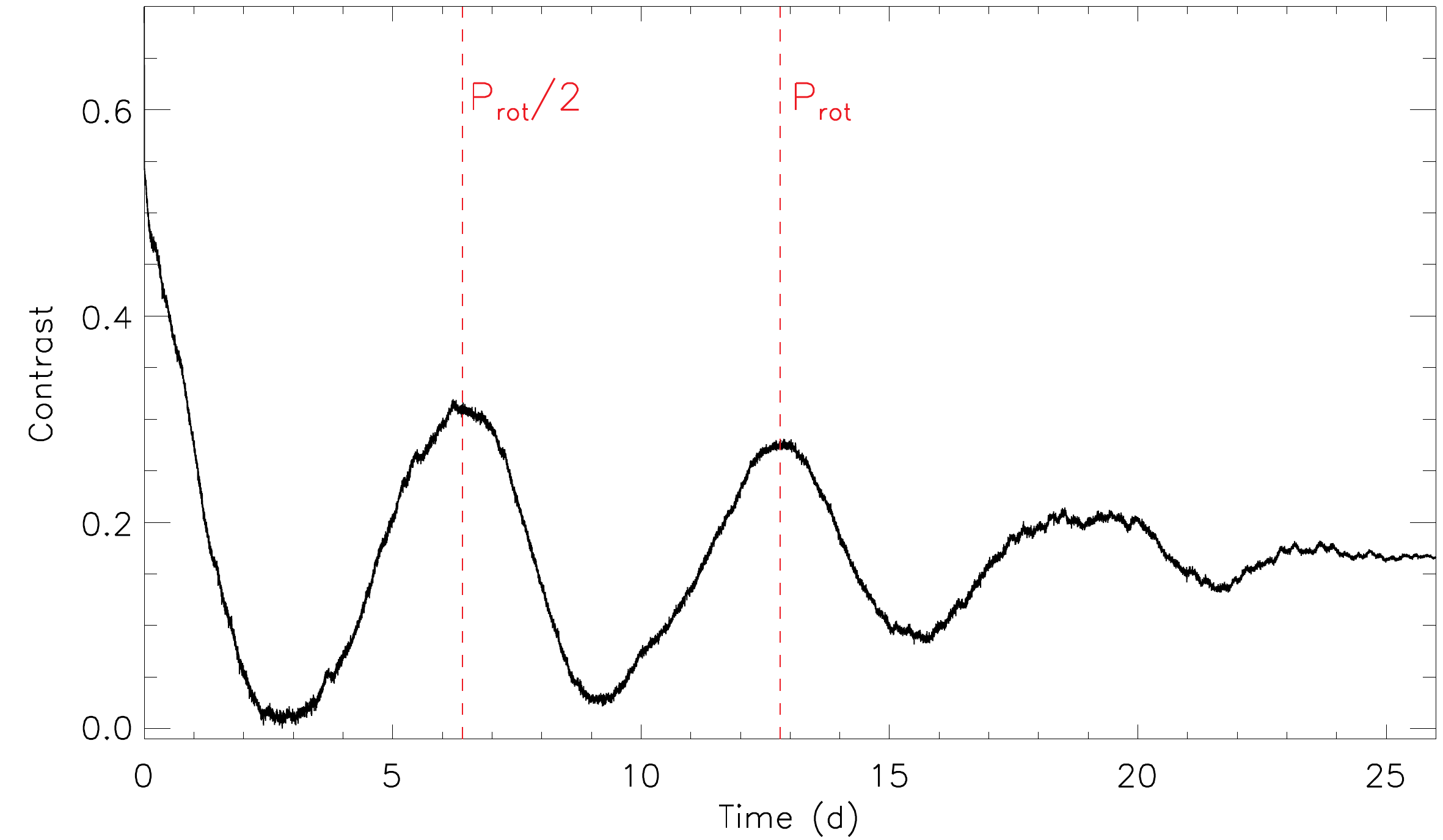}
    \caption{Study of the \target{A} rotational period using \tess\ data. \emph{Upper panel}. Lomb-Scargle periodogram of \target\,A light curves. The red dashed line marks the first harmonic of the rotation frequency of the star. \emph{Lower panel}: Autocorrelation function of the \tess\ light curve. The red dashed lines mark the rotation period of the star and its first harmonic.}
    \label{fig:Prot}
\end{figure}

The Lomb-Scargle periodogram \citep{Lomb76,Scargle82} of the \tess\ time series displays its most significant peak at 0.159\,d$^{-1}$, which corresponds to a period of $\sim$6.3\,d (Fig.~\ref{fig:Prot}, upper panel). We computed its false alarm probability (FAP) following the bootstrap method described in \citet{Murdoch1993} and \citet{Hatzes2019}, and found it to be well below FAP\,$<$\,10$^{-6}$. The auto-correlation function \citep[ACF;][]{McQuillan13} of the \tess\ light curve displays its first two correlation peaks at $\sim$6.4 d and $\sim$2\,$\times$\,6.4\,=\,12.8\,d (Fig.~\ref{fig:Prot}, lower panel). A visual inspection of the \tess\ light curve (Fig.~\ref{Tess_data}, first panel) hints that the flux pattern repeats every $\sim$12.8\,d, suggesting that the stellar rotation period is P$_\mathrm{rot}$\,$\approx$\,12.8\,d. We attribute the peak at 6.4\,d to the presence of active regions at opposite longitudes on the stellar photosphere. From the peak of the ACF and its half full width at half maximum we estimated a rotation period of P$_\mathrm{rot}$\,=\,12.8\,$\pm$1.8\,d. Assuming that the star is seen equator on, the projected rotation velocity ($\mathrm{v} \sin \mathrm{i}_{\star}$\,=\,5.9\,$\pm$\,0.8\,km\,s$^{-1}$) and the stellar radius ($\mathrm{R}_{\star}$\,=\,1.043\,$\pm$\,0.009\,R$_{\odot}$) translate into a rotation period of $8.9^{+1.4}_{-1.1}$\,d, which agrees with our finding within\footnote{Here $\sigma$ has been obtained by quadratically adding 1.4 and 1.8\,d.} 1.7\,$\sigma$.

\subsection{Mass and age}

By adopting $\mathrm{T}_{\mathrm{eff}}$, [Fe/H], and $\mathrm{R}_{\star}$ as input parameters, we then determined the isochronal stellar age $\mathrm{t}_{\star}$ and mass $\mathrm{M}_{\star}$ following two different methods. The first method uses the isochrone placement technique implemented by \citet{bonfanti15,bonfanti16}, taking further benefit of the stellar rotation period determined from the \tess\ photometry $\mathrm{P}_{\mathrm{rot}}=12.8 \pm 1.8$\,d (Sect.~\ref{rotation}). The knowledge of the rotation period of the star improves the convergence of the fitting procedure since it gives an additional constraint via the gyrochronological relation from \citet{barnes10gyro}, as thoroughly discussed in \citet{bonfanti16}. The isochrone placement interpolates the input parameters within pre-computed grids of stellar isochrones and tracks obtained from the \emph{PA}dova \& T\emph{R}ieste \emph{S}tellar \emph{E}volutionary \emph{C}ode \citep[\texttt{PARSEC} v1.2S,][]{marigo17} to infer the stellar age and mass. The second method uses the Code Liègeois d'Évolution Stellaire \citep[\texttt{CLES},][]{scuflaire08} to directly compute the evolutionary track that fits the input parameters and derive age and mass following the Levenberg-Marquardt minimization scheme \citep{salmon21}. After checking the consistency of the two results through a $\chi^2$-based criterion \citep[see][for further details]{Bonfanti21}, we finally combined our age and mass estimates by summing the respective probability density distributions and obtained the following median values (1\,$\sigma$ confidence level): $\mathrm{t}_{\star}=1.4_{-0.4}^{+0.8}$ Gyr and $\mathrm{M}_{\star}$\,=\,$1.109_{-0.043}^{+0.041}$~$\mathrm{M}_{\odot}$. 

As a sanity check, we further employed the python module\footnote{Available at \url{https://github.com/RuthAngus/stardate}.} \texttt{stardate} \citep{Angus19}. The code combines the isochrone fitting within the MESA Isochrones \& Stellar Tracks \citep[\texttt{MIST}][]{dotter16,choi16} with the gyrochronological relation empirically calibrated by \citet{Angus19} in order to infer the ages of F, G, K, and M type stars. The approach is the same as in the isochrone placement of \citet{bonfanti16}, but the theoretical models and the gyrochronological relation are different. Here we used as input the spectroscopic parameters and the photometric band values of \target\,A listed in Table~\ref{tab:stellar_values}. For the parallax we adopted the value in Table~\ref{companion}, while for the stellar rotation we injected the period of $\mathrm{P}_{\mathrm{rot}}=12.8 \pm 1.8$\,d we estimated in Sect.~\ref{rotation}.
We then ran an MCMC simulation, with 100000 iterations and discarding the first 10000 as burn in, to create the posterior distributions from which we infer the best-fit parameters. We obtained $\mathrm{t}_{\star, 1} = 1.94^{+0.34}_{-0.40}$~Gyrs, which is consistent with the result obtained by combining \texttt{PARSEC} and \texttt{CLES} outcomes and suggests that the star is younger than the Sun.

While we have no reason to prefer one method over the other, we adopted the results obtained with the \texttt{PARSEC} and \texttt{CLES} stellar models. Results are listed also in Table~\ref{tab:stellar_values}.

\subsection{Interstellar extinction}

We determined the interstellar extinction A$_\mathrm{v}$ along the line of sight to the star following the procedure described in \citet{Gandolfi2008}. Briefly, we fitted the spectral energy distribution of \target\,A using the synthetic magnitudes obtained by convolving the low-resolution \texttt{BT-NextGen} model spectrum \citep{Allard12} having the same spectroscopic parameters as the star, with the transmission functions of the magnitudes listed in Table~\ref{tab:stellar_values}. Using the \citet{Cardelli1989} extinction law and assuming a total-to-selective extinction ratio of R$\mathrm{v}$\,=\, A$\mathrm{v}/\mathrm{E(B-V})$\,=\,3.1, we found that A$_\mathrm{v}$\,=\,0.10\,$\pm$\,0.05, as expected given the relatively short distance to the star ($\sim$82.5 pc).  

\subsection{Position in galaxy}
We used the systemic radial velocity reported in Table~\ref{RV_table} and the proper motion and parallax in Table~\ref{companion} to determine the local standard of rest (LSR) U, V, and W space velocities of \target\,A following the methodologies in \citet{Johnson87}. We did not subtract the Solar motion and computed the UVW values in the right-handed system. We obtained: $\mathrm{U} = -33.03\,\pm\,0.04\,\mathrm{km}\,\mathrm{s}^{-1}$, $\mathrm{V} = -21.42\,\pm\,0.03\, \mathrm{km}\,\mathrm{s}^{-1}$, $\mathrm{W} = -5.06\,\pm\,0.03\, \mathrm{km}\,\mathrm{s}^{-1}$. Adopting the formalism presented in \citet{Reddy06}, we used these estimates to compute the probability that \target\,A belongs to the galactic thin disk, thick disk, or halo stellar population. Briefly, we adopted an MC approach with 100\,000 samples using the velocity dispersion standards described in \citet{Chen21}, \citet{Reddy06}, \citet{Bensby14}, and \citet{Bensby03}, and estimated the probability of the star belonging to a certain kinematic galactic family as a weighted average. We obtained that \target\,A has a 98.7\,\% probability of belonging to the thin disk, while it is highly unlikely that it is part of the thick disk (1.3\,\%) or of the halo ($\sim$0\,\%). We additionally used the \gaia\ EDR3 position, proper motion, parallax, and the stellar radial velocity to compute the galactic orbit with the package \texttt{galpy} \citep[integrating over 5 Gyrs,][]{Bovy15}, from which we retrieved a galactic eccentricity of 0.09 and a high galactic Z-component of the specific relative angular momentum of $\sim$1686\,kpc\,\kms. These results, together with our derived $\mathrm{[Fe/H]}$, well support the kinematic membership of the star to the galactic thin disk.

\section{Joint analysis of the \tess\ and \cheops\ transit light curves}
\label{Joint}

We performed a joint analysis of the detrended \tess\ and \cheops\ light curves (Sects.~\ref{search_for_transits} and \ref{CHEOPS_detrend}). We excluded the \lcogt\ data due to their poor quality when compared to the space-based \tess\ and \cheops\ photometry. For the joint fit, we used the software suite \texttt{pyaneti} \citep{Barragan19,Barragan21b}, which couples a Bayesian framework with an MCMC sampling to produce posterior distributions of the fitted parameters. \texttt{pyaneti} uses the quadratic limb-darkening transit model of \citet{Mandel02} following the $\mathrm{q}_1$ and $\mathrm{q}_2$ parametrization proposed by \citet{Kipping13}. We set Gaussian priors on $\mathrm{q}_1$ and $\mathrm{q}_2$ using the limb darkening coefficients derived by \citet{Claret17} and \citet{Claret21} for the \tess\ and \cheops\ passbands, respectively. We imposed a conservative standard deviation of 0.1 on both the linear and quadratic terms. We sampled for the mean stellar density $\rho_\star$ and recovered the scaled semi-major axis for each planet (a/R$_\star$) using Kepler’s third law \citep{Winn2010}. In detail, we imposed a Gaussian prior on $\rho_\star$ using the stellar mass and radius derived in Sect.~\ref{star}. We adopted wide uninformative uniform priors for the remaining transit parameters (see Table\,\ref{table_priors}), with the exception of the eccentricity and the argument of  periastron, which were fixed to zero and $90\degr$, respectively.

We explored the parameter space with 500 chains and checked for convergence using the Gelman-Rubin statistics \citep{Gelman92}. Once the chains converged, we used the last 500 iterations and saved the chain states every 10 iterations, generating a posterior distribution of 250\,000 points for each fitted parameter. The inferred transit parameters and their uncertainties are defined as the median and the 68\,\% region of the credible interval of the corresponding posterior distributions (Table~\ref{table_results}). The transit depths of \depthbTESS\ and \depthcTESS\ of the 1.04\,d and 3.65\,d signal imply planetary radii of R$_\mathrm{b}$\,=\, \rpbTESS\ (3\% precision) and R$_\mathrm{c}$\,=\,\rpcTESS\ (1.7\% precision), respectively. 

We checked whether the \tess\ and \cheops\ light curves provide consistent transit depths repeating the procedure described above, but independently modelling the transit depths within the two data-sets. Results of this sanity check are reported in Table~\ref{transit_depths}. For the 3.65\,d candidate \tess\ gives a transit depth of $\mathrm{D}_{\mathrm{c}}$\,=\,\depthcTESSone, while the one measured by \cheops\ is \depthcCHEOPSone. For the 1.04\,d candidate we found that \tess\ gives a transit depth of $\mathrm{D}_{\mathrm{b}}$\,=\,\depthbTESSone, while \cheops\ provides \depthbCHEOPSone. For both signals, the transit depths measured with the two instruments are consistent within 1\,$\sigma$. The \cheops\ bandpass is a broad optical bandpass very similar to that of the \gaia\ Gmag, providing coverage at wavelengths bluer than the red-optical \tess\ one. Given the different passbands, this provides evidence that the transits are achromatic, unrelated to instrumental effects. 

We also tested whether the stellar density would significantly change with respect to the spectroscopic value when adopting uniform priors between 0 and 1 on the limb darkening coefficients and between 0 and 6 $\mathrm{g}\,\mathrm{cm}^{-3}$ on the stellar density. This test gave a stellar density of $1.42_{-0.29 }^{+0.15}\,\mathrm{g}\,\mathrm{cm}^{-3}$ consistent with the spectroscopic density \densspb, as well as planetary parameters compatible within 1\,$\sigma$ with those reported in Table~\ref{table_results}. This implies that the stellar density is also well constrained by the transit light curves.

\begin{table}
  \footnotesize
  \caption{Priors used in the joint analysis of the \tess\ and \cheops\ transit light curves. \label{table_priors}}  
  \centering
  \begin{tabular}{lc}
  \hline
  \hline
  \noalign{\smallskip}
  Parameter & Prior  \\
  \noalign{\smallskip}
  \hline
  \noalign{\smallskip}
  \multicolumn{2}{l}{\emph{ \bf Model Parameters for \target\,A\,b}} \\
  \noalign{\smallskip}
    Orbital period P$_{\mathrm{b}}$ (days)  & $\mathcal{U}[1.0390, 1.0392]$ \\
    Transit epoch T$_{0\,\mathrm{b}}$ (BJD - 2457000)  & $\mathcal{U}[1901.2277, 1901.3277]$ \\
    Scaled planetary radius R$_{\mathrm{b}}/\mathrm{R}_{\star}$ &  $\mathcal{U}[0,0.03]$ \\
    Impact parameter, $\mathrm{b}$ &  $\mathcal{U}[0,1]$ \\
    \noalign{\smallskip}
    \multicolumn{2}{l}{\emph{ \bf Model Parameters for \target\,A\,c}} \\
    \noalign{\smallskip}
    Orbital period P$_{\mathrm{c}}$ (days)  &  $\mathcal{U}[3.6441 , 3.6461]$ \\
    Transit epoch T$_{0\, \mathrm{c}}$ (BJD - 2457000)  & $\mathcal{U}[1902.7740 , 1902.9740]$ \\
    Scaled planetary radius R$_{\mathrm{c}}/\mathrm{R}_{\star}$ & $\mathcal{U}[0,0.035]$ \\
    Impact parameter, b$_{\mathrm{c}}$ & $\mathcal{U}[0,1]$  \\
    \noalign{\smallskip}
    \multicolumn{2}{l}{\emph{ \bf Other system parameters}} \\
    \noalign{\smallskip}
    Stellar density, $\rho_{\star}$ & $\mathcal{N}[1.380,0.064]$ \\
    Limb darkening coefficient $\mathrm{q}_1$ \tess\  & $\mathcal{N}[0.3100 , 0.1000]$  \\
    Limb darkening coefficient $\mathrm{q}_2$ \tess\ & $\mathcal{N}[0.2500 , 0.1000]$ \\
    Limb darkening coefficient $\mathrm{q}_1$ \cheops\  & $\mathcal{N}[0.4200 , 0.1000]$  \\
    Limb darkening coefficient $\mathrm{q}_2$ \cheops\ & $\mathcal{N}[0.2900 , 0.1000]$  \\
    \noalign{\smallskip}
    \hline
   \end{tabular}
\end{table}

\begin{table*}
  \footnotesize
  \caption{\target\,A system parameters. \label{table_results}}  
  \centering
  \begin{tabular}{lc}
  \hline
  \hline
  \noalign{\smallskip}
  Parameters & Values from \tess+\cheops\ joint fit  \\
  \noalign{\smallskip}
  \hline
  \noalign{\smallskip}
  \multicolumn{2}{l}{\emph{ \bf Model parameters for \target\,A\,b}} \\
  \noalign{\smallskip}
    Orbital period P$_{\mathrm{b}}$ (days) & \Pb[] \\
    \noalign{\smallskip}
    Transit epoch T$_{0\,\mathrm{b}}$ (BJD - 2457000) & \Tzerob[] \\
    Planet-to-star radius ratio R$_{\mathrm{b}}/\mathrm{R}_{\star}$ & \rrbTESS[] \\
    \noalign{\smallskip}
    Impact parameter $\mathrm{b}$ & \bb[] \\
    \noalign{\smallskip}
    Scaled semi-major axis a$_{\mathrm{b}}/\mathrm{R}_{\star}$ & \arb[] \\
  \noalign{\smallskip}    
  \multicolumn{2}{l}{\emph{ \bf Model parameters for \target\,A\,c}} \\
  \noalign{\smallskip}
    Orbital period P$_{\mathrm{c}}$ (days) & \Pc[] \\
    \noalign{\smallskip}
    Transit epoch T$_{0\, \mathrm{c}}$ (BJD - 2457000) & \Tzeroc[] \\
    \noalign{\smallskip}
    Planet-to-star radius ratio R$_{\mathrm{c}}/\mathrm{R}_{\star}$ & \rrcTESS[] \\
    Impact parameter b$_{\mathrm{c}}$ & \bc[] \\
    Scaled semi-major axis, a$_{\mathrm{c}}/\mathrm{R}_{\star}$ & \arc[] \\
  \noalign{\smallskip}
  \multicolumn{2}{l}{\emph{ \bf Other system parameters}} \\
  \noalign{\smallskip}
    Limb darkening coefficient $\mathrm{q}_{1, \tess}$ & \qoneTESS[] \\
    Limb darkening coefficient $\mathrm{q}_{2, \tess}$ & \qtwoTESS[] \\
    \noalign{\smallskip}
    Limb darkening coefficient $\mathrm{q}_{1, \cheops}$ & \qoneCHEOPS[] \\
    Limb darkening coefficient $\mathrm{q}_{2, \cheops}$ & \qtwoCHEOPS[] \\
    \noalign{\smallskip}
    Jitter term $\sigma_{\tess\ }$ & \jtrTESS[]\\
    Jitter term $\sigma_{\cheops\ }$ & \jtrCHEOPS[]\\
 \noalign{\smallskip}
    \hline
 \noalign{\smallskip}
 \multicolumn{2}{l}{\textbf{Derived parameters for \target\,A\,b}} \\
 \noalign{\smallskip}
    Planetary radius R$_{\mathrm{b}}$ ($\mathrm{R}_{\oplus}$) & \rpbTESS[] \\
    Semi-major axis a$_{\mathrm{b}}$ (AU) & \ab[] \\
    Orbital inclination i$_{\mathrm{b}}$ (deg) & \ib[] \\
    Transit duration W$_\mathrm{b}$(hours) & \ttotb[] \\
    Transit depth D$_\mathrm{b}$(ppm) & \depthbTESS[] \\
    Equilibrium temperature T$_{\rm eq, b}$ (K) & \Teqb[] \\
    Insolation F$_{\rm p, b}$ (F$_{\oplus}$) & \insolationb[] \\
  \noalign{\smallskip}
  \multicolumn{2}{l}{\textbf{Derived parameters for \target\,A\,c}} \\
  \noalign{\smallskip}
    Planetary radius R$_{\mathrm{c}}$ ($\mathrm{R}_{\oplus}$) & \rpcTESS[] \\
    Semi-major axis a$_{\mathrm{b}}$ (AU) & \ac[] \\
    Orbital inclination i$_{\mathrm{c}}$ (deg) & \ic[] \\
    Transit duration W$_{\mathrm{c}}$(hours) & \ttotc[] \\
    Transit depth D$_{\mathrm{c}}$(ppm) & \depthcTESS[] \\
    Equilibrium temperature T$_{\rm eq, c}$ (K) & \Teqc[] \\
    Insolation F$_{\rm p, c}$ (F$_{\oplus}$) & \insolationc[] \\
 \noalign{\smallskip} \multicolumn{2}{l}{\textbf{Other derived parameters}}\\
 \noalign{\smallskip}
    Limb darkening coefficient u$_{1, \tess\ }$ & \uoneTESS[] \\
    Limb darkening coefficient u$_{2, \tess\ }$ &  \utwoTESS[] \\
    Limb darkening coefficient u$_{1, \cheops\ }$ & \uoneCHEOPS[] \\
    Limb darkening coefficient u$_{2, \cheops\ }$ &  \utwoCHEOPS[] \\
   \noalign{\smallskip}
   \hline
   \end{tabular}
\end{table*}

\begin{table}
  \footnotesize
  \caption{Scaled planetary radii, estimated planetary radii and transit depths retrieved from the photometric joint fit that accounts for the difference in pass-bands between \tess\ and \cheops. \label{transit_depths}}  
  \centering
  \begin{tabular}{lcc}
  \hline
  \hline
  \noalign{\smallskip}
  Parameter & \tess\ & \cheops\  \\
  \noalign{\smallskip}
  \hline
  \noalign{\smallskip}
    Scaled radius $\mathrm{R}_{\mathrm{b}} /\mathrm{R}_{\star}$ & \rrbTESSone[] & \rrbCHEOPSone[]\\
    Scaled radius R$_{\mathrm{c}}/\mathrm{R}_{\star}$ & \rrcTESSone[] & \rrcCHEOPSone[]\\
    \noalign{\smallskip}
    Radius R$_{\mathrm{b}}$ ($\mathrm{R}_{\oplus}$) & \rpbTESSone[] & \rpbCHEOPSone[]\\
    Radius R$_{\mathrm{c}}$ ($\mathrm{R}_{\oplus}$)  & \rpcTESSone[] & \rpcCHEOPSone[] \\
    \noalign{\smallskip}
    Transit depth D$_\mathrm{b}$(ppm) & \depthbTESSone[] & \depthbCHEOPSone[]\\
    Transit depth D$_{\mathrm{c}}$(ppm) & \depthcTESSone[] & \depthcCHEOPSone[] \\
    \noalign{\smallskip}
    \hline
   \end{tabular}
\end{table}

\begin{figure*}
    \centering
    \includegraphics[width=0.49\linewidth]{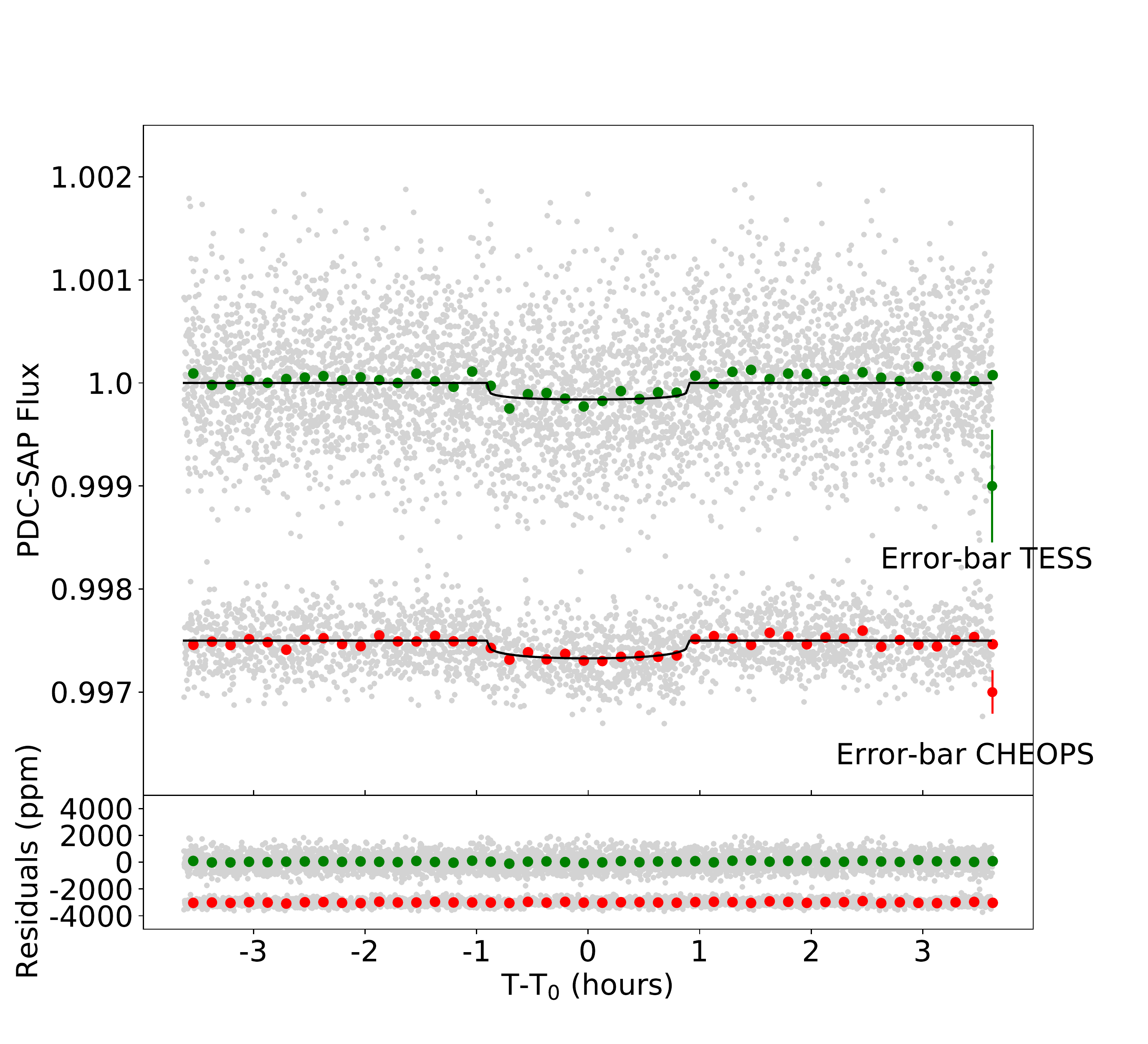}
   \includegraphics[width=0.49\linewidth]{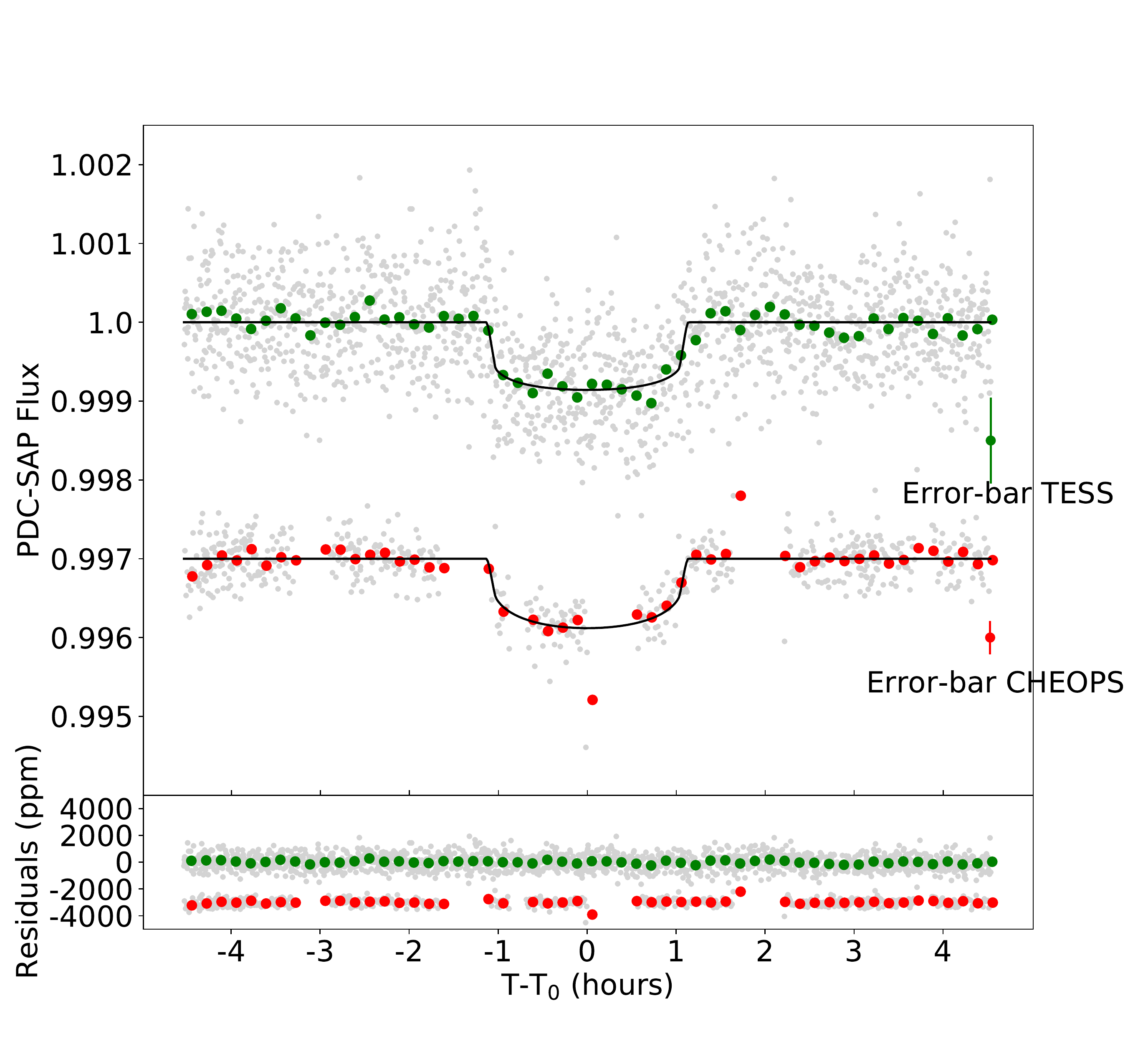}
    \caption{Phase-folded light curve of planets \target\,A\,b (left panel), and c (right panel). \textit{Upper panels}: photometric measurements are shown with light grey circles, along with the 10-minute binned data  (green circles for \tess, and red for \cheops), and the best-fitting transit model (solid black line). On the right-hand side of the plot, we also show the size of \tess\ and \cheops\ error-bars, respectively in green and red. \textit{Lower panels}: residuals, colour-coded as above.}
    \label{planetb_phase_folded}
\end{figure*}

\section{Validation of the two transiting planets}
\label{validation}

\begin{figure}
    \centering
    \includegraphics[width = 8cm]{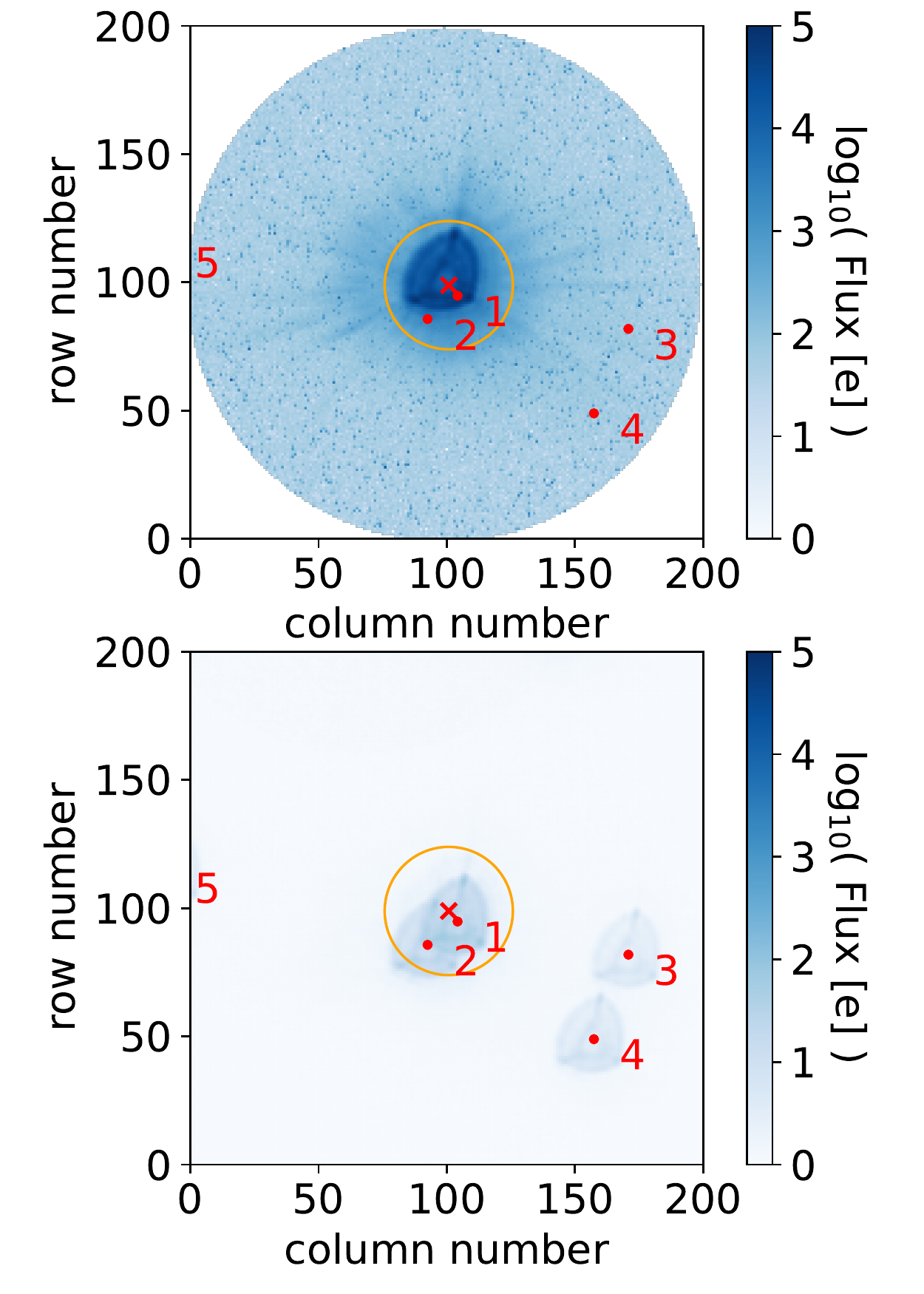}
    \caption{\cheops\ images of \target\ FoV. \textit{Upper panel}: An example of a \cheops\ subarray science image (from visit CH\_PR110045\_TG002001\_V0200).
    \textit{Lower panel}: \texttt{DRP} simulated FoV only with the background stars (therefore with the target removed), used to estimate the contamination level induced by these stars in the photometric aperture. \textit{Both panels}: The orange circle represents the \texttt{DEFAULT} photometric aperture and the red numbered dots indicate the location of nearby stars. Due to the large difference in brightness, the background stars are barely detected in the \cheops\ science images.}
    \label{FoV}
\end{figure}

\begin{figure}
    \centering
    \includegraphics[width = 8cm]{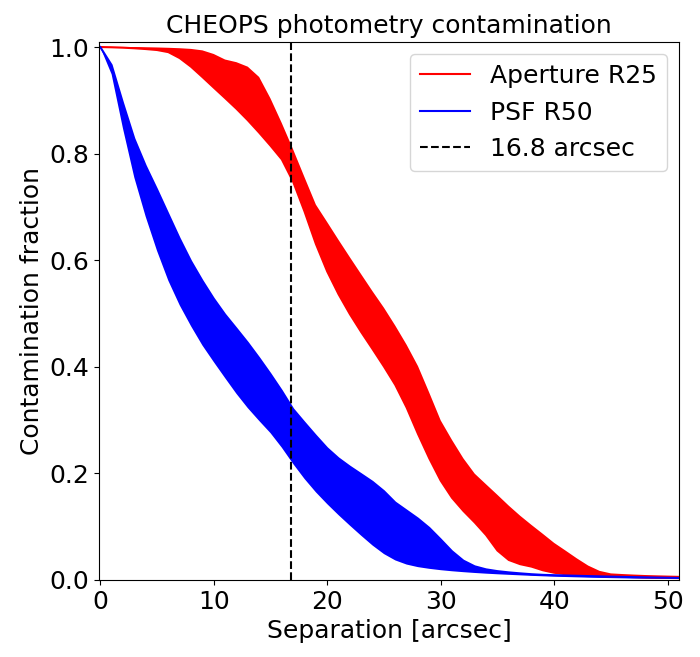}
    \caption{The fraction of flux of a contaminant that affects the photometry for aperture and PSF photometry as a function of separation. The width of the lines reflect the angular dependence due to the asymmetry of the \cheops\ PSF. R25 and R50 indicate the aperture radii expressed in pixels. The dashed vertical line at 16.8\arcsec\ corresponds to the separation of the most significant potential contaminant.}
    \label{fig:comparison}
\end{figure}

\target\,A hosts two non-grazing transiting planet candidates with periods of 3.65\,d (TOI) and 1.04\,d (non-TOI). The validation of the system requires ruling out the following possible false positive scenarios  \citep[see, e.g.][]{Daylan21,Wilson2022}. The first possibility is that the transit-like events (hereafter transits in this section) might be the result of instrumental effects. The second suggests that the host star is an eclipsing binary, whose eclipses ar being misinterpreted as transit features. The third scenario is that the transits might be caused by planets or stars orbiting \target\,B whose light leaks into the photometric mask of \target\,A. Finally there is a risk that planets or stars might be orbiting a background or foreground star whose light leaks into the photometric mask of \target\,A, generating a transit like feature.

We can exclude that the transits are due to systematics because they have been detected by two instruments (\tess\ and \cheops) with the same periods and durations. Moreover, both signals are achromatic, with \tess\ and \cheops\ transit depths being compatible well within 1\,$\sigma$ (Sect.~\ref{Joint}), as expected from bona fide transiting planets. The measured stellar density, as derived from the modelling of the transit signal, is consistent with the spectroscopic density (Sect.~\ref{Joint}), and the impact parameters and scaled semi-major axes imply that the orbits of the two planets are co-planar ($\mathrm{i}_{b} = $ \ib\ and $\mathrm{i}_\mathrm{c} = $ \ic; Table~\ref{table_results}), providing additional evidence that the two series of transit-like dips are actually transits of \target\,A.

\subsection{Contaminants in the \cheops\ field of view}
\label{contamination}

An analysis of the \cheops\ FoV indicates that the level of contamination is low. Figure~\ref{FoV} shows the \texttt{DRP} simulated image of the FoV of \target\,A, with the red circles marking the potential sources of contamination. Given the high galactic latitude of the star ($\sim$63$^\circ$), the number of contaminants is relatively small. Within the \cheops\ FoV there are 6 known \gaia\ sources brighter than \gaia\ G-band magnitude 19.5 (Table~\ref{companion}), with only 2 contaminants being inside the photometric aperture. Four out of the 6 contaminants are too faint ($\mathrm{G}\,>\,17.5$) and too separated from \target\,A to be the source of the transit signals (see below). The closest contaminant, marked as 1 in Fig.~\ref{FoV}, is the companion \target\,B (see Sect.~\ref{imaging_palomar}). In the following section, we show that the transit signals cannot happen on the companion \target\,B. The star marked with 2 in Fig.~\ref{FoV}, is DR3 729899902062379776, a G\,=\,17.5 star located 16.8$\arcsec$ away from \target\,A. This is the only contaminant potentially capable of generating a false positive detection.

A contaminant can affect the light curve in two ways: firstly, flux from the contaminant dilutes any transit signal. Secondly, if the contaminant is an eclipsing binary (EB), it can generate a false transit detection. The dilution is already strongly mitigated, because \texttt{PIPE} removes contributions due to contaminants by subtracting a synthetic image produced using the PSF and parameters from the EDR3 catatalogue for the contaminant stars. This correction does not account for variability, so an EB may still induce a transit signal. The contribution of a detectable contaminant star (within 0.5\arcsec) to the PSF photometry of the target is different from the case of aperture photometry, depending on the target-contaminant separation. We can therefore do an a priori test of false positive detections for the two transits, by comparing the variability in the photometry extracted with the two methods; if the transit signal comes from the target, both methods should give consistent transit depths.

To quantify the effects of contamination on PSF and aperture photometry, we used an empirical \cheops\ PSF and computed the effects of contamination for all position angles in steps of 5 degrees. We included the angular dependence because the \cheops\ PSF is strongly asymmetric, so the contamination depends on the orientation of the PSF with respect to the contaminant. In Fig.~\ref{fig:comparison} we plot the contamination fraction, which is the fraction of flux that affects the flux estimate, as a function of separation for the two photometric extraction methods and the full range of position angles. In the \texttt{DRP}, the assumed aperture has the default radius 25\,pixels. Since the PSF drops steeply at larger radii, the dependence on its defined radius is insignificant as long as it is greater than the contaminant separation.

From the plot, we see that the contamination fraction for the two extraction methods have different dependencies on separation, but that the contamination in PSF extraction is typically a few times lower than in aperture extraction. Among the sources listed in Table~\ref{companion}, we confirm that the strongest contaminant is DR3 729899902062379776. At this separation, the contamination in the extracted \texttt{PIPE} photometry should be about 40\,\% weaker than in the \texttt{DRP} photometry.  A comparison of the extracted photometry from PIPE and the DRP shows about a 3\,\% difference for both planets,  consistent within the estimated measurement uncertainty of 7\,\% and 4\,\% for the 1.04\,d and 3.65\,d signals, respectively. This confirms that the transits are unlikely to be due to the contaminating star. For the 3.65\,d signal, this can also be inferred from the fact that the observed transit depth of 805\,ppm is stronger than the signal even a 100\,\% occulting EB would induce, which would be 690\,ppm in an R25 aperture and 480\,ppm for the PSF extraction.

An additional confirmation that the 1.04\,d candidate is not a false positive has been provided by the photometric in-transit observations carried out with MuSCAT\,2 (Sect.~\ref{MUSCAT_2}). Although the transit is too shallow to be detected from the ground, the analysis of the light curve of the contaminant star DR3\,729899902062379776 rules out a false positive detection due to a contaminating eclipsing binary scenario. Finally, the \lcogt\ observations showed that the transit of the 3.65\,d period planet happens on \target\,A, therefore excluding the signal is a false positive caused by NEBs.

\subsection{No transits on \target\,B}
\label{TOI1797B}

\target\,A has a gravitationally bound stellar companion at $\sim$5.9$\arcsec$ south-east (Sect.~\ref{imaging_palomar}). We now want to understand whether the planets transit the bright component, \target\,A, or the faint companion, \target\,B. As mentioned in Section~\ref{imaging_palomar}, \target\,B is an M5\,V type star, that is a star with $\mathrm{T}_\mathrm{eff, B}$\,$\approx$\,3060\,K, $\mathrm{M}_{\star,\mathrm{B}}$\,$\approx$\,0.162$\mathrm{M}_{\odot}$, and $\mathrm{R}_{\star,\mathrm{B}}$\,$\approx$\,0.196\,$\mathrm{R}_{\odot}$ \citep{Mamajek08}.

If \target\,B were totally occulted by an inflated gas giant planet or a second M-dwarf, the diluted depth (estimated taking into account the difference of 7.9 between the G-band magnitudes of \target\,A and \target\,B) would be $\sim$700\,ppm, which is too shallow to account for the \depthcTESS\ depth of the transit signal of planet c. For a 3.65-d orbit, the transit duration would be $\sim$1.2\,h for a Jupiter mass companion and shorter for a stellar companion, in disagreement with the observed duration of \ttotc. If \target\,B were totally occulted by an orbiting companion having the same size as the star, the transit would be v-shaped, while the \tess\ and \cheops\ transits are u-shaped.

However, the depth of the 1.04\,d transit signal (\depthbTESS) could be explained by an object transiting \target\,B. Based on its predicted radius of 0.196\,$\mathrm{R_{\odot}}$, only a
Jupiter-size object with a radius of $\sim$0.09\,$\mathrm{R_{\odot}}$ would be able to account for the detected transit depth. Yet, given the stellar radius and the scaled semi-major axis $\mathrm{a}_\mathrm{b}/\mathrm{R}_\mathrm{B}$\,$\approx$\,12, the transit duration would be $\sim$0.9\,h, which is significantly shorter than the measured duration of \ttotb. A highly-eccentric orbit could still give a longer transit duration if the transit occurred at the apocentre. On the other hand, with a period of just 1\,d it is highly unlikely that the orbit is eccentric.

We conclude that the two series of transit signals do not happen on \target\,B given their shapes, depths, and durations. This is also corroborated by the fact that the stellar density derived from the modelling of the transit light curves is consistent with the spectroscopic density of \target\,A (Sect.~\ref{Joint}).

\subsection{Ruling out false positive scenarios}
\label{Triceratops}

We used the Tool for Rating Interesting Candidate Exoplanets and Reliability Analysis of Transits Originating from Proximate Stars  \citep[\texttt{TRICERATOPS};][]{Giacalone21} to further verify that the transits are not due to contaminating eclipsing binaries. \texttt{TRICERATOPS} is a Bayesian tool that uses the phase-folded primary transit, any pre-existing knowledge about the stellar host and contaminants, and the understanding of planet occurrence and stellar multiplicity. For each planet, the code requires the photometric phase-folded light curve (with phase equal to zero corresponding to the transit centre), the target \tess\ input catalogue ID and the \tess\ sector in which the target was observed. The code refers to the MAST database to recover the list of stars within $10$ pixels (each pixel corresponds to $21\arcsec$) from the target. \texttt{TRICERATOPS} uses these inputs to calculate the contribution of nearby stars to the observed flux, and identifies those that are bright enough to produce the observed transit. The accounted for scenarios of false positive detections are the following: an unresolved bound companion with a planet transiting the secondary star, an unresolved foreground or background star hosting a planet, a foreground or background eclipsing binary, a transit on a nearby star, and a nearby eclipsing binary (NEB). The code uses a Bayesian framework to estimate the probability associated with each scenario, by fitting transit and eclipsing binary models accounting for the input orbital period and twice the period. 

We ran \texttt{TRICERATOPS} for both candidates. The pipeline identified a total of 23 stars within 10$\arcsec$ from the target star, although only 2 nearby stars were considered to be potential NEBs: the gravitationally bound stellar companion DR3 729899906357408640, and the star DR3 729899902062379776. For DR3 729899906357408640, the stellar companion, we assumed the typical mass and radius of an M5 dwarf \citep[according to][]{Mamajek08}, while for DR3 729899902062379776 we had no specific assumption (specifying the spectral type does not affect the results of the analysis). We first injected phase-folded \tess\ data, using for $\mathrm{T}_0$ and $\mathrm{P}$ the values listed in Table~\ref{table_results}. We then ran \texttt{TRICERATOPS} on the combination of the \tess\ and the \cheops\ light curves also using the values of T0 and P listed in Table~\ref{table_results}. For both candidates we found that the signals are very likely associated with planets transiting \target\,A. We obtained a false positive probability of FPP\,$\sim$\,0.009 and a negligible nearby false positive probability (NFPP) for the 3.65\,d planet, which is therefore validated according to the \texttt{TRICERATOPS} criteria (FPP\,$<$ \,0.015 and NFPP\,$<$\,0.001 for a planet to be validated). This result represents a notable improvement with respect to that obtained by \citet{Giacalone21}. \citet{Giacalone21} ran this analysis only on the \tess\ data, obtaining an FPP\,$\sim$\,0.3, according to which the candidate is not validated, but must be classified as ``likely a planet''. In the case of the 1.04\,d transit signal, the code provides an FPP\,$\sim$\,0.09 and an NFPP\,$\sim$\,0.006. Based on the \texttt{TRICERATOPS} criteria this test tells us that this object is ``likely a planet'' but there is a low probability of false positives due to binaries. Since planet c is already confirmed, folding in the FPP of planet c to the validation analysis of planet b would likely decrease the FPP of planet b. Furthermore, based on multiplicity statistics determined from \kepler\ discoveries \citep{Lissauer2012}, it is highly likely that an additional transit signal in a system with a validated planet is also planetary in nature. We therefore conclude that both planets are validated.

\section{Radial velocity analysis}
\label{RV_analysis}
\begin{figure*}
    \centering
    \includegraphics[width=\linewidth]{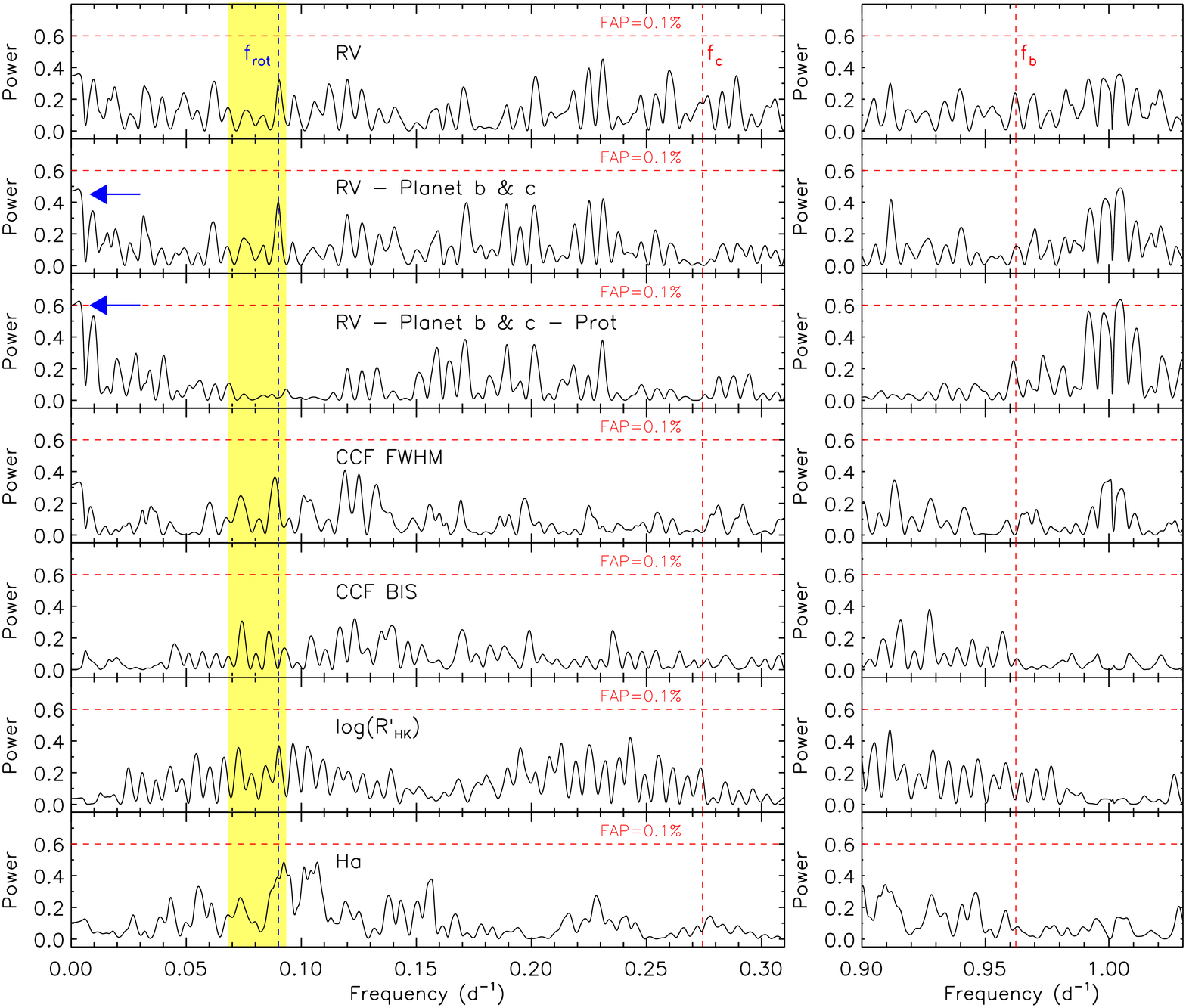}
    \caption{Generalized Lomb-Scargle periodograms of the \sophie\ RV measurements and activity indicators. The right and left columns cover two frequency ranges encompassing the orbital frequencies of \target\,A\,b and c ($f_\mathrm{b}$\,=\,0.96\,d$^{-1}$ and $f_\mathrm{c}$\,=\,0.27\,d$^{-1}$; vertical dashed red lines) and the 68.3\,\% credible interval of the rotation frequency of the star (yellow area), as estimated from the \tess\ light curve. The vertical dashed blue line marks the rotation frequency of the star ($f_\mathrm{rot}$\,=\,0.09\,$^{-1}$, which implies P$_\mathrm{rot}$ $\sim$11.1\,d), as derived from the \sophie\ data. From top to bottom: RV data; RV residuals after subtracting the signals of the two transiting planets; RV residuals after subtracting the signals of the star and of the two transiting planets; FWHM and BIS of the CCF; log\,R$^\prime_\mathrm{HK}$; H$\alpha$. The horizontal dashed red lines mark the 0.1\,\% false alarm probability. The blue arrow marks the excess of power at low frequencies detected in the \sophie\ RVs.}
    \label{fig:RV_periodogram}
\end{figure*}

We performed a frequency analysis of the \sophie\ RV measurements and activity indicators to search for the Doppler reflex motions induced by \target\,A\,b and c, and unveil the presence of possible additional RV signals arising from other orbiting planets and-or stellar activity. We estimate the expected masses of the two planets using the empirical mass-radius relations reported in \citet{Otegi20}. With a radius of $\sim$1.3\,R$_\oplus$, \target\,A\,b is very likely a rocky planet with an expected planetary mass of $\mathrm{M}_{\mathrm{b}}$\,$\approx$\,2.4~$\mathrm{M}_{\oplus}$ and an induced Doppler semi-amplitude of $\mathrm{K}_\mathrm{b}$\,$\approx$\,1.6\,\ms. With a radius of $\sim$3.2\,R$_\oplus$, \target\,A\,c might be a volatile-rich sub-Neptune ($\rho_\mathrm{c}$\,<\,3.3\, $\mathrm{g}\,\mathrm{cm}^{-3}$), or a highly dense ($\rho_\mathrm{c}$\,>\,3.3\, $\mathrm{g}\,\mathrm{cm}^{-3}$) planet rich in iron, silicates and water, with a thin gaseous envelope accounting for no more than a few per cent of the total mass, similarly to TOI-849\,b \citep{Armstrong20}. Therefore, its mass may range from $\mathrm{M}_{\mathrm{c}}$\,$\approx$\,10 to 50~$\mathrm{M}_{\oplus}$, while the predicted radial velocity semi-amplitude ranges from $\mathrm{K}_\mathrm{c}$\,$\approx$\,5 to 23~\ms. Given the median uncertainty of the \sophie\ RV measurements (3\,\ms; Table~\ref{tab:sophie_rvs}), we should be able to detect the Doppler reflex motion induced by \target\,A\,c.

\target\,A is a magnetically active solar-like star with a mean Ca\,{\sc II}\,H\,\&\,K chromospheric activity indicator of log\,R$^\prime_\mathrm{HK}$\,=\,$-4.63$\,$\pm$\,0.05, as derived from the \sophie\ spectra (Table~\ref{tab:sophie_rvs}), and a peak-to-peak photometric variability of $\sim$0.2\,\%, as measured by \tess. It is well known that the presence of magnetically active regions (spots and plages) coupled to stellar rotation can induce periodic and quasi-periodic Doppler signals at the stellar rotation frequency and its harmonics. Using the code \texttt{SOAP2} \citep{Dumusque2014}, we estimated the RV semi-amplitude of the activity-induced signal -- the so-called activity-induced RV jitter -- from the effective temperature, radius, rotation period, and photometric variability of the star. We found that the predicted semi-amplitude of the RV jitter is $\sim$5-10\,\ms. Given the precision of the \sophie\ measurements, the stellar signal is also expected to be detected in our Doppler data.   

\begin{figure*}
    \centering
    \includegraphics[width=\linewidth]{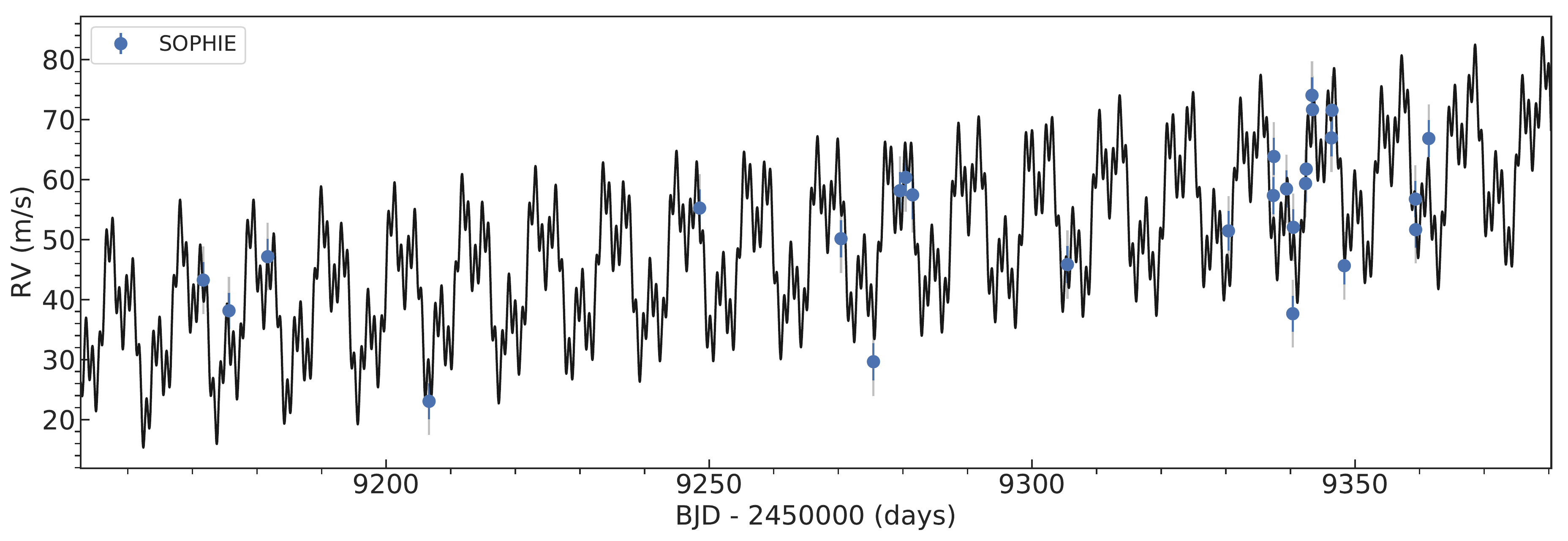}
    \includegraphics[width=0.48\linewidth]{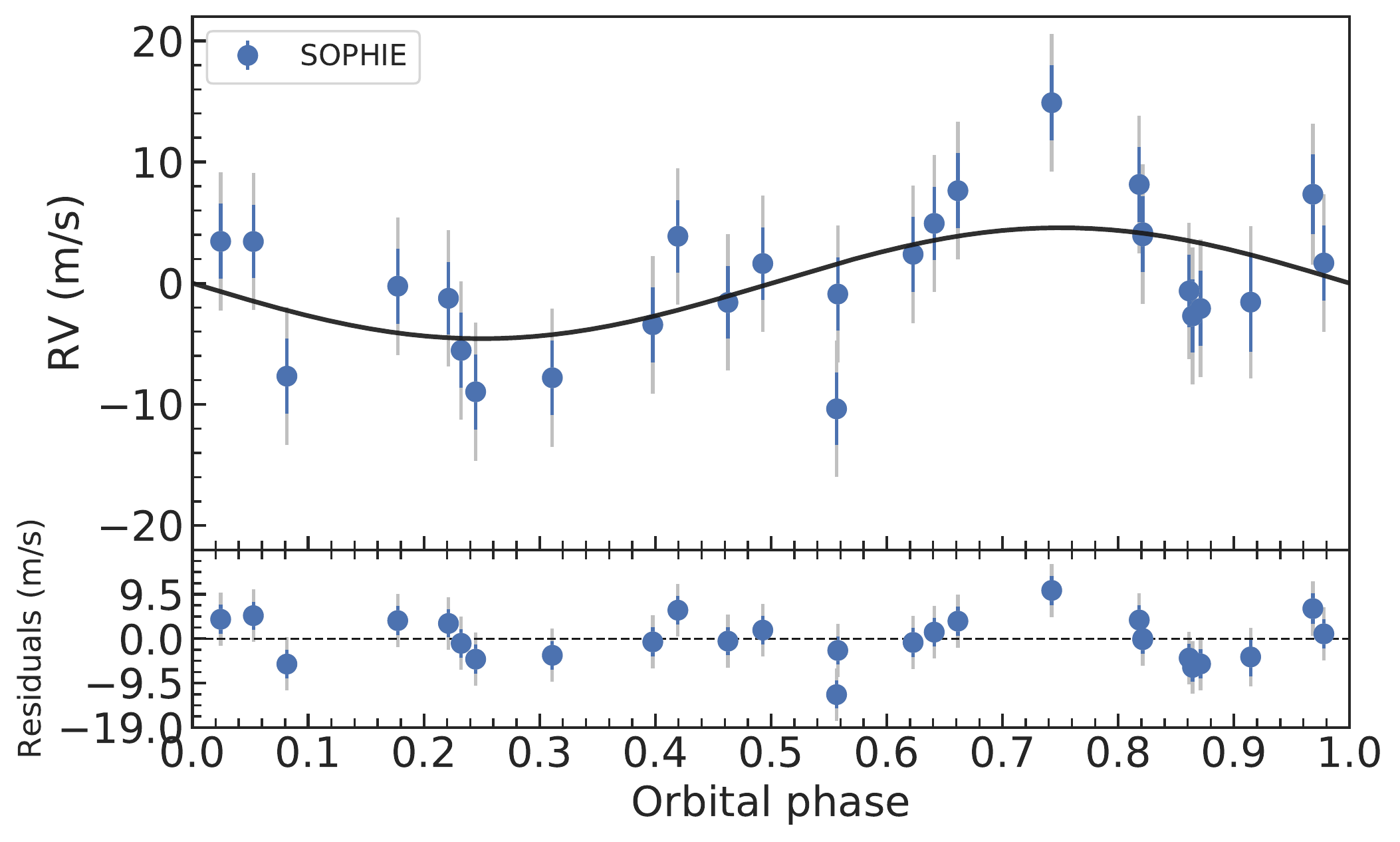}
    \includegraphics[width=0.48\linewidth]{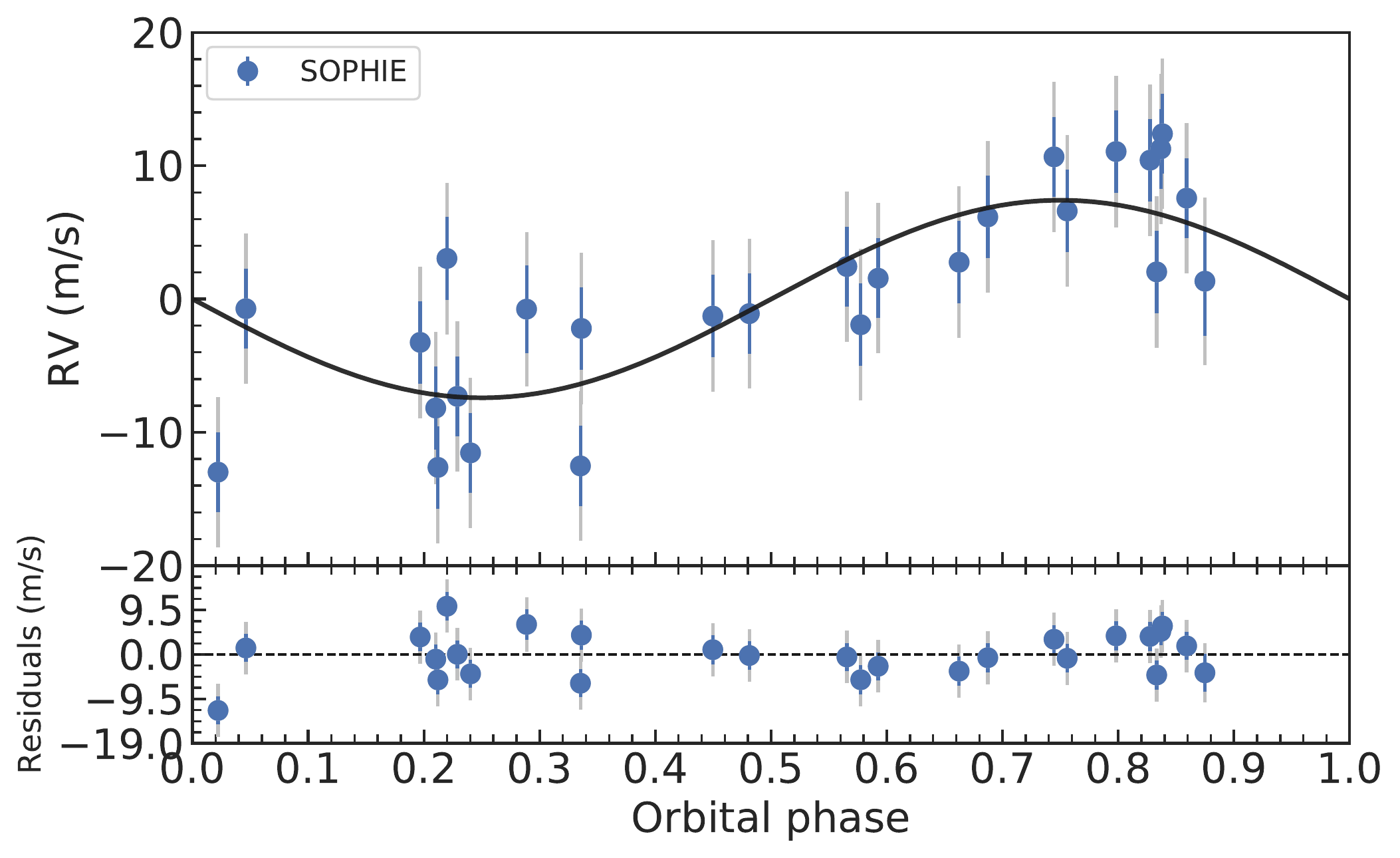}
    \caption{\emph{Upper panel}: \sophie\ RV time series and best fitting model. \emph{Lower panels}: phase-folded RV curve of \target\,A\,b (left), \target\,A\,c (right) and best-fitting Keplerian model. The vertical grey lines mark the error bars including the RV jitter}
    \label{fig:RV_curves}
\end{figure*}

The upper left and right panels of Figure~\ref{fig:RV_periodogram} display the generalized Lomb-Scargle periodogram \citep{Zechmeister09} of the \sophie\ RV measurements for two different frequency ranges encompassing the orbital frequencies of the two planets ($f_\mathrm{b}$\,=\,0.96\,d$^{-1}$ and $f_\mathrm{c}$\,=\,0.27\,d$^{-1}$; vertical dashed red lines). The yellow region marks the 68.3\,\% (1\,$\sigma$) credible interval of the rotation frequency of the star, as estimated from the \tess\ light curve (Sect~\ref{rotation}). The horizontal dashed red line marks the false alarm probability (FAP) of 0.1\,\%, which was estimated using the bootstrap method described in \citet{Murdoch1993} and \citet{Hatzes2019}. The periodogram of the \sophie\ RVs shows no significant peak (FAP\,$<$\,0.1\,\%)\footnote{We adopted a FAP\,$<$\,0.1\,\% for a signal to be significant.} and weak powers at $f_\mathrm{b}$ and $f_\mathrm{c}$. The null detection of the RV signal induced by planet~c might be a consequence of the relatively small number of measurements (27), aliasing effects due to the sparse sampling of our \sophie\ RVs, and the presence of additional, non-accounted for Doppler signals in the data. We note that the periodogram of the \sophie\ RVs shows a peak at 0.09\,d$^{-1}$ ($\sim$11\,d), which falls within the 68.3\,\% credible interval of the rotation frequency of the star, and that might arise from stellar activity (vertical dashed blue line). This is corroborated by the fact that the 0.09\,d$^{-1}$ peak seems to have a possible counterpart in the periodogram of the FWHM of the cross-correlation function (CCF), as well as in the periodograms of log\,R$^\prime_\mathrm{HK}$ and H$\alpha$ (Fig.~\ref{fig:RV_periodogram}). 

\begin{table}
  \footnotesize
  \caption{Bayesian information criterion (BIC) and jitter term of the three different models considered for our RV-only analysis.\label{RV_models}}  
  \centering
  \begin{tabular}{lcc}
  \hline
  \hline
  \noalign{\smallskip}
  Model & Jitter term & BIC  \\
        & (\ms) &    \\
  \noalign{\smallskip}
  \hline
  \noalign{\smallskip}
 two planets & $11.5_{-1.6}^{+2.1}$ &  $-140.4$  \\
  \noalign{\smallskip}  
 two planets + stellar rotation & $9.4_{-1.5}^{+2.0}$ &  $-144.2$  \\
  \noalign{\smallskip}
 two planets + stellar rotation + linear trend & $4.8_{-1.0}^{+1.3}$ & $-171.4$  \\
\noalign{\smallskip}
    \hline
   \end{tabular}
\end{table}

As \target\,A\,b and c are two bona fide, validated planets (see Sect.~\ref{validation}) and the Doppler reflex motion of planet c should be detectable with \sophie, we modelled our RV measurements with two Keplerians using the code \texttt{pyaneti}. We adopted Gaussian priors for the orbital periods and the times of mid-transit, as derived from the joint analysis of the \tess\ and \cheops\ light curves presented in Sect.~\ref{Joint}, and uniform uninformative priors for the K-amplitudes and systemic velocity. We assumed circular orbits and added an ``RV jitter term'' to account for noise not included in the nominal uncertainties. While \target\,A\,b and c remain undetected (1\,$\sigma$ and 2.4\,$\sigma$ level, respectively), we found a high RV jitter term of $\sim$11.5\,\ms\ (Table~\ref{RV_models}), which is $\sim$4 times the median uncertainty of our measurements, suggesting that additional Doppler signals might be present in the data. 

The panels in the second row of Fig.~\ref{fig:RV_periodogram} display the periodogram of the RV residuals. Although we found no peak with FAP\,$<$\,0.1\,\%, there are two important points to note. First, there seems to be an excess of power at frequencies lower than 0.006\,d$^{-1}$, the spectral resolution of our \sophie\ data\footnote{Defined as the inverse of the time base of our observations ($\sim$180\,d).}, suggesting the presence of a long-term trend (blue arrow in Fig.~\ref{fig:RV_periodogram}). Second, the significance of the stellar signal at 0.09\,d$^{-1}$ has increased. We thus proceeded by modelling the  activity-induced RV variation at the star’s rotation period adding a coherent sine-like curve whose period was constrained with a uniform prior centred at P$_\mathrm{rot}$\,=\,12.8\,d, and having a width of 4\,d. For the  phase and semi-amplitude of the activity signal we adopted uniform priors. While we acknowledge that this simple model might not fully reproduce the quasi-periodic variations induced by evolving active regions, it has proven to be effective in accounting for the stellar signal of active and moderately active stars \citep[see, e.g.][]{Hatzes2010, Barragan18, Gandolfi2017, Gandolfi2019}.

\begin{table}
  \footnotesize
  \caption{Best fitting parameters and planetary masses as derived from the analysis of the \sophie\ RV measurements.\label{RV_table}}  
  \centering
  \begin{tabular}{lc}
  \hline
  \hline
  \noalign{\smallskip}
  Parameters & Values  \\
  \noalign{\smallskip}
  \hline
  \noalign{\smallskip}
    \multicolumn{2}{l}{\textbf{Parameters of \target\,A\,b}} \\
    $\mathrm{K}_{\mathrm{b}}$ (\ms) & \kb[] \\
    M$_{\mathrm{b}}$ ($\mathrm{M}_{\oplus}$) & \mpb[] \\
    \noalign{\smallskip}
    \multicolumn{2}{l}{\textbf{Parameters of \target\,A\,c}} \\
    $\mathrm{K}_{\mathrm{c}}$ (\ms) & \kc[] \\
    M$_{\mathrm{c}}$ ($\mathrm{M}_{\oplus}$) & \mpc[] \\
    \noalign{\smallskip}
    \multicolumn{2}{l}{\textbf{Model parameters of activity induced signal}} \\
    \noalign{\smallskip}
    P$_\mathrm{rot}$ (days) & \Pd[] \\
    \noalign{\smallskip}
    $\mathrm{k}_\star$ (\ms) & \kd[] \\
    \noalign{\smallskip}
    \multicolumn{2}{l}{\textbf{Other system parameters}} \\
    Linear trend (m\,s$^{-1}$\,d$^{-1}$) & \ltrend[] \\
    \noalign{\smallskip}
    RV jitter term (\ms) & \jSOPHIE[] \\
    \noalign{\smallskip}
    Systemic velocity (\kms) & \SOPHIE[]\\
    \noalign{\smallskip}
    \hline
   \end{tabular}
\end{table}

The panels in the third row of Fig.~\ref{fig:RV_periodogram} show the periodogram of the RV residuals following the subtraction of the signals arising from two planets and the star. With an FAP\,$<$\,0.1\,\%, we found that the excess of power at low frequencies has become significant (blue arrow in Fig.~\ref{fig:RV_periodogram}). We therefore added a linear trend to the last RV model, which includes two Keplerians plus one sine curve, and refitted the \sophie\ data with \texttt{pyaneti}. Based on the Bayesian information criterion (BIC) and the jitter term, we found that the model with the linear trend is preferred over the other two previous models (Table~\ref{RV_models}).

Figure~\ref{fig:RV_curves} shows the \sophie\ RV time series and the phase-folded RV curves of \target\,A\,b and c, along with the best fitting models. The inferred parameters are given in Table~\ref{RV_table}. With an RV semi-amplitude variation of K$_\mathrm{c}$\,=\,7.4$\pm$1.6\,\ms, \target\,A\,c is significantly detected in the \sophie\ measurements (4.6\,$\sigma$), while \target\,A\,b is not significantly detected in our data (K$_\mathrm{b}$\,=\,4.6$\pm$1.9\,\ms; 2.6\,$\sigma$). The stellar RV signal has an amplitude of 10.4\,$\pm$\,2.1\,\ms, as predicted by \texttt{SOAP2}. We also found evidence of the presence of a linear trend in the data with an acceleration of 0.13$\pm$0.02\,\ms\,d$^{-1}$. Although the periodogram of the FWHM shows an excess of power at low frequencies, the corresponding peak is not significant to claim that this signal arises from stellar activity. The source and periodicity are hard to pinpoint given the $\sim$180-d baseline of our radial-velocity campaign with \sophie. Longer baseline RV campaigns should be performed in order to unveil the true nature of this long period signal.

\section{Discussion and conclusions}
\label{system}

\begin{figure}
    \centering
    \includegraphics[width = 10cm]{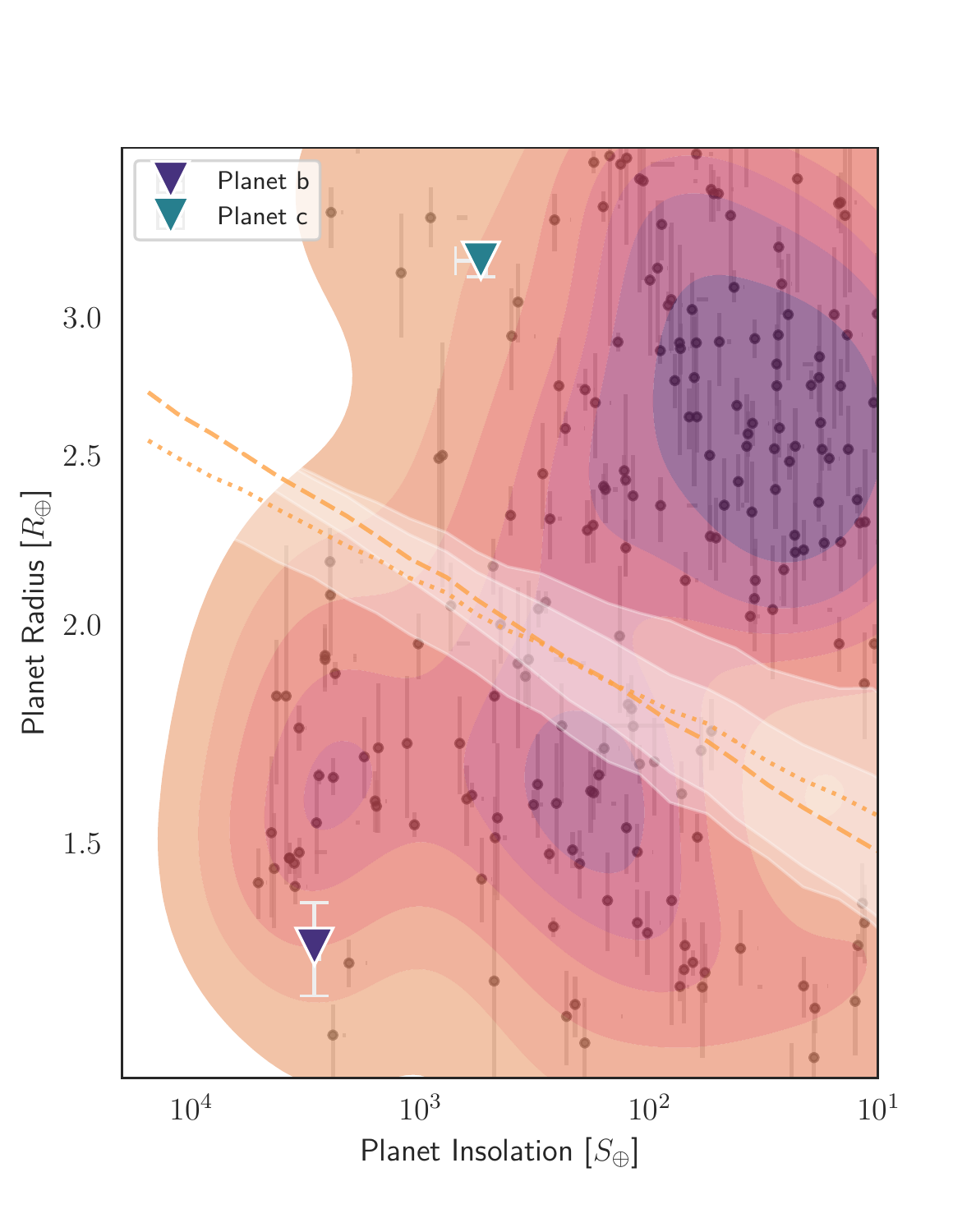}
    \caption{\target\,A\,b \& c (filled triangles) as a function of planetary radius and insolation, compared with the population of exoplanets. Colours represent a kernel density estimation (KDE) applied to small ($\mathrm{R}_\mathrm{p} < 4 \mathrm{R}_{\oplus}$), transiting planets retrieved from the NASA Exoplanet Archive \citep{akeson2013nasa}. The dashed and dotted lines (with associated 1\,$\sigma$ white error bands) show estimates for the position of the evaporation valley from \citet{martinez2019spectroscopic} and \citet{VanEylen18} respectively.}
    \label{insolation}
\end{figure}

\begin{figure}
    \centering
    \includegraphics[width=\linewidth]{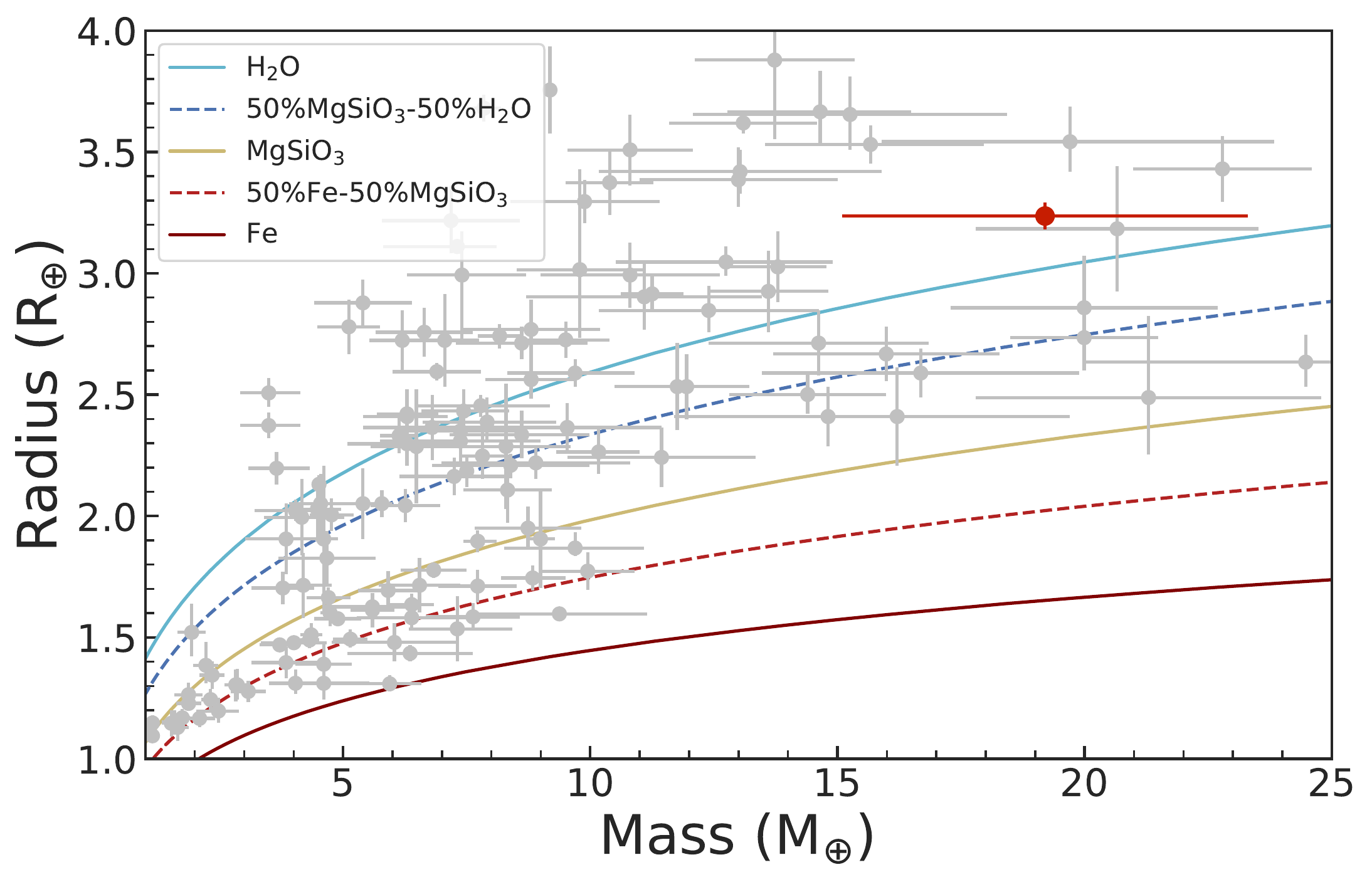}
    \caption{Mass-radius diagram for small planets (R$_\mathrm{p}$\,<\,4\,R$_\oplus$), as retrieved from the catalogue for transiting planets \texttt{TEPCat} \citep{Southworth11}. Planets whose masses and radii are known with a precision better than 22\% are plotted with grey circles. The solid red circle marks the position of \target\,A\,c. Planets in the literature are marked with grey circles. Composition models from \citet{Zeng2016} are displayed with different lines and colours.}
    \label{fig:mass-radius}
\end{figure}

\target\,A is a northern ($\delta$\,$\approx$\,+25$^\circ$), young ($\sim$1.4\,Gyr), visual binary at 82.5\,\,pc from the Sun consisting of a bright (G=9.0) G0\,V star and a $\Delta$G\,=\,7.8 magnitudes fainter M-type dwarf. The two stars have an angular separation of $\sim$5.9$\arcsec$, which corresponds to a sky-projected separation of $\sim$484\,AU. 

The brightest component has a rotation period of P$_\mathrm{rot}$\,=\,12.8\,$\pm$\,1.8\,d (as estimated from the \tess\ light curve), a mass of $\mathrm{M}_{\star} = $\smass, a radius of $\mathrm{R}_{\star} = $\sradius, an effective temperature of T$_\mathrm{eff}$\,=\,\stemp, and an iron content of [Fe/H]\,=\,0.10\,$\pm$\,0.04. \target\,B is an M5\,V star, with an estimated mass of $\sim$0.16\,M$_{\odot}$ and radius of $\sim$0.20\,R$_{\odot}$. If we assume that the orbit of the smaller companion is seen face on and that the eccentricity is zero, we can use Kepler's third law to estimate that the binary has an orbital period of about 8875\,years. 

\target\,A hosts two transiting small planets, namely, \target\,A\,b and \target\,A\,c. The inner planet, \target\,A\,b, is a hot super-Earth with an orbital period of $\sim$1.04\,d, a radius of R$_\mathrm{b}$ =\rpbTESS, a predicted mass of $\mathrm{M}_\mathrm{b}$\,$\approx\,$ 2.4\,$\mathrm{M}_{\oplus}$ \citep[according to the mass-radius empirical relation from][]{Otegi20}, and an equilibrium temperature of $\mathrm{T}_{\mathrm{eq},\mathrm{b}}$\,=\,\Teqb\ (assuming zero albedo). Based on its orbital period and radius, \target\,A\,b can be classified as a small ultra-short period (USP) planet, the first USP planet discovered and validated by \tess\ and \cheops. The outer planet, \target\,A\,c, is a sub-Neptune with an orbital period of $\sim$3.65\,d, a radius of R$_\mathrm{c}$\,=\,\rpcTESS, and an RV-determined mass of $\mathrm{M}_\mathrm{c}$\,=\,19.2\,$\pm$\,4.1\,$\mathrm{M}_{\oplus}$, implying a mean density of $\rho_\mathrm{c}$\,=\,3.1\,$\pm$\,0.7\,g\,cm$^{-3}$. Given the vicinity to its host star, \target\,A\,c is also hot, with an equilibrium temperature of $\mathrm{T}_{\mathrm{eq},\mathrm{c}}$\,=\,\Teqc. With inclinations of $\mathrm{i}_\mathrm{b} = $ \ib\ and $\mathrm{i}_\mathrm{c} = $ \ic, the orbits of the two planets might be coplanar.

\begin{figure*}
    \centering
    \includegraphics[width =0.35 \linewidth]{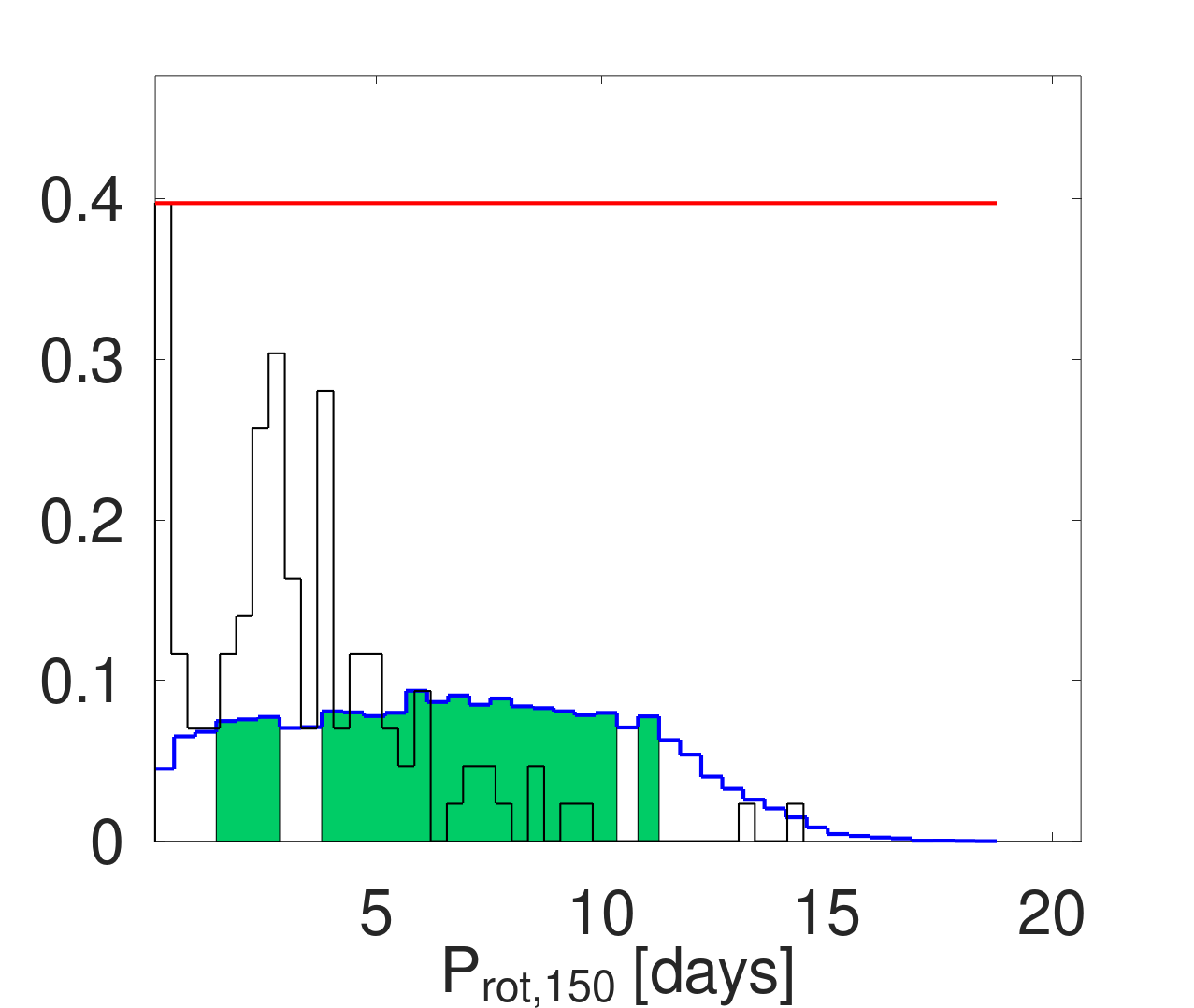}
    \includegraphics[width =1.0 \linewidth]{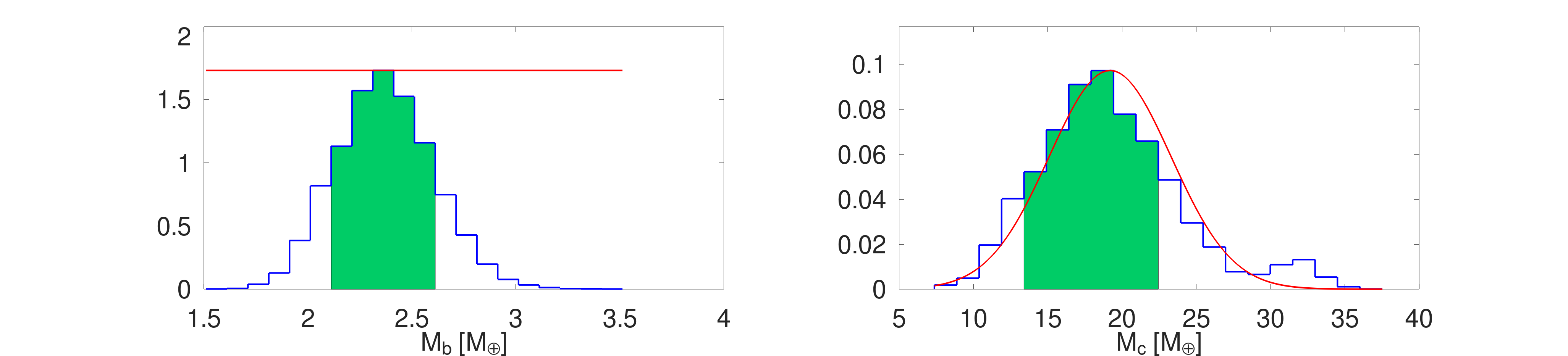}
    \includegraphics[width =1.0 \linewidth]{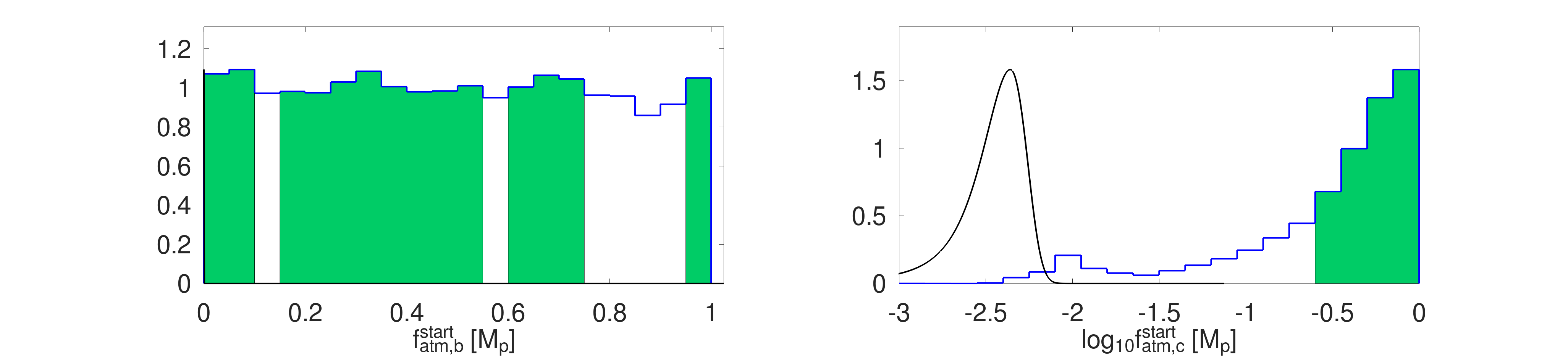}
    \caption{Results of the \texttt{PASTA} modelling. \textit{Upper panel:} Posterior-distribution of the stellar rotation rate at an age of 150\, Myr. The black line represents the distribution measured from a sample of open cluster stars of comparable mass and age \citep{johnstone2015}. \textit{Middle and Lower panels:} Posterior-distributions of planetary mass (middle) and initial atmospheric mass fraction (bottom) for \target\,A\,b and c. The red lines represent the imposed priors (uniform or Gaussian), while in the bottom-right panel the black line shows the current atmospheric mass fraction estimated by \texttt{PASTA} on the basis of the measured planetary mass and radius and estimated equilibrium temperature. We note that the x axis of the bottom-left panel is in linear scale, while that of the bottom-right panel is in logarithmic scale.}
    \label{PASTA}
\end{figure*}

In Fig.~\ref{insolation}, we show the position of the two planets in the planetary radius-insolation scatter plot. \target\,A\,b and c fall on opposite sides of the so-called \emph{radius valley}, a gap identified in the bi-modal distribution of small exoplanet radii \citep{Fulton17, VanEylen18}. This gap is predicted by photo-evaporation models, which suggest that the loss of H/He atmospheres in close-in planets is a consequence of the strong stellar irradiation \citep{Lopez14}. The two planets might have experienced drastically different evolution processes. Due to its semi-major axis $\mathrm{a}_{\mathrm{b}} = $ \ab, which implies a strong irradiation from the star ($\sim$2900\,F$_\oplus$, see Table~\ref{table_results}), the atmosphere of \target\,A\,b probably underwent photo-evaporation \citep{owen2020} and now the planet has no gaseous content. For \target\,A\,c, the position in the radius-insolation plot is insufficient to conclude on the possible presence of an atmosphere. We therefore refer to the mass-radius diagram in Fig.~\ref{fig:mass-radius}, where the planet is marked as a red circle. We notice that \target\,A\,c is located in a low populated area of the diagram. The only close-by planets are TOI-132\,b \citep[$\mathrm{M}_\mathrm{p}$\,=\,22.8\,$\mathrm{M}_{\oplus}$, R$_\mathrm{p}$\,=\,3.4\,$\mathrm{R}_{\oplus}$, P$_\mathrm{p}$\,=\,2.1~d; ][]{Diaz20}, K2-25\,b \citep[$\mathrm{M}_\mathrm{p}$\,=\,24.5\,$\mathrm{M}_{\oplus}$, R$_\mathrm{p}$\,=\,3.4\,$\mathrm{R}_{\oplus}$, P$_\mathrm{p}$\,=\,3.5~d;][]{Stefansson20}, NGTS-4\,b \citep[$\mathrm{M}_\mathrm{p}$\,=\,20.6\,$\mathrm{M}_{\oplus}$, R$_\mathrm{p}$\,=\,3.2\,$\mathrm{R}_{\oplus}$, P$_\mathrm{p}$\,=\,1.3~d;][]{West19} and TOI-824\,b \citep[$\mathrm{M}_\mathrm{p}$\,=\,18.5 $\mathrm{M}_{\oplus}$, R$_\mathrm{p}$\,=\,2.9\,$\mathrm{R}_{\oplus}$, P$_\mathrm{p}$\,=\,1.4~d;][]{Burt20}, for which an atmosphere rich in volatiles has been predicted. We can therefore suppose that \target\,A\,c is surrounded by a primary (therefore H/He-dominated) atmosphere, an hypothesis corroborated by the mean density of the planet ($\rho_\mathrm{c}$\,=\,3.1\,$\pm$\,0.7\,g\,cm$^{-3}$) being significantly lower than that of Earth and by its position in the mass-radius diagram above the 100\,\% H$_2$0 composition. 

To confirm our predictions on the atmospheric content of both planets and to derive an independent constraint on the mass of \target\,A\,b, we employed the tool \texttt{P}lanetary \texttt{A}tmospheres and \texttt{S}tellar Ro\texttt{T}ation R\texttt{A}tes \citep[\texttt{PASTA};][]{bonfanti21b}. Starting from the measured system parameters, \texttt{PASTA} returns posterior probability distributions for the initial atmospheric mass fraction of each planet and for the evolution of the stellar rotation rate. To this end, \texttt{PASTA} uses the connection existing between the rotation rate and high-energy (mainly X-ray and EUV) emission of late-type stars and, through this, traces the impact of the stellar high-energy emission on planetary atmospheric escape throughout the system lifetime.

We found that \target\,A\,c can only host an atmosphere if at young ages the stellar rotation period was shorter than the average value measured for stars in young open clusters \citep{johnstone2015}. This is shown in the top panel of Fig.~\ref{PASTA}, which presents the posterior probability distribution of the stellar rotation rate at an age of 150\,Myr derived by \texttt{PASTA}. Although this posterior does not have a narrow peak, it clearly favours rotation evolutionary tracks slower than those measured for stars member of young open clusters that we use as a comparison: if at a young age the host star was rotating fast (emitting a larger amount of X-ray and EUV radiation), \target\,A\,c would have completely lost its atmosphere, in contradiction to its measured average density. 

Using the constraint on the evolution of the stellar rotation rate, \texttt{PASTA} simultaneously provides a constraint on the mass of \target\,A\,b. We find that the posterior probability distribution for the mass of \target\,A\,b is centred around 2.4\,$\mathrm{M}_{\oplus}$ (left-middle panel of Figure~\ref{PASTA}). Combined with the measured planetary radius, this value leads to an Earth-like density, and thus it confirms our prediction that \target\,A\,b has completely lost its primary atmosphere through escape.

Finally, \texttt{PASTA} enables one to constrain the initial atmospheric mass fraction ($f^{\rm start}_{\rm atm}$), which is the mass of a planetary atmosphere (in planetary masses) at the time of the dispersal of the protoplanetary nebula. The posterior probability distributions of $f^{\rm start}_{\rm atm}$ for both planets are shown in the bottom panels of Figure \ref{PASTA}. A \target\,A\,b has completely lost its primary atmosphere at some point in the past, \texttt{PASTA} is unable to constrain the initial atmospheric mass fraction. For \target\,A\,c, instead, \texttt{PASTA} derives an initial atmospheric mass fraction peaking around 0.15 planetary masses, which is about 50 times higher than the estimated current atmospheric content (about 0.004 planetary masses; black line in the bottom-right panel of Figure \ref{PASTA}).

Using the stellar photospheric abundances of Mg, Si, and Fe and the stoichiometric model presented in \citet{Santos15, Santos17}, we estimated the iron mass fraction ($f_{\mathrm{iron}}$) of the planet building blocks assuming that the host star composition reflects the composition of the protoplanetary disk. For $f_{\mathrm{iron}}$ we obtained a value of 34.4\,$\pm$\,2.8\,\%, which is larger than that of our Earth $\sim$32\,\% \citep{McDonough03} and of the planet building blocks of our Solar System (33.2\,$\pm$\,1.7\,\%) estimated with the aforementioned stoichiometric model. Recently, \citet{Adibekyan21a} showed that there is a statistically significant correlation between $f_{\mathrm{iron}}$ and the density of terrestrial planets normalized to the density of a planet with an Earth-like composition \citep{Dorn17}. The normalization accounts for the dependence of planet's density on the planet mass for the same composition. This finding was further confirmed by \citet{Adibekyan21b} based on a slightly larger sample. Using the aforementioned relation between the normalized density ($\rho / \rho_{\mathrm{Earth-like}}$) and $f_{\mathrm{iron}}$ we estimated the $\rho / \rho_{\mathrm{Earth-like}}$ of HD\,93963\,b to be 1.01$\pm$0.34. This value combined with the planet radius allowed us to estimate the mass of 2.9$^{+1.3}_{-1.2}$ $M_{\mathrm{\oplus}}$, which is slightly larger but within the uncertainties consistent with the one computed with \texttt{PASTA} and the mass predicted using the mass-radius relation of \citet{Otegi20}. We could not constrain the iron composition of the two planets, although in \citet{Adibekyan21a} the authors found that the iron mass fraction of potentially rocky planets is usually larger than that expected from the host star composition.

\target\,A\,b belongs to the family of
small (R$_\mathrm{p}$\,=\,1\,--\,2\,R$_\oplus$) USP planets, high-density rocky objects ($>$\,4.5\,g\,cm$^{-3}$) with occurrence rates ranging from $\sim$1.1\,\% for M-type dwarfs to $\sim$0.15\,\% for F-type dwarfs \citep{Winn18}. Small USP planets are thought to be stripped cores of former warm Neptunes and mini-Neptunes with radii between 2 and 4 R$_\oplus$, which underwent photo-evaporation processes \citep{Lundkvist16, Sanchis14}. As we demonstrated with our analysis, this is likely the case for \target\,A\,b, which is expected to have completely lost its primary atmosphere. In addition, similarly to most USP small planets, \target\,A\,b belongs to a packed multi-planet system, which includes a second planet on a 3.65\,d orbit, and our \sophie\ RV measurements show a linear trend (Sect~\ref{RV_analysis}), suggesting the possible presence of an additional, long-period planet orbiting the star.

Ultra-short period planets likely formed further out in the system, before migrating to the close-in orbit where it is now. \citet{Petrovich19} proposed a secular migration scenario in which the planets of the systems are initially on highly eccentric and highly mutually inclined ($>$\,30$^{\circ}$) orbits. As migration begins, the eccentricities are damped out on timescales from kyrs to Myrs (although the current value could still be well different from zero), due to tidal dissipation within the planets, while the orbital inclinations are damped out on timescales of Gyrs by the dissipation inside the host star \citep{Winn18}. As a result the planets of the system reached stabilization on highly mutually inclined and-or highly eccentric orbits. This model could work for \target\,A if the planets were on highly eccentric orbits ($\gtrsim 0.1$) and it has been shown to work for many systems hosting USP planets \citep{Dai18, Kamiaka}. On the other hand, \target\,A might be a coplanar system, and it therefore might have migrated through another secular process, as reported by \citet{Petrovich19} in their Section~5.2.2. According to this migration channel, the planets were in orbits with low eccentricities and low mutual inclinations when they started migrating. The migration process happens in a non-violent way, with the eccentricities damped to zero by tidal forces and small inclination fluctuations. After a few Gyrs the planets in the system should be nearly coplanar, comprising a USP planet and a second companion with period shorter than 10\,d. Only a small number of systems with USP planets have reached these final conditions: TOI-561 \citep{Lacedelli20}, TOI-500 \citep{Serrano22}, CoRoT-7 \citep{Leger09, Hatzes2010, Haywood, Faria}. Unlike this small subset of systems, \target\,A is a young star, meaning that for this model to work the planets should have already been on coplanar or almost coplanar orbits when they started migrating. Compared to other multi-planet systems with USP planets, \target\,A has a relatively close spacing between the orbits of the inner two planets (comparable to CoRoT-7 and Kepler-42). This limits the possible extent of the migration of planet~b, both because stability considerations set an outer limit for the orbit of planet~b (it must have been far enough from \target\,A\,c at 3.65\,d), and because the available angular momentum deficit in the system that could drive tidal migration would have been limited. The latter problem can be ameliorated if additional planets exist on exterior orbits.

In this context, it is also interesting to consider the possible influence of the binary companion on the system dynamics. While the system at present is extremely hierarchical, an inclined binary companion would attempt to impose Lidov--Kozai cycles on the planets \citep{Lidov62,Kozai62}. If the binary is on a circular orbit at its projected separation of 484\,AU, the timescale for these to act on \target\,A\,c is \citep[see][Eq.~9.48]{Valtonen06} $\sim$1\,Gyr, comparable to the system age, and would be longer if the binary were currently at pericentre on a wider orbit. Because of the stronger coupling between the two planets, Lidov--Kozai eccentricity cycles should be suppressed, with the planets maintaining coplanarity while becoming misaligned with the stellar equator, as in the similar system of 55 Cnc \citep{Kaib11}. The timescale for this misalignment to be generated would be considerably reduced if additional undetected planets existed in the system. A measurement of misalignment between the stellar spin and the planetary orbits, via the Rossiter--McLaughlin effect, would therefore suggest the presence of one or more planets beyond \target\,A\,c.

\target\,A\,b is the first small USP planet confirmed and validated by \tess\ and \cheops. 
Further observations of the \target\,A system have the potential to yield new discoveries. It is a compact planetary system around a bright star that is easily accessible from the northern hemisphere. For instance, it can be observed from November until June with HARPS-N, the high-precision spectrograph mounted at the Telescopio Nazionale Galileo in La Palma (Spain). An intensive RV campaign with a spectrograph capable of reaching a precision of $\sim$1\,\ms\ would improve the precision on the mass of \target\,A\,c, measure the mass of \target\,A\,b, constrain the eccentricities of both planets, detect additional planetary companions, and monitor the long-term trend in order to unveil its origin (stellar or planetary). Such observations would enable the study of the system architecture, providing the  ``ingredients'' necessary to constrain the migration process. Since \target\,A is a relatively young star ($\sim$1.4\,Gyr) hosting at least two small planets, we would be able to impose stronger constraints on the time scale of the migration process.

The precise RV follow up would finally allow us to add \target\,A to the relatively small sample of multi-planet systems hosting a small USP planet with both masses and radii measured, which currently counts 14 members \citep[K2-141, TOI-561, Kepler-42, K2-106, K2-229, Kepler-407, 55\,Cnc, LTT\,3780, WASP-47, Kepler-32, Kepler-10, \corot\ -7, HD\,3167 and TOI-500; see, e.g. the \texttt{TEPCat} catatalogue,][for references]{Southworth11}.
Increasing this sample is fundamental to better constrain the theory of formation and migration of exotic objects such as USP planets and, more in general, of tightly packed multi-planet systems. Especially for \target\,A, the measurement of the planetary masses and eccentricities will allow us to perform stability analysis and understand whether the low-eccentricity migration process described in Sect.~\ref{system} could explain its current architecture.

\begin{acknowledgements}
CHEOPS is an ESA mission in partnership with Switzerland with important contributions to the payload and the ground segment from Austria, Belgium, France, Germany, Hungary, Italy, Portugal, Spain, Sweden, and the United Kingdom. The CHEOPS Consortium would like to gratefully acknowledge the support received by all the agencies, offices, universities, and industries involved. Their flexibility and willingness to explore new approaches were essential to the success of this mission. We acknowledge the use of public TESS data from pipelines at the TESS Science Office and at the TESS Science Processing Operations Center. Resources supporting this work were provided by the NASA High-End Computing (HEC) Program through the NASA Advanced Supercomputing (NAS) Division at Ames Research Center for the production of the SPOC data products.This research has made use of the NASA Exoplanet Archive, of the Exoplanet Follow-up Observation Program website, which are operated by the California Institute of Technology, under contract with the National Aeronautics and Space Administration under the Exoplanet Exploration Program. 
LMS \& DG gratefully acknowledges financial support from the CRT foundation under Grant No. 2018.2323 ``Gaseous or rocky? Unveiling the nature of small worlds''. 
SH gratefully acknowledges CNES funding through the grant 837319. 
ABr was supported by the SNSA. 
S.G.S. acknowledge support from FCT through FCT contract nr. CEECIND/00826/2018 and POPH/FSE (EC). 
Some of the observations in the paper made use of the High-Resolution Imaging instrument ‘Alopeke obtained under Gemini LLP Proposal Number: GN/S-2021A-LP-105. ‘Alopeke was funded by the NASA Exoplanet Exploration Program and built at the NASA Ames Research Center by Steve B. Howell, Nic Scott, Elliott P. Horch, and Emmett Quigley. Alopeke was mounted on the Gemini North (and/or South) telescope of the international Gemini Observatory, a programme of NSF’s OIR Lab, which is managed by the Association of Universities for Research in Astronomy (AURA) under a cooperative agreement with the National Science Foundation. on behalf of the Gemini partnership: the National Science Foundation (United States), National Research Council (Canada), Agencia Nacional de Investigación y Desarrollo (Chile), Ministerio de Ciencia, Tecnología e Innovación (Argentina), Ministério da Ciência, Tecnologia, Inovações e Comunicações (Brazil), and Korea Astronomy and Space Science Institute (Republic of Korea). 
ACC and TW acknowledge support from STFC consolidated grant number ST/M001296/1. 
MF and CMP gratefully acknowledge the support of the Swedish National Space Agency (DNR 65/19, 174/18). 
This project was supported by the CNES. 
unding for the Stellar Astrophysics Centre is provided by The Danish National Research Foundation (Grant agreement no.: DNRF106). 
This work makes use of observations from the LCOGT network. LCOGT telescope time was granted by NOIRLab through the Mid-Scale Innovations Program (MSIP). MSIP is funded by NSF. 
YA and MJH acknowledge the support of the Swiss National Fund under grant 200020\_172746. 
We acknowledge support from the Spanish Ministry of Science and Innovation and the European Regional Development Fund through grants ESP2016-80435-C2-1-R, ESP2016-80435-C2-2-R, PGC2018-098153-B-C33, PGC2018-098153-B-C31, ESP2017-87676-C5-1-R, MDM-2017-0737 Unidad de Excelencia Maria de Maeztu-Centro de Astrobiologí­a (INTA-CSIC), as well as the support of the Generalitat de Catalunya/CERCA programme. The MOC activities have been supported by the ESA contract No. 4000124370. 
S.C.C.B. acknowledges support from FCT through FCT contracts nr. IF/01312/2014/CP1215/CT0004. 
XB, SC, DG, MF and JL acknowledge their role as ESA-appointed CHEOPS science team members. 
PC thanks the LSSTC Data Science Fellowship Program, which is funded by LSSTC, NSF Cybertraining Grant  Number 1829740, the Brinson Foundation, and the Moore Foundation. 
her participation in the programme has benefited this work. 
A.De. acknowledges support from the European Research Council (ERC) under the European Union's Horizon 2020 research and innovation programme (project {\sc Four Aces}, grant agreement No. 724427), and from the National Centre for Competence in Research ``PlanetS'' supported by the Swiss National Science Foundation (SNSF). 
The Belgian participation to CHEOPS has been supported by the Belgian Federal Science Policy Office (BELSPO) in the framework of the PRODEX Program, and by the University of Liège through an ARC grant for Concerted Research Actions financed by the Wallonia-Brussels Federation. L.D. is an F.R.S.-FNRS Postdoctoral Researcher. 
This work was supported by FCT - Fundação para a Ciência e a Tecnologia through national funds and by FEDER through COMPETE2020 - Programa Operacional Competitividade e Internacionalizacão by these grants: UID/FIS/04434/2019, UIDB/04434/2020, UIDP/04434/2020, PTDC/FIS-AST/32113/2017 \& POCI-01-0145-FEDER- 032113, PTDC/FIS-AST/28953/2017 \& POCI-01-0145-FEDER-028953, PTDC/FIS-AST/28987/2017 \& POCI-01-0145-FEDER-028987, O.D.S.D. is supported in the form of work contract (DL 57/2016/CP1364/CT0004) funded by national funds through FCT. 
B.-O.D. acknowledges support from the Swiss National Science Foundation (PP00P2-190080). 
This project has received funding from the European Research Council (ERC) under the European Union’s Horizon 2020 research and innovation programme (project {\sc Four Aces} grant agreement No 724427). 
ZG was supported by the VEGA grant of the Slovak Academy of Sciences number 2/0031/22, and by the Slovak Research and Development Agency - the contract No. APVV-20-0148. 
M.G. is an F.R.S.-FNRS Senior Research Associate. 
CXH's work was funded by the Australian Government through the Australian Research Council. 
KGI is the ESA CHEOPS Project Scientist and is responsible for the ESA CHEOPS Guest Observers Programme. She does not participate in, or contribute to, the definition of the Guaranteed Time Programme of the CHEOPS mission through which observations described in this paper have been taken, nor to any aspect of target selection for the programme. 
This work was granted access to the HPC resources of MesoPSL financed by the Region Ile de France and the project Equip@Meso (reference ANR-10-EQPX-29-01) of the programme Investissements d'Avenir supervised by the Agence Nationale pour la Recherche. 
ML acknowledges support of the Swiss National Science Foundation under grant number PCEFP2\_194576. 
PM acknowledges support from STFC research grant number ST/M001040/1. 
This work is partly supported by JSPS KAKENHI Grant Number JP17H04574, JP18H05439, JST CREST Grant Number JPMJCR1761, the Astrobiology Center of National Institutes of Natural Sciences (NINS) (Grant Number AB031010). This article is based on observations made with the MuSCAT2 instrument, developed by ABC, at Telescopio Carlos Sánchez operated on the island of Tenerife by the IAC in the Spanish Observatorio del Teide. 
This work was also partially supported by a grant from the Simons Foundation (PI Queloz, grant number 327127). 
Acknowledges support from the Spanish Ministry of Science and Innovation and the European Regional Development Fund through grant PGC2018-098153-B- C33, as well as the support of the Generalitat de Catalunya/CERCA programme. 
GyMSz and ZG acknowledges the support of the Hungarian National Research, Development and Innovation Office (NKFIH) grant K-125015, a PRODEX Institute Agreement between the ELTE E\"otv\"os Lor\'and University and the European Space Agency (ESA-D/SCI-LE-2021-0025), the Lend\"ulet LP2018-7/2021 grant of the Hungarian Academy of Science and the support of the city of Szombathely. 
V.V.G. is an F.R.S-FNRS Research Associate. 
\end{acknowledgements}
\bibliography{Submitted}

\begin{appendix}

\section{\sophie\ data}

In Table~\ref{tab:sophie_rvs} we report the radial velocity data of \target\,A obtained with \sophie\ instrument. The last two rows correspond to the observations acquired using the high-efficiency mode of the instrument. 

\begin{table*}[]
    \centering
    \small
    \caption{Radial velocity measurements of \target\,A obtained with the \sophie\ spectrograph.}
    \label{tab:sophie_rvs}
    \begin{tabular}{cccccccccccc}
    
\hline
\hline
 BJD$_{\text{UTC}}$ & RV & $\sigma_{\text{RV}}$	& FWHM & CCF & BIS & BERV & S/N$_{35}$\tablefootmark{(a)} & log\,R'$_{\text{HK}}$ & $\sigma_{\text{logR'}_{\text{HK}}}$ & H$\alpha$ & $\sigma_{H\alpha}$ \\
 
 -2\,450\,000 & [km\,s$^{-1}$] & [km\,s$^{-1}$] & [km\,s$^{-1}$]& Contrast &[km\,s$^{-1}$] & [km\,s$^{-1}$] & --- & [dex] & [dex] & & \\
\hline     
59171.68254 	&13.2679	&0.0030	&8.6855	&34.1512	&0.0012		&28.7697			&53.9   & -4.6654   	&0.0196 &  0.13196   &   0.00166 \\
59175.66388 	&13.2628	&0.0030	&8.6765	&34.0729	&0.0247		&28.9868			&53.6   & -4.7554   	&0.0226     &  0.12919   &   0.00166\\
59181.65817 	&13.2718	&0.0030	&8.6754	&34.0541	&-0.0094	&29.0167			&53.2   & -4.5595   	&0.0152 &  0.12723   &   0.00167 \\
59206.64221 	&13.2477	&0.0030	&8.6524	&34.3134	&-0.0117	&25.7727			&52.7   & -4.6746 &0.0173      &  0.13081   &   0.00170 \\
59248.53304 	&13.2799	&0.0031	&8.7397	&33.7943	&-0.0121	&10.2290			&52.5   & -4.6587   	&0.0165     &  0.13047   &   0.00170\\
59249.63983 	&13.3083	&0.0063	&8.7450	&36.0736	&-0.0380	&9.5008				&23.6   & -4.7387   	&0.0689      &  0.14718   &   0.00448\\
59270.42489 	&13.2748	&0.0031	&8.7155	&33.8769	&0.0037		&-0.5020			&53.5   & -4.6352   	&0.0170      &  0.13084   &   0.00162\\
59275.45107 	&13.2543	&0.0031	&8.6965	&33.9194	&0.0067		&-3.0603			&53.8   & -4.6000   	&0.0149    &  0.13073   &   0.00161\\
59279.54811 	&13.2828	&0.0031	&8.7407	&34.0541	&-0.0019	&-5.2748			&54.6   & -4.5855   	&0.0152      &  0.13238   &   0.00158\\
59280.42807 	&13.2850	&0.0031	&8.7600	&34.0278	&0.0103		&-5.4861			&55.7   & -4.5926 	&0.0167      &  0.13302   &   0.00157\\
59281.51218 	&13.2821	&0.0041	&8.6802	&33.8415	&0.0305		&-6.1773			&39.7   & -4.6969   	&0.0339     &  0.13615  &   0.00237\\
59305.47805	&13.2705	&0.0031	&8.7279	&33.8180	&0.0068		&-16.8856	&52.6   & -4.6080   	&0.0153      &  0.13069   &   0.00170\\
59330.40744	&13.2761	&0.0033	&8.7347	&32.6086	&0.0019		&-24.7656	&52.2   & -4.6377  	&0.0179  &  0.12923  &   0.00178\\
59337.36266	&13.2820	&0.0031	&8.7330	&34.2586	&-0.0013	&-26.1512	&54.2   & -4.6061 	&0.0157   &  0.13331   &   0.00163\\
59337.37065	&13.3472	&0.0242	&8.8127	&29.2515	&0.1733		&-26.1674	&8.1    & --- 	& ---   &  0.19168   &   0.01843\\
59337.44689	&13.2885	&0.0031	&8.7328	&34.0806	&0.0004		&-26.3059	&55.3   & -4.6052 	&0.0176   &  0.13198   &   0.00159\\
59339.40069	&13.2831	&0.0031	&8.7437	&34.1450	&0.0162		&-26.5831	&54.1   & -4.7147  	&0.0205 &  0.13187  &   0.00158\\
59340.37111	&13.2623	&0.0030	&8.7039	&34.2688	&-0.0040	&-26.6942	&54.4   & -4.6096  	&0.0173 &  0.12764   &   0.00156\\
59340.45879	&13.2767	&0.0030	&8.6820	&34.1901	&0.0133		&-26.8428	&54.3   & -4.6907   	&0.0192     & 0.12710   &   0.00156\\
59342.35182	&13.2840	&0.0030	&8.6893	&34.3234	&-0.0050	&-26.9677	&53.8   & -4.6020 	&0.0183   &  0.12749   &   0.00154\\
59342.45045	&13.2864	&0.0030	&8.7028	&34.1657	&-0.0008	&-27.1381	&54.9   & -4.6747   	&0.0179   &  0.12910   &   0.00159\\
59343.34600	&13.2987	&0.0030	&8.6969	&34.2927	&-0.0046	&-27.1000	&53.5   & -4.6417 	&0.0197  &  0.12641 &   0.00153\\
59343.42224	&13.2963	&0.0030	&8.7109	&34.2236	&-0.0212	&-27.2383	&55.0   & -4.6762   	&0.0178      &  0.12947   &   0.00160\\
59346.35068	&13.2916	&0.0031	&8.7590	&33.9897	&0.0240		&-27.4908	&53.3   & -4.5538  	&0.0135 &  0.13407   &   0.00168\\
59346.44088	&13.2962	&0.0031	&8.7341	&33.9766	&-0.0063	&-27.6404	&53.8   & -4.5034   	&0.0138    &  0.13278   &   0.00164\\
59348.34707	&13.2703	&0.0031	&8.7316	&34.1953	&0.0122		&-27.6960	&53.6   & -4.6247  	&0.0163  &  0.13109   &   0.00164\\
59359.35052	&13.2814	&0.0030	&8.7334	&34.0813	&-0.0156	&-28.2705	&54.6   & -4.6766  	&0.0201  &  0.13350   &   0.00158\\
59359.39217	&13.2763	&0.0030	&8.6973	&34.1965	&0.0069		&-28.3322	&54.6   & -4.6081 	&0.0175    &  0.12921   &   0.00156\\
59361.42563	&13.2915	&0.0031	&8.7418	&33.5236	&0.0081		&-28.3681	&54.9   & -4.6372   	&0.0235     &  0.12542  &   0.00155\\
\hline
59267.42645\tablefootmark{(b)}	&13.2457	&0.0031	&10.3382	&30.3026	&0.0021	&1.0045		&127.8	  &-4.5529 &0.0032   & 0.12614    &  0.00061\\
59269.41278\tablefootmark{(b)}	&13.2503	&0.0026	&10.3402	&30.4552	&0.0047	&0.0255		&156.1	  &-4.5655 &0.0023  & 0.12439    &  0.00048\\
 \hline
    \end{tabular}
 \tablefoot{\tablefoottext{a}{Signal-to-Noise ratio at order 35. } \tablefoottext{b}{Observations acquired using high-efficiency mode. }  } 
\end{table*}

\end{appendix}

\end{document}